\documentclass[dissertation, copyright, numsections, gsmodern, final]{uothesis}
\usepackage[english,UKenglish]{babel}
\usepackage{afterpage,geometry}
\usepackage{amsmath,amssymb,amsthm,amsfonts,dsfont,ragged2e}

\newcommand{\phiclub}{\phi^{\mbox{\fontsize{5}{0}$\clubsuit$}}}
\newcommand{\had}{\hat{a}^\dagger}

\newcommand{\ha}{\hat{a}}
\newcommand{\hbd}{\hat{b}^\dagger}
\newcommand{\hb}{\hat{b}}

\newcommand{\vac}{|{\rm vac}\rangle}
\newcommand{\vacl}{|{\rm vac}\rangle \langle{\rm vac}|}
\newcommand{\ket}[1]{\left| #1 \right\rangle}
\newcommand{\bra}[1]{\left\langle #1 \right|}
\newcommand{\braket}[2]{\left\langle #1 | #2 \right\rangle}
\newcommand{\ain}{a_{{\rm in}}}
\newcommand{\bin}{b_{{\rm in}}}
\newcommand{\bout}{b_{{\rm out}}}
\newcommand{\aout}{a_{{\rm out}}}
\newcommand{\adin}{a^\dagger_{{\rm in}}}
\newcommand{\bdin}{b^\dagger_{{\rm in}}}
\newcommand{\Proj}[1]{| #1\rangle\!\langle #1 |}

\newcommand{\expect}[1]{\left\langle#1\right\rangle}
\newcommand{\eea}{\end{eqnarray}}
\newcommand{\bea}{\begin{eqnarray}}
\newcommand{\ee}{\end{equation}}
\newcommand{\be}{\begin{equation}}

\newcommand{\sch}{Schr\"{o}dinger }
\newcommand{\lettersection}[1]{\emph{#1.---}}
\newcommand{\appropto}{\mathrel{\vcenter{
  \offinterlineskip\halign{\hfil$##$\cr
    \propto\cr\noalign{\kern2pt}\sim\cr\noalign{\kern-2pt}}}}}

    \newcommand{\idh}{\hat{\mathds{1}}}
\DeclareUnicodeCharacter{000E}{~}

\covertitle{Fundamental Limits to Single-Photon Detection}
\abstracttitle{Fundamental Limits to Single-Photon Detection}
\author{Tzula B. Propp}
\department{Department of Physics}
\narrowdepartment{Department of Physics}
\degreetype{Doctor of Philosophy}
\degreemonth{September}  
\degreeyear{2020}
\advisor{Steven J. van Enk}
\chair{Benjamin McMorran}
\committee{Brian Smith} 
\committeetwo{Miriam Deutsch}
\committeethree{David Wineland}
\committeefour{Jeff Cina}

\graddean{Kate Mondloch}
\abstract{Quantum mechanics cements the intimate relationship between the nature of light and its detection. Historically, quantum theories of photodetection have generally fallen into two categories: the first tries to determine what quantum field observable is measured when photoelectrons are detected, laying the theoretical groundwork for photodetection being possible. The second type are phenomenological theories, which take great care to model the details of specific photodetectors. In this dissertation, we fill in the gap between these two models in the modern literature on photodetection by constructing a fully quantum-mechanical and sufficiently realistic model that includes all stages of the photodetection process. We accomplish this within the framework of quantum information theory using the language of positive operator valued measures (POVMs). 

A POVM provides the most general description of a quantum measurement. In the context of single-photon detection, the photodetector POVM provides a complete characterization of the single-photon detector (SPD) from which all figures of merit can be calculated. Each element of the POVM is comprised of a weighted sum over single-photon state projectors. To construct SPD POVMs, we identify the states projected onto by a measurement and the associated weights. We first accomplish this for a simple time-dependent system, laying out how to efficiently project onto arbitrary single-photon states. Then, we describe the three stages of a realistic SPD: transmission, amplification, and measurement.

In the transmission stage of photodetection a photon enters a network of discrete energy levels, its energy propagating through that network and finally exiting into another output continuum of modes, a process entirely described by a single complex function.  In the amplification stage of photodetection, a single excitation is amplified into a macroscopic signal via a nonlinear amplification mechanism. In the final stage, an inefficient ``classical'' measurement is made on the macroscopic signal. By combining these three stages we form a chain of inference from the final ``click'' to the input photon and construct a POVM projecting onto the input Hilbert space. Lastly, we discuss limits and tradeoffs that arise at each stage and implications for photodetection applications.

This dissertation contains previously published and unpublished material.

}

\school{University of Oregon, Eugene, Oregon}
\school{College of Wooster, Wooster, Ohio}

\degree{Doctor of Physics, 2020, University of Oregon}
\degree{Bachelor in Physics, Philosophy, 2015, College of Wooster, Ohio}

\interests{Theoretical Quantum Optics and Quantum Information Theory}

\position{Graduate Teaching Assistant, University of Oregon, 2015-2019}
\position{Graduate Research Assistant, University of Oregon, 2016-2020}
\position{Science Literacy Program Fellow, University of Oregon, Fall 2017 \& 2019}
\position{Adjunct Faculty Instructor, Lane Community College, 2019-2020}
\position{Northstar Project Teacher, University of Oregon 2016-2019}

\position{Queer Caucus Chair, University of Oregon GTFF, 2018-2020}
\position{LGBT+ in STEM Executive Board, University of Oregon, 2017-2020}

\award{John R. Moore Scholarship, University of Oregon, 2020}
\award{Most Informative Talk, University of Oregon Women in Graduate Science Science Slam, 2018}
\award{First Place, Oregon Statewide 3 Minute Thesis Championship, 2017}
\award{First Place, University of Oregon 3 Minute Thesis Competition, 2017}
\award{Physics Department Teaching Award, University of Oregon, 2017}

\publication{Propp, \relax{Tz}. B \& \relax{van Enk}, S. J. (2019). On nonlinear amplification: improved quantum limits for photon counting. \emph{Optics Express} 27, \textbf{16}, 23454-23463.}

\publication{Propp, \relax{Tz}. B \& \relax{van Enk}, S. J. (2019). Quantum networks for single photon detection. \emph{Physical Review A}, \textbf{100}, 033836.}

\publication{Propp, \relax{Tz}. B \& \relax{van Enk}, S. J. (2020). How to project onto an arbitrary single-photon wavepacket. Submitted to \emph{Physical Review A}.}

\acknowledge{I would like to thank my advisor Dr. Steven van Enk for his mentorship in completing this dissertation, for supporting me through the hardest years of graduate school, and for inspiring me to strive to be the best and happiest theoretical physicist I can possibly be. I also want to thank my former advisor and committee chair Dr. Ben McMorran for believing in me and helping instill in me the value of outreach, along with PhD committee member Dr. Dave Wineland for invaluable revisions. I would like to thank Andrea Goering, Brandy Todd, and Benjam\'{i}n Alem\'{a}n for supporting me, providing mentorship, and enabling my physics outreach endeavors. Similarly, I want to thank Dr. Elly Vandegrift who opened my eyes to the science of science education, and whose mentorship has been instrumental to my development as both teacher and learner. I would like to acknowledge the leadership of the Graduate Teaching Fellows Federation, as well as University of Oregon Lesbian, Gay, Bisexual, and Trans+ Education Support Services (LGBTESS) coordinator Haley Wilson for their institutional support. Special thanks are due to Saumya Biswas who--in his own words--did not get in the way, along with our collaborators in the DARPA DETECT program for thought-provoking discussions. I would like to acknowledge the Kalapuya tribe, on whose stolen land this research was conducted. Lastly, I want to thank my queer chosen family and friends for being there for me through the toughest times, particularly V\'{e} Gulbransen, Hales Wilson, and Liana Clark as well as Deepika Sundarraman, Hayley Shapiro, Christianna Hannegan, Alice Greenberg, Abby Pauls, Caden Valencia, and many others. This research was supported financially in part by DARPA contract No. W911NF-17-1-0267 and the Emanuel OMQ Scholarship. }

\dedication{

\phantom{blank filler text.}

I dedicate this dissertation to my beloved cat King Ubu who is also known as Small Bean Black Bean Moon Bean Sun Bean Bean-O'Noire Tiabanie Beanie Baby Sneaky Bean and The Bean.

\vspace{10em}

\vspace{10em}

\noindent This dissertation is the product of unionized labor as part of the Graduate Teaching Fellows Federation, AFT Local 3544.
}

\graphicspath{{./figures/}}

\begin{document}
\maketitle
\chapter{Introduction}

\section{History}

\begin{figure}[ht]
\centering
\includegraphics[width=.5\textwidth]{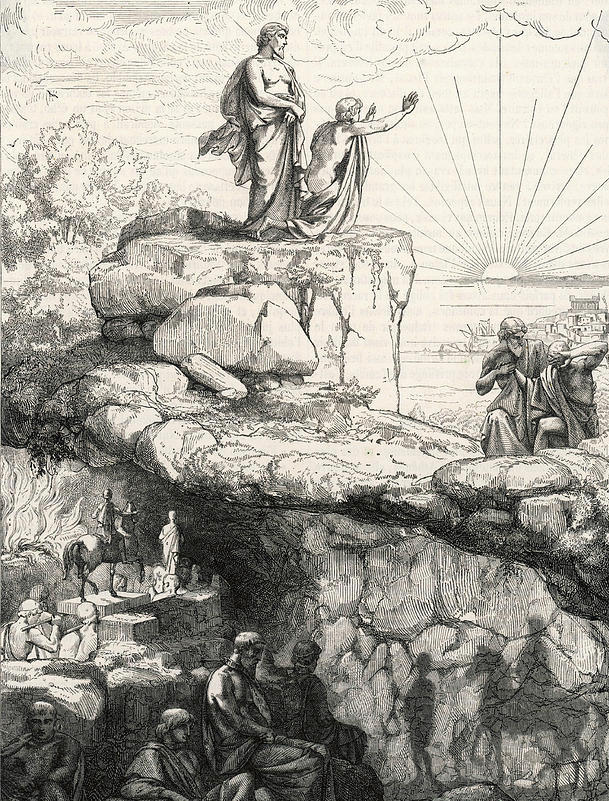}
\caption{\label{cave}Plato's allegory of the cave \cite{plato}, as illustrated on the cover of \emph{Diagnosis: interpreting the shadows} \cite{Croskerry2017}. By improving our measurements, we can improve our understanding of the universe even if the ultimate level of direct access to reality (for Plato, looking directly at the sun) is unattainable.}
\end{figure}

Measurements are our window to the universe. Since the ancient Greeks, we have accepted that we only have access to our perceptions, not direct access to the \emph{phusis} or physical reality. The statement that we can only know what we can measure has withstood the test of time, and is now enshrined and made explicit in quantum theory's separation of measurement and the quantum state. Yet even though no measurement can reveal the complete underlying reality (if such an underlying reality even exists), we can improve our measurements to better reveal physical reality in analogy to Plato's famous Allegory of the Cave  \cite{plato}, depicted in Fig. \ref{cave}; in quantum theory, it is as if we are trying to learn about light (the quantum state) by studying the shadows different objects cast (projective measurements). This is necessary, as there is a real sense in which our measurements constitute the most foundational layer of the universe to which we have direct access. In this way, if we wish to learn more about the structure and constituents of physical reality, our study should include consideration of our measurement schemes themselves.

The study of light and its measurement date back to at least $500$ to $300$ BCE \cite{Euclid}. These early works by Ptolemy and Euclid were preserved and expanded upon in the Islamic world with the works of Ibn Sahl on curved mirrors and lenses in $984$ CE \cite{Hogendijk2003}, Ibn Al-Haytham on the correct identification of vision as a three-part process involving a light source, a reflected object, and the eye between $1028$ and $1038$ CE \cite{Jacobs2012}, and Kamāl al-Dīn al-Fārisī on the mathematically rigorous description of the rainbow phenomena in $1309$ CE \cite{khan2007}. These concepts were introduced to Europe during the Renaissance, where the well-developed science of spectacle construction (curved lenses) was used by Hans Lippershay in $1608$ to create the first telescope, an invention which was subsequently refined by Johannes Kepler and modified to magnify nearby objects (that is, the invention of the microscope) in the subsequent three years \cite{King1955}. These developments in our ability to measure and manipulate light paved the way for a series of early experiments in the mid-1600's, most famously by Isaac Newton and Christiaan Huygens, trying to elucidate the nature of light which was believed to be a particle by the former and a wave by the latter. These experiments were built upon in the following century and a half by Thomas Young (wave theory: interference), Etienne Louis Malus (wave theory: polarization), Augustin-Jean Fresnel (wave theory: polarization and interference), Simeon-Denis Poisson (particle theory: shadows and attempting to refute Fresnel) and Francois Arago (wave theory: diffraction and refuting Poisson). During this same time period, parallel research by  Hans Christian {\O}rstedand and later Michael Faraday in unifying electricity and magnetism. This all culminated in the unifying work by James Clerk Maxwell, who proved in 1864 that light is an electromagnetic wave with a speed exactly determined by the electromagnetic permittivity and permeability of free space \cite{Zubairy2016}. 

Despite the rigorous theoretical foundation of the wave theory of light and the tremendous predictive success of classical physics\footnote{In 1900, Lord Kelvin is quoted as saying, “There is nothing new to be discovered in physics now. All that remains is more and more precise measurement” \cite{Zubairy2016}.}, the wave theory of light alone failed to explain several phenomena including the photoelectric effect\footnote{The photoelectric can in fact be explained semiclassically, but this was discovered later \cite{Bosanac1998}.} and the Rayleigh–Jeans ultraviolet catastrophe for blackbody radiation. The resolution of these mysteries required the elucidation of wave-particle duality, wherein both the particle and wave properties of light are understood as differing aspects of a single quantum theory \cite{Zubairy2016}. The definitive evidence for wave-particle duality was the discovery of the Compton effect in 1923, wherein an X-ray changes momentum after colliding with a target \cite{Compton1923}. While the wave nature of light was already well established as essential to quantum theory, the Compton effect made it clear that light must not only be quantized but also exhibit the properties of classical particles as well at sufficiently small wavelengths. 

The development of the modern understanding of light and its detection is simultaneous with the advent of quantum theory in the early 1900's \cite{Einstein}. In the low-energy limit (that is, below $1.02 MeV$ where a photon can spontaneously produce an electron and positron), it is convenient to represent light in the Fock basis so that it is comprised of photons, the quanta of the electromagnetic field\footnote{At higher energies, interaction terms lead to non-commutativity between number operators and the Hamiltonian so that Fock states cease to be energy eigenstates \cite{Peskin1995}.}. Not all quantum states of light have a definite integer number of photons, either because of unknown correlations with an environment (for instance, a thermal distribution of photons) or because of quantum necessity (for instance, coherent states of light, such as those produced by a laser). In these cases, the state projected onto is not a pure Fock state but a mixture (or a coherent superposition) of Fock states\footnote{Similarly, a measurement that retrodicts a non-Fock input state may rely on a Fock state as an intermediate stage, such as counting photons in homodyne and heterodyne measurements \cite{Yuen1978,Collett1987}.}. However as the quantum degree of freedom of the electromagnetic field, any pure measurement that projects onto Fock states will only retrodict\footnote{To retrodict is to utilize present information or ideas to infer or explain a past event or state of affairs \cite{MerriamWebster2009}.} an integer number of photons even though the internal state of the photodetector may be partially unknown (mixed); that is, there may be \emph{many} internal photodetector states corresponding to a single detector outcome.

The lowest non-zero integer is one, so single-photon detection is the intensity-frontier limit to detection of photon Fock states. For this reason, we expect the fundamental quantum limits to general photodetector figures of merit to be manifest in quantum descriptions of single-photon detectors (SPDs). In this dissertation, we will construct what is (to our knowledge) the first completely quantum description of the entire photodetection process in a SPD in order to identify the fundamental limits and tradeoffs inherent to single-photon detection.

In addition to the pure intellectual merit, answering the question “What are the fundamental limits to single-photon detection?” is important for industry applications of SPD-based technology. Recently, there has been progress in achieving single photon detection with unprecedented performances in standard photodetector figures of merit \cite{hadfield2009}: timing jitters or temporal uncertainties in the tens of picoseconds in superconducting nanowire SPDs \cite{Rosfjord2006,sunter2018}, dark count rates on the order of a single dark count per day \cite{marsili2013,wollman2017uv}, and robustness to thermal fluctuations nearing room temperature performance \cite{dowling2018,leonard2019}. This is in no small part due to the work done by the DARPA DETECT collaboration, one of the aims of which is to jumpstart the next generation of photodetection technology.  Exploration of fundamental limits and tradeoffs inherent to single-photon detection is essential to guiding this work, so experimentalists can have ideas of what they will come up against and decisions can be made about which areas of tradespace are important to explore. Furthermore, they enable accountability; if a research group claims their photodetector can simultaneously perform well in metric X and metric Y, and we know metric X and Y are not simultaneously attainable from our generic model, we have good reason to question their claims. While the DARPA DETECT collaboration has resulted in a plethora of device-specific phenomenological theories of photodetection  \cite{young2018,sunter2018,dowling2018,Zhang2018,jahani2020}, the universal limits to SPDs are not explicit in such models (even though they are present). To construct a model where the limits themselves are revealed (our goal), we must answer the following two questions: What are the essential components of an SPD? What are the fundamental limits of each component's performance? We will answer these two questions and construct a model where the fundamental limits to SPD performance are manifest. However, it first behooves us to review photodetection theory in the modern era.

The advent of a quantum theory of photodetection is simultaneous with the advent of quantum theory, and has its origins in Einstein's Nobel prize-winning study of the photoelectric effect \cite{Einstein}. Here for the first time the notion that light can only be absorbed (and thus detected) in discrete packets of energy or quanta was rigorously incorporated into the foundation of a theory, which was accomplished by building upon the more ad hoc treatment by Planck \cite{planck1914}. With this, Einstein was able make statistical arguments for rates of atomic transitions in the two-level approximation in the presence and absence of an external electromagnetic field. In this way, Einstein was able to describe the known processes of spontaneous emission and absorption, as well discover stimulated emission \cite{Einstein2}. This was done without a relativistic quantum treatment of the electromagnetic field itself, the methods for which did not exist at the time. Instead, Einstein made use of knowledge of the Planck blackbody spectrum in order to demand agreement with quantum theory. In a photodetector approximated as a two level system, the Einstein rate equations are directly connected to the efficiency with which a photon is absorbed and with how long it stays in the detector before leaking back out (which can be connected to a gain factor in an amplification scheme such as electron shelving \cite{dehmelt1975}). 

Although Einstein was able to correctly predict the rates of emission and absorption for a two-level system, it was not until work by Dirac in 1927 (and, throughout the next two decades, Wignerl Oppenheimer, Fermi, Bloch, Weisskopf, Tomonoga, Schwinger, and Feynman) formulating a theory of quantum electrodynamics (QED) that a complete quantum treatment of the electromagnetic field interacting with matter (here, only the electron) was developed \cite{dirac1927,Wigner1927,Oppenheimer1930,fermi1932,Bloch1937,Weisskopf1939,Tomonaga1946,Schwinger1948,Feynman1949}. One of the key developments was a quantum treatment of the vacuum, clarifying spontaneous emission as stimulated emission due to vacuum fluctuations by Wigner and Weisskopf \cite{Weisskopf1930}. Another key development was a quantum method for treating the continuum of electromagnetic field modes using Dirac-normalized creation and annihilation operators. These well-known physicists, along with others, contributed to the extension of QED in the development (and predictive success) of relativistic quantum field theory, laying the foundation for modern particle physics which has dominated the popular news spotlight ever since. Meanwhile and behind the scenes, another revolution was brewing as the the initial question of an electromagnetic field interacting with matter was further explored with the new methods of QED. 

In 1959, Hanbury Brown and Twiss published a controversial paper showing that, for a few-photon signal collected from the star Sirius, an interference effect was observable in intensity from which the angular size of Sirius was calculable \cite{hanburrybrowntwiss1956}.  Purcell was quick to point out that this interference can be understood as a manifestation of boson statistics applied to counting photons \cite{Purcell1956}. However, the underlying mechanism behind the Hanbury Brown-Twiss effect was partial coherence, a property of classical waves that had yet to be understood at the quantum level. 

In his seminal work in 1963 \cite{glauber1963}, Glauber connected the quantized formulation of the electromagnetic field to experimentally measurable correlations using the language of QED, introducing the field intensity correlation as a measure of quantum coherence and calculating transition probabilities correlated with the detection of individual photons. We will illustrate the method now, focusing on a single polarization $\mu$ of the electromagnetic field and following \cite{steckquoptics}. The first step is to separate the quantized (that is, operator-valued) electric field into positive and negative frequency components 

\bea
\hat{E}_\mu(\vec{r},t)= \hat{E}_\mu^+(\vec{r},t) + \hat{E}_\mu^-(\vec{r},t)\label{electric1}
\eea where $\mu$ denotes the polarization we are describing and we note $\hat{E}_\mu^+(\vec{r},t) ^\dagger = \hat{E}_\mu^-(\vec{r},t)$. We can represent the operator  $\hat{E}_\mu^+(\vec{r},t)$ as a sum over discrete   plane wave modes with frequencies $\omega_j$ 

\bea
\hat{E}_\mu^+(\vec{r},t) = - \sum_j  \sqrt{\frac{\hbar \omega_j}{2\epsilon_0}} e^{\vec{k}_j\cdot \vec{r} - i \omega_j t}\hat{\epsilon}_\mu \hat{a}_j  \label{electric2} 
\eea with $\hat{\epsilon}_\mu$ the component of the polarization vector, $\vec{k}_j = \vec{p}/\hbar$ the wave number proportional to the momentum $\vec{p}$, $\epsilon_0$ the permittivity of free space, $\hbar$ Planck's reduced constant, $i=\sqrt{-1}$, and $\hat{a}_j$ the annihilation operator corresponding to the mode $j$. Since $[\hat{a}_j,\hat{a}^\dagger_j]=1$, the positive and negative electric field components will not commute. 
We can associate a photodetection event with a transition in the state of the field $\ket{i}\rightarrow\ket{f}$. For idealized single photon detection, the difference between the initial and final state is that one photon in mode $j$ was removed through an interaction at location $\vec{r}$ and time $t$. For each particular initial field configuration, the transition probability is described by unitary evolution 

\bea\label{electric3}
\bra{f} \hat{U}(i\rightarrow f) \ket{i} &\propto& \bra{f} \hat{E}_\mu^+(\vec{r},t)   \ket{i} \nonumber \\ 
&\propto&   \bra{i} \hat{E}_\mu^-(\vec{r},t)  \hat{E}_\mu^+(\vec{r},t)   \ket{i} 
\eea where proportionality with the full sum in (\ref{electric2}) achieved due to only a single term contributing to the to the matrix element and the sum over polarizations $\mu$ is implicit. Since the initial state $\ket{i}$ is arbitrary, we can carry out a sum over initial states and perform an average so that the full transition probability is proportional to

\bea\label{electric4}
P(t) \propto   \langle \hat{E}_\mu^-(\vec{r},t)  \hat{E}_\mu^+(\vec{r},t) \rangle
\eea where the angled brackets indicate an average or expectation value. This expression for the photodetection probability motivated Glauber to define a field correlation function. (\ref{electric4}) is a special case of the first-order coherence

\bea\label{coherence1}
G^{(1)}(\vec{r}_1,t_1;\vec{r}_2,t_2) = \langle \hat{E}_\mu^-(\vec{r}_1,t_1)  \hat{E}_\mu^+(\vec{r}_2,t_2) \rangle
\eea

By rewriting the electric field in terms of positive and negative frequency components promoted to operators, Glauber was able to calculate measures or degrees of coherence of the quantum field in terms of a series of coherence functions, each measuring correlations at some order of the electric field operators. Both (\ref{electric4}) and (\ref{coherence1}) are normally ordered, so that, for the vacuum state, $P(t)=G^{(1)}(\vec{r}_1,t_1;\vec{r}_2,t_2) = 0$. First-order coherence described by (\ref{coherence1}) is analogous to classical coherence in standard optics, but Glauber showed that quantum systems can also exhibit higher order correlations and coherences, making rigorous the theoretical foundation for the maser and laser. Critically, Glauber also introduced the coherent state representation for quantum states. These coherent states are minimum uncertainty states in the quantum field quadratures and are denoted by a single complex number identifying both the amplitude and phase of the electromagnetic field (which are precisely what defines a particular laser field). As the ``most classical'' states, the coherent states form a bridge between quantum and classical descriptions of light. 

To this day, the Glauber model of photodetection, given in terms of field correlations and point-like interactions between a quantized electromagnetic signal and a photoionization system (or other quantum electrical system) as in (\ref{electric4}), provides the foundation for modern photodetection theory. Although it is mathematically rigorous, the Glauber model is highly idealized as a practical theory of measurement; critically, Glauber merely posits that one is able to count photoelectrons without describing any subsequent mechanisms. This work was soon expanded to incorporate less-idealized descriptions of the electromagnetic field-photodetection interaction \cite{Mandel1959,kelley1964}, with later additions to the theory also incorporating the back action of the detector on the detected quantum field \cite{scully1969,yurke1984,ueda1999,schuster2005,clerk2010}. Throughout, a challenge to photodetection theory has been describing the interaction between a discrete system (from a photon's perspective, a system of discrete bosonic states) and the full continuum of bosonic states of the electromagnetic field\footnote{See, for instance, the Jaynes-Cummings model, describing only a single mode of the electromagnetic field's interaction with a two-level system \cite{Jaynes1963}.}. In the intermediate years, numerous quantum optical techniques have been developed for describing the continuum of states, some of which we will discuss in detail in the next chapter. 

We will end this section with a brief survey of modern photodetecting platforms amenable to single photon detection. (For in-depth review, see Refs.~\cite{hadfield2009,Eisaman2011}.)

Photoionization, the production of a current due to incident radiation, is the method of photodetection longest studied quantum mechanically and has as its fundamental theory the photoelectric effect \cite{MandelWolf}. First constructed by Hertz in 1887 \cite{Hertz1887}, the photomultiplier tube (PMT) is a cathode and vacuum-tube device that uses the photoelectric effect coupled with secondary emission to amplify signals as small as an single incident photon into a macroscopic signal. Despite its quantum mechanical foundation and use in counting photons, it is possible to generate a ``classical'' theory of a PMT where the external quantum field is treated as a perturbation to which the PMT responds, achieving a new steady-state. In this case, it is straightforward to calculate the statistics of photodetection events \cite{Mandel1959}; the probability of detecting $n$ photons between a time $t$ and $T$ is calculated

\bea\label{mandelform}
P(n,t,T) = \int_0^\infty \frac{1}{n!} W^n e^{-W} {\cal P}(W) dW 
\eea with 

\bea\label{mandelform2}
W = \eta \int_t^{t+T} I(t') dt' 
\eea where $\eta$ is the photodetection efficiency, $I(t)$ is the intensity of the electromagnetic field, and ${\cal P}(W)$ is the quasi-probability distribution of the integrated intensity. While highly idealized, this expression can be included into more realistic semiclassical models \cite{Nadarajah2007}, and alone rapidly gives useful predictions (even for a somewhat generic photodetector). For instance, for a thermal distribution of light one finds that (\ref{mandelform}) yields a Bose-Einstein distribution for photodetection events \cite{MandelWolf}. However, this semiclassical formula does not describe the truly quantum nature of multi-photon detection as made explicit in Hanburry Brown and Twiss' work \cite{hanburrybrowntwiss1956}.  

Another widespread SPD platform is the avalanche photodiode (APD), invented by Jun-ichi Nishizawa in 1952 \cite{japan2011}. Like the PMT, APDs function based on the photoelectric effect. Whereas in an light-emitting diode (LED) electrons and holes annihilate converting current into light, in an APD an incident photon with energy greater than the material band gap generates an electron-hole pair, resulting in a current \cite{Cova1996}. Generally, APDs can only operate in single-photon detection mode at the expense of significant reset or dead time as the circuit must be quenched unless an array is used \cite{Cova1996,Zappa2007}, so for an individual APD pixel within an array Mandel's formula (\ref{mandelform}) only applies for $n\leq 1$ (higher order counts are highly suppressed). However, for single photon detection we are indeed interested in the $n\leq 1$ regime and Mandel's formula will have validity.

Other single photon detecting platforms are less straightforward, but all share a generic feature; the presence of a photon changes the state of the system, either by shifting the equilibrium state or temporarily moving the system out of equilibrium. Examples of the former include quantum dot field transistor detectors \cite{Rao2005} and superconducting transition-edge sensors (TES) \cite{Lita2008}, both of which involve a measurable change in system conductance due to absorption of a photon. (It is especially drastic for TES, the system is taken out of the superconducting state!) Since the equilibrium state has changed, resetting these mechanisms is resource intensive and SPDs built on these principles have a longer dead time ($\sim 10\mu s$) compared to newer technologies \cite{hadfield2009}. An interesting example of temporarily moving a system out of equilibrium are superconducting nanowire single photon detectors (SNSPDs) \cite{Goltsman2001,marsili2013,wollman2017uv,Korzh2020}, which in recent years have seen a rising popularity due to their low dark count rates ($\sim 10 Hz$) and dead times ($\sim 1 ns$). However, while theoretical modeling based on time-dependent Ginzburg-Landau theory has produced qualitative agreement with experiment, quantitate models for SNSPDs remain elusive \cite{Vodolazov2019}. In part, this is due to the complexity of system dynamics; in an SNSPD a photon is collectively absorbed by many electrons and phonons, resulting in a hot spot in which vortex anti-vortex pairs form with local variations in the superconductivity. This results in a measurable voltage when the vortex anti-vortex pair crosses the wire. 

It is worth noting that one method of detecting single photons has been around for thirty million years, namely, the African clawed frog \emph{Xenopus laevis} \cite{frog}. This species can detect single photons of visible light with moderate ($~30\%$) efficiency and, as living organisms, they do this at room temperature. This has motivated other participants in the DARPA DETECT program to develop bio-inspired SPD platforms \cite{leonard2019}. Although these will suffer from larger dark count rates due to the increased thermal noise at higher temperatures, the hope is that they still may have high enough signal-to-noise ratio to be useful in room temperature experiments.

\section{Very Brief Introduction to Quantum Information and POVMs}

During the late 20th century and on into the 21st century, a revolution has been taking place in quantum science; technological advancements are allowing for increasing control over environmentally isolated single quantum systems \cite{Wineland1979,Diedrich1989,Itano1990,Warren1993,Monroe1995,lloyd2000,Rabitz2000,Chu2002,Dowling2003,Vandersypen2005,haroche2013,wineland2013}. Simultaneously, physicists were increasingly realizing the limits of classical computation both in terms of scalability and applicability to quantum systems \cite{Feynman1982}; Moore's law describing the size of transistors \emph{must} break down when we approach the size of single atoms. And if we are interested in simulating quantum systems, shouldn't we utilize a quantum system itself to do so \cite{Feynman1982}? Early pioneers of quantum information science were quick to discover problems where a quantum algorithm provided an exponential speedup \cite{1992Deutsch,1994Shor} and the so-called second quantum revolution was born \cite{Dowling2003}. In many quantum algorithms, an essential ingredient is measurement of an internal two-level system or qubit whose state encodes information (generically, the answer to some input query with some known probability). Thus the need for measurement theory, the rigorous theoretical foundation for describing quantum measurement.

Measurement is a ubiquitous concept in quantum theory: examples include measurements of an electron's position or momentum (limits to the simultaneous performance of which are enforced by Heisenberg's uncertainty principle \cite{Heisenberg1927}), measurement of the population of a multi-level system (such as an atom, simple harmonic oscillator, optical cavity, or quantum dot), or detection of light through the use of a photo detector (as we will explore in this thesis). The most general quantum description of a measurement is in terms of a positive operator-valued measure (POVM), a set of positive operators $\hat{\Pi}_k$ that sum to the identity, where each $k$ corresponds to a different measurement outcome.  Given an arbitrary quantum state $\hat{\rho}$ the probability to obtain outcome $k$ is given by the Born rule
\be\label{bornrule}
P(k)={\rm Tr}(\hat{\rho}\hat{\Pi}_k)
\ee where ${\rm Tr}(\hat{O})$ is the trace of the operator $\hat{O}$.

Generically, each POVM element $\hat{\Pi}_k$ can be written as a weighted sum over projectors onto orthogonal quantum states
\bea \hat{\Pi}_k = \sum_i w_{i}^{(k)} \ket{\phi_i^{(k)}}\bra{\phi_i^{(k)}} \label{POVM1}
\eea reducing to an ideal projective von Neumann measurement only when the sum reduces to a single term with its weight $w^{(k)}$ equal to unity \cite{von1932}. 
The weight $w_i^{(k)}$ equals the conditional probability to obtain measurement outcome $k$ given input state $\ket{\phi_i^{(k)}}$. The \emph{posterior} conditional probability that, given an outcome $k$, we project onto input $i$ is given by Bayes' theorem  \cite{Bayes63}
\bea P(i|k)= \frac{w_{i}^{(k)}  P(i)}{P(k)} \label{bayes}
\eea 
with $P(k)=\sum_i w_{i}^{(k)}$. Here, $P(k)$ is \emph{a priori} probability for observing measurement outcome $k$, and and $P(i)$ is the \emph{a priori} probability for the $i$th input state $\ket{\phi_i^{(k)}}$ to be present \cite{kraus1983}. Through Bayes' theorem, an experimentalist is able to \emph{retrodict} the likely input state or states; they can update the probability distribution over the possible (past) inputs \emph{if and only if they know what measurement they actually perform}.

In the next chapter, we discuss applications of POVMs to single-photon detection. First, we will review general properties of POVMs in the remainder of this section, before giving an overview of the structure, format, and goals of this dissertation in the final section of this chapter. 

The POVM has several interesting properties it is worthwhile to review and note: firstly, a POVM forms a partition of unity. If one sums over all measurement outcomes (including the null outcome), from the Born rule one finds $\sum_k {\rm Tr}(\hat{\rho}\hat{\Pi}_k) = 1$ for any state $\hat{\rho}$; in other words, regardless of the input quantum state we can be sure that \emph{some} outcome will occur since the probabilities of all outcomes necessarily sum to unity! 

Although the POVM is a complete description of a measurement, we note that alone it is insufficient to determine the post-measurement quantum state. For this we must additionally define Kraus operators $\hat{A}_k$ satisfying $\hat{A}_k^\dagger \hat{A}_k = \hat{\Pi}_k$ \cite{kraus1983}. Note that the POVM does not uniquely specify $\hat{A}_k$, as for any unitary $\hat{U}$ the substitution $\hat{A}_k \rightarrow \hat{U} \hat{A}_k$ leaves the POVM unchanged. Having defined Kraus operators, the normalized post-measurement state $ \hat{\rho}'$ is calculated

\bea \hat{\rho}'=\frac{\hat{A}_k \hat{\rho}\hat{A}_k^\dagger}{{\rm Tr}(\hat{A}_k \hat{\rho}\hat{A}_k^\dagger)}.\label{poststate}
\eea We note that repeated measurement will not guarantee the same outcome (unless the measurements are von Neumann measurements), as the measurement outcomes are not orthogonal.

Non-orthogonality is a general feature of POVM elements (except in the case of the idealized von Neumann measurement, which always project onto orthogonal [and pure] quantum states). Of particular interest in quantum information theory are symmetric, informationally complete (SIC)-POVMs; the POVM is considered symmetric if each of the elements of these POVMs share the same mutual overlap ${\rm Tr} (\hat{\Pi_i} \hat{\Pi_j}) = \frac{d \delta_{ij} + 1}{d+1}$ with $d$ the dimension of the Hilbert space and $\delta_{ij}$ the Kronecker delta ($0$ when the indices are not the same, unity when the indices are the same). Additionally, the SIC-POVM is informationally complete if there are $d^2$ elements in the POVM, where $d^2-1$ are linearly independent. While they will not be discussed further in this dissertation, they are of particular importance to two applications of our work which we will discuss, quantum state tomography \cite{Caves2002} and quantum cryptography \cite{fuchs2003}, where they illuminate the most efficient measurement scheme for quantum state characterization. Furthermore, there are interesting connections to be made between SIC-POVMs and mutually unbiased bases of quantum states, with the latter playing an essential role in quantum communication and cryptographic protocols \cite{Wootters2004}.

\section{Overview and Glossary}

Having laid the theoretical foundation for our work in this first chapter, we now review the structure of this dissertation. We also provide a glossary of terms ubiquitous in this thesis at the end of this section (see Tbl. \ref{glossary}).

The motivation and goal of this work is to uncover the fundamental limits inherent to single photon detection across physical platforms, both current and future. To accomplish this, we approach photodetection from the perspective of information theory, recasting the process of detecting a photon into the interpretation of a photodetection outcome in the absence of priors. Interpretation of a measurement outcome requires sufficient information about the measurement apparatus to identify the POVM implemented. As such, it is necessary to develop a model of photodetection that is realistic in describing all parts of the photodetection process, but sufficiently idealized and general so that the limits to each part of the process are fundamental limits across all platforms. 

In the next chapter of this dissertation, we begin the construction of such a model starting with a POVM corresponding to the time-dependent two-level system. Here a detection event occurs when one ``checks'' whether the two-level system is in the excited state at a particular time $T$. The two-level system is an essential component to a description of an SPD, as it ensures the measurement projects onto the Hilbert space of single-photon states. As we show, the time-dependent nature of this system is sufficient to project onto arbitrary single-photon wavepackets (with arbitrarily high efficiency), including Gaussian wavepackets so that Heisenberg-limited simultaneous measurements of time and frequency are implemented. At the end of the second chapter, we introduce our three-stage model of the photodetection process: transmission, amplification, and measurement (Fig. \ref{detectbox2}), and clarify the role of the time-dependent two-level system in this model as the trigger for the amplification mechanism.

In the third chapter of this dissertation, we analyze transmission functions $T(\omega)$ describing the quantum network structure of the transmission stage of photodetection. In the transmission stage, the photon has to interact with one or more charged particles, its excitation energy will be converted into other forms of energy which will later lead to a macroscopic signal. In this chapter, we discuss the tradeoffs inherent to different network structures and their implications for photodetecting systems. 

In the fourth chapter of this dissertation, we analyze the amplification process itself. In particular, we identify commutator-preserving transformations that implement different nonlinear amplification schemes outperforming linear amplification in terms of signal-to-noise ratios (SNRs) as illustrated in Fig. \ref{ampsum}. Additionally, we demonstrate that amplification into a single bosonic mode not only outperforms all previously analyzed amplification schemes, but provides the fundamental limit to amplification of single and many-photon Fock states. 

In the fifth chapter of this dissertation, we introduce the final measurement stage and construct realistic SPD POVMs that include all three stages of photodetection. In particular, we demonstrate that an SPD POVM retrodicts a relatively pure external state of the electromagnetic field, even though the internal state of the detector is highly mixed in general. In the remainder of this chapter, we discuss applications to quantum cryptographic protocols, quantum sensing, and quantum information science. 

In the sixth and final chapter of this dissertation, we discuss the fundamental limits and tradeoffs inherent to photodetection encapsulated by the three-stage model of photodetection. We also discuss implications of this work for the future development of novel SPD platforms. 


\afterpage{%
\begin{table}[h!]
\centering
\includegraphics[width=\textwidth]{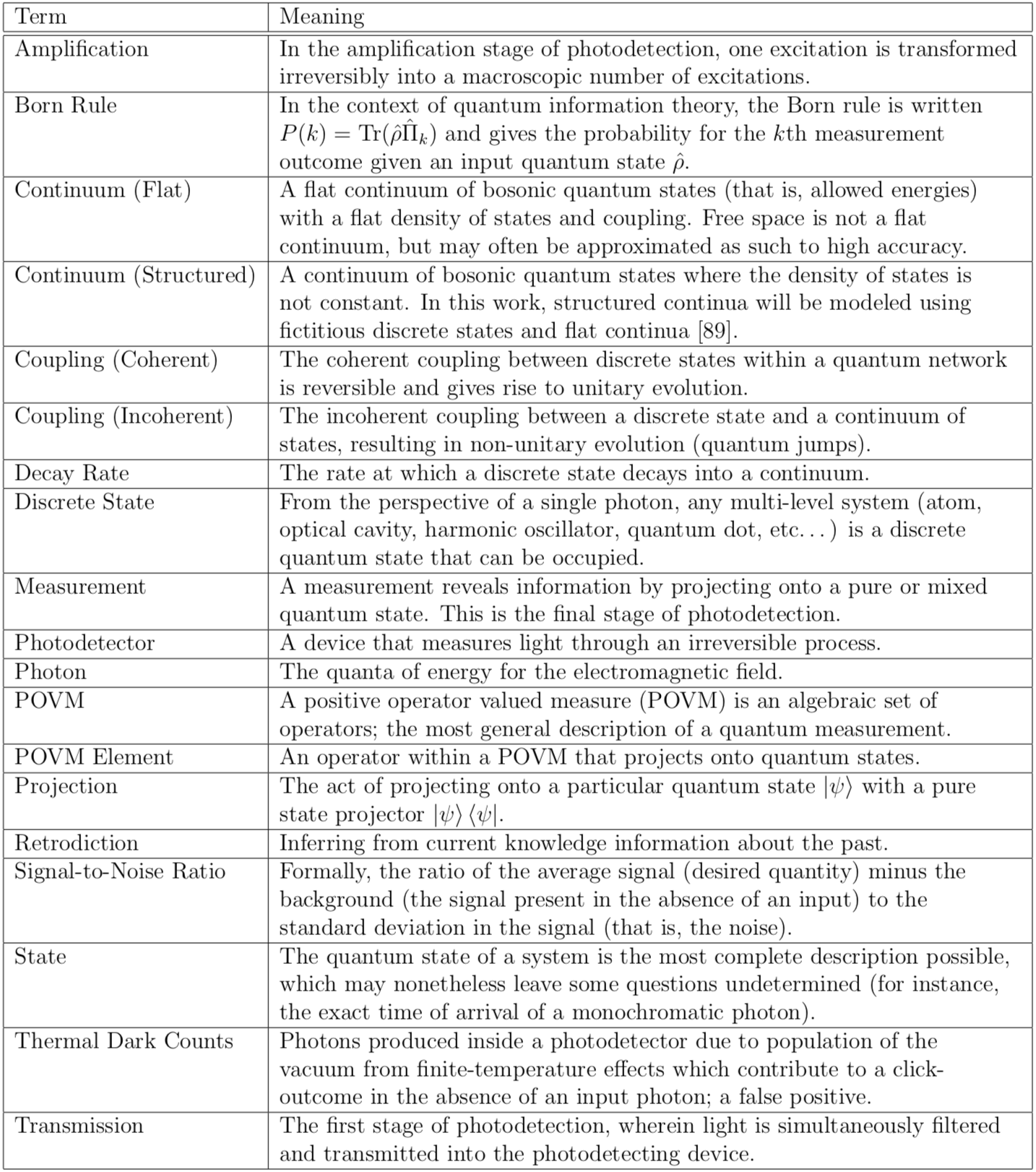}
\caption{Glossary of terms.}\label{glossary}
\end{table}
\clearpage
}

\chapter{POVMs for Single-Photon Detection}


In this chapter, we review techniques in quantum information theory amenable to single-photon detection, in particular, the definition of photodetector figures of merit. Then, we construct a simplified single-photon detector (SPD) positive operator valued measures (POVM) capable of projecting onto arbitrary single-photon wavepackets. Lastly, we discuss the three stages of photodetection universal to all detector schemes, each of which we will analyze in detail in its own chapter.

\section{Quantum Information for Photodetection}
Photodetection is at its core an information theoretic process; a measurement outcome---a click---reveals information about the outside world quantifiable in bits \cite{shannon1948}. In the case of a single-photon detector (SPD), a click is correlated (imperfectly) with the presence of a photon in a particular single-photon state. (Here, the quantum state refers to the spectral, spin angular momentum, and orbital angular momentum quantum degrees of freedom at every point in spacetime). In this way, a detection reveals information about the presence of photons in that particular state, along with whatever else in the world is correlated with that photon state. For an example from fluorescence microscopy, detection of ultraviolet light when a sample is illuminated with red light is positively correlated with the presence of fluorine \cite{stokes1852}. For single photons, measurements that detect light are naturally described through a SPD POVM.

Knowledge of the POVM is essential for both gaining information from a measurement device and characterizing detector performance, hence the experimental need for detector tomography \cite{luis1999,goltsman2005,coldenstrodt2009,lundeen2009,dauria2011,brida2012}. In detector tomography, known quantum states are input to a device to determine the quantum states projected onto by different outcomes (along with their associated weights) and calculate the full measurement POVM. Commercial photodetectors are characterized by industry-standard figures of merit \cite{hadfield2009}, which can be calculated from a POVM (for an in-depth review, see Ref.~\cite{vanenk2017}). Here we will concern ourselves mostly with two standard figures of merit, detection efficiency and time-frequency uncertainty:

\lettersection{Efficiency}The maximum efficiency with which an SPD outcome $k$ (for instance, a single click) can be triggered by input single-photon states is exactly the maximum relative weight in (\ref{POVM1}): $\eta_{\max} = {\rm Max}_{i} [w_{i}^{(k)} ] $. This maximum efficiency is achieved only when the input quantum state is the single-photon state the measurement maximally projects onto. This follows directly from the Born rule; $P(k|i) = {\rm Tr} [ \hat{\Pi}_k \hat{\rho}] \rightarrow w_{i}^{(k)}  $ if and only if $\hat{\rho}=\ket{\phi_i^{(k)}}\bra{\phi_i^{(k)}} $. 

\lettersection{Time-Frequency Uncertainty}The spectral uncertainty (that is, the uncertainty in measurements of photon frequency) and the input-independent timing jitter (uncertainty in measurements of photon time-of-arrival) are determined entirely by the spectral and temporal distributions of the single-photon states projected onto by the measurement outcome $k$ \cite{helstrom1974}, which form a retrodictive probability distribution. For any continuous variable $X$ (here either time $t$ or frequency $\omega$), we find it less convenient to use the variance as measure of uncertainty and instead define the uncertainty entropically \cite{helstrom1974,epi1975,oppenheim2010,vanenk2017,wild2019} 
\bea\label{uncert}
\Delta X^{(k)} = 2^{H^{(k)}_X} \delta X.
\eea Here $H^{(k)}_X$ is the Shannon entropy defined as
\bea\label{entropy} 
H_X^{(k)} =-\sum_j p(j|k) {\rm log}_2 p(j|k)
\eea with the sum over discretized $X$-bins of size $\delta X$. $p(j|k)$ is the \emph{a posteriori} probability for the detected photon to be in bin $j$ given outcome $k$, and is calculated as
 \bea p(j|k) = \int_{(j-1)\delta X}^{j\delta X} dX \sum_i P(i|k) | \phi_i(X) |^2
 \label{postprob}
\eea where we have defined a normalized distribution over $X$ given by the norm squared of the amplitude of the quantum state $ | \phi_i(X) |^2$ (where we have defined $\phi_i(X) \equiv \braket{X}{\phi_i}$ for the single-photon state $\ket{\phi_i}$). The conditional probability $P(i|k) $ is precisely the one from Bayes theorem (\ref{bayes}); $P(i|k)$ reduces to $w_i^{(k)} / \Omega^{(k)}$ in the case of a uniform prior\footnote{For inclusion of priors in updating information about the quantum state, see Ref.~\cite{Holevo}.}, where $\Omega^{(k)} = \sum_i w_i^{(k)} $ is the channel bandwidth defined in Ref.~\cite{vanenk2017} (which characterizes the number of states the measurement efficiently projects onto). Critically, $\Delta X^{(k)}$ is independent of the bin size $\delta X$ in the small-bin limit, even though the entropy $H_X^{(k)}$ is strongly dependent on the bin size. One can verify that this definition of uncertainty yields an uncertainty relation   \cite{epi1975}
\be
\Delta\omega\Delta t \geq e\pi.
\ee

The construction of measurements projecting onto arbitrary single-photon states is critical in quantum optical and quantum communication experiments. Mismatch between the single-photon state generated and the state projected onto by the measurement induces an irreversible degradation in efficiency. (This is a direct consequence of the Born rule (\ref{bornrule}).) Furthermore, the capacity to efficiently project onto single-photon states with orthogonal wavepackets enables a wide range of quantum information and quantum optical applications, as we will revisit in a later chapter.  From a foundational perspective, a procedure to build measurements projecting onto minimum-uncertainty Gaussian single-photon wavepackets paves the way for future tests of fundamental quantum theory.

%

\section{Simple SPD POVMs}

We will now discuss how to construct a simple POVM that efficiently projects onto an arbitrary single-photon wavepacket\footnote{A wavepacket is a localized envelope of wave actions, which can be decomposed into an infinite set of complex sinusoidal waves with differing wave numbers. Here we define the wavepacket $\psi(t)$ as the representation of the single-photon state $\ket{\psi}$ in the temporal basis $\psi_(t) \equiv \braket{t}{\psi}$.}. To aid us, we will now make four simplifying assumptions. First, we will consider only the time-frequency degree of freedom of the electromagnetic field, as the other degrees of freedom (e.g. polarization) can be efficiently sorted prior to detection in a pre-filtering process \cite{osullivan2012,Bouchard2018,Fontaine2019}. Second, we consider only a single photon incident to the photodetector. Multiple photons can always be efficiently multiplexed to achieve a photon number resolution using SPD pixels \cite{Nehra20}. Third, we will not model a continuous measurement (as briefly discussed in the appendix of \cite{proppnet}), but instead a discretized measurement where at a particular time $T$ we ascertain whether or not a photon has interacted with the SPD, ending the measurement. Lastly, we will consider only a binary-outcome photodetector, ``click'' or ``no click.'' This simplifies the POVM so that it only contains the two elements $\hat{\Pi}_T$ and $\hat{\Pi}_{0}$, both projecting onto the Hilbert space of  single-photon states and the vacuum state. Generalizations to non-binary-outcome SPDs are straightforward: one can concatenate binary-outcome POVMs to generate non-binary-outcome experiments.

We now begin construction of the POVM $\{\hat{\Pi}_T,\,\hat{\Pi}_0\}$ in earnest. We begin in the rotating frame by considering a two-level system with time-dependent detuning (that is, the transition frequency in the rotating frame) $\Delta(t)$\footnote{In the rotating frame, we define $\Delta(t)=\omega_e(t)-\omega_g(t)$, with $\omega_e(t)$ and $\omega_g(t)$ the time-dependent resonance frequencies of the excited and ground states.} and time-dependent coupling to a Markovian external electromagnetic continua of states $\kappa(t)$ \footnote{Non-markovianity of the external continua can be included via couplings to fictitious discrete states or pseudomodes, see Refs.~\cite{pseudomodes2,pseudomodes3,pseudomodes4,pseudomodes1,pseudomodes5}.}. Experimentally, a time-dependent decay rate $\kappa(t)$ is induced by a rapid variation of density of states \cite{Mart2004,Thyrrestrups2013}\footnote{Since the rate of incoherent coupling to the continuum is $\sqrt{\frac{\kappa(t)}{2\pi}}$, a time-dependent decay rate also induces a time-dependent coupling.}, and a time-dependent resonance $\Delta(t)$ can be varied with a time-dependent external electric field (Stark effect, \cite{Stark1914}) or through a two-channel Raman transition \cite{Eberly1996}.

The general state of the two-level system and the continuum can be written in the \sch picture $\ket{\psi(t)} = C_0 (t) \ket{0} + C_1 (t) \ket{1}$, where the first term corresponds to the ground state of the two-level system and the second term corresponds to the excited state of the two-level system. In the quantum trajectory picture, there are two types of evolution of $\ket{\psi(t)} $: Schr\"{o}dinger-like smooth evolution with a non-Hermitian effective Hamiltonian and quantum jumps (at random times) \cite{gardiner1992,carmichael1993}. A quantum jump will always correspond to the excitation leaking out of the system and so, in the absence of a dark counts, we only need consider the Schr\"{o}dinger-like evolution of the system. In the quantum jump picture, the quantum state of the two-level system remains will remain pure. We proceed to calculate this in the rotating frame under the rotating wave approximation, following the methods for time-dependent systems introduced in Refs.~\cite{Gheri2004,Circac2004} and letting $\hbar = 1$\footnote{For similar treatments of universal quantum memory and, more recently, a quantum scatterer, see see Ref.~\cite{lukin2007} and Ref.~\cite{Molmer2020}, respectively.}. 

Since we are working the Schr\"{o}dinger picture, we begin the calculation by invoking an additional system: a time-dependent optical cavity with time-dependent detuning in the rotating frame $\delta(t)$ and time-dependent decay rate $\Gamma(t)$ containing exactly one photon in the infinite past. This will populate the continuum driving the two-level system with an (arbitrary) photon wavepacket. The population of this optical cavity is described by a single bosonic mode amplitude $C_b(t)$ which is in turn equal to $C_0(t)$; due to our assumption of a single photon, the ground state acts as a bosonic mode.

In the Schr\"{o}dinger picture, the evolution of the bosonic mode population and two-level system population are given

\bea\label{simpleC0}
\dot{C}_b(t)&=& -i \delta(t)C_b (t) -\frac{\Gamma(t)}{2} C_b (t)\nonumber \\
\dot{C}_1(t)&=& -i \Delta(t)C_1 (t)  -\frac{\kappa(t)}{2} C_1 (t) - \sqrt{\kappa(t)\Gamma(t)} C_b(t).
\eea 

The decay rate $\kappa(t)$ appears in the second line of (\ref{simpleC0}) twice due to its dual role: acting as the system's time-dependent decay rate and determining the time-dependent coupling $\sqrt{\frac{\kappa(t)}{2\pi}}$ to the external electromagnetic continuum of state. Let us consider a time in the infinite past when there was no coupling to the external continua ($\kappa(-\infty)=0$), the excited state is unpopulated ($C_1(-\infty)=0$), and bosonic mode is fully populated ($C_b(-\infty)=1$), so that we can solve the first line of (\ref{simpleC0}) for $C_b(t)$ directly

\bea\label{simpleC0t}
C_b(t)= \exp\left[-\int_{-\infty}^t dt' (i \delta(t') +\frac{\Gamma(t')}{2})\right].
\eea 

Defining a driving term $f(t)=\sqrt{\kappa(t)} C_b(t)$ and substituting (\ref{simpleC0t}) into the second line of (\ref{simpleC0}), we find the time-dependent excited state amplitude $C_1(t)$ will have the form of a Langevin equation 

\bea\label{simpleC}
\dot{C}_1(t)= -\frac{\kappa(t)}{2} C_1 (t) -i\Delta(t) C_1 (t) + \sqrt{\kappa(t)} f(t)
\eea with decay (the first term), resonance frequency (the second term), and driving from the continuum (the last term). 

We can now interpret $f(t)$ as the normalized input photon wavepacket corresponding to the state 

\be 
\ket{f} = \int dt f(t) \adin (t) \vac 
\ee with $\adin(t)$ the standard creation operator for the bosonic continuum of input states (as we will use extensively in the input-output theory treatments to come in this thesis) \cite{input1985}. 
We can solve this equation exactly with the result
\be
C_1(t)=\int_{T_0}^t dt'' f(t'')\sqrt{\kappa(t'')}\exp\left[-\int_{t''}^t dt' D(t')\right],
\ee
where we have defined $T_0$ to be a (finite) time in the distant past where our photodetector was still off so that there is no coupling to the external continua ($\kappa(T_0)=0$ [this could be a smooth transition as explored in Fig. \ref{rainbow2}]) and thus the excited state is unpopulated ($C_1(T_0)=0$), and we have also defined 

\be
D(t)=i\Delta(t)+\frac{\kappa(t)}{2}.
\ee

Our measurement consists in checking if the system is in the excited state at time $t=T$.
The probability to obtain a positive result (corresponding to detecting the incident photon wavepacket)
is $|C_1(T)|^2$. We can write
\be
C_1(T)=\int_{T_0}^T dt \Psi^*(t)f(t) 
\ee
with 
\be\label{timeCSolPack}
\Psi^*(t)=\sqrt{\kappa(t)}\exp\left[-\int_{t}^T dt' D(t')\right].
\ee

  \begin{figure}[h!] 
  \centering
	\includegraphics[width=.8\linewidth]{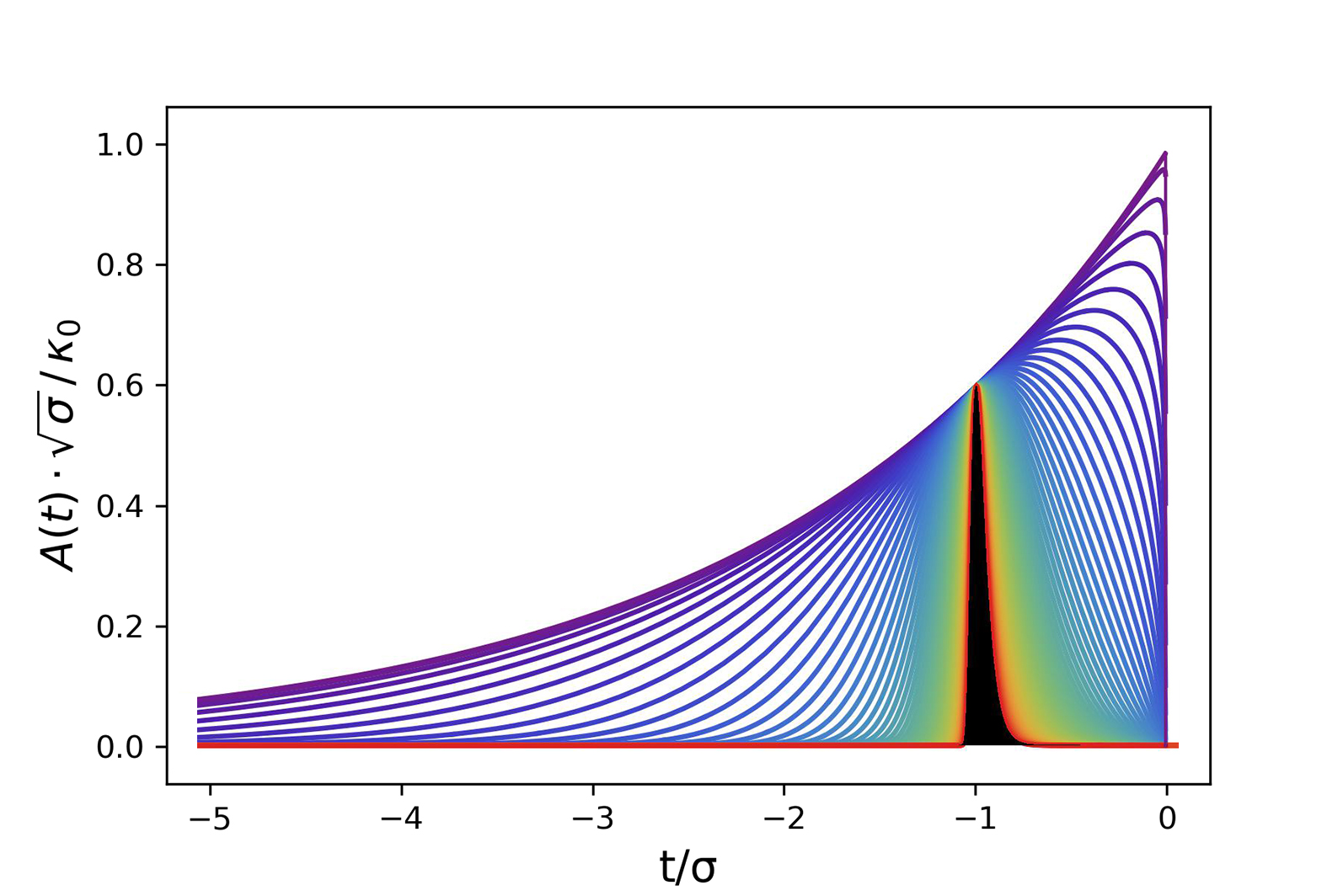} 
	\caption{The retrodictive probability amplitudes ${A(t)=|\Psi(t)|}$ defined in (\ref{timeCSolPack}) are plotted for polynomial decays of the form ${\kappa(t) =\kappa_0 \left(\frac{T-t}{\sigma}\right)^n}$ with polynomial order $n$ varying from $0$ (violet, top) to $40$ (red, bottom) denoted by color. The polynomial decays correspond to systems where the decay rate drops to zero precisely at the time of detection. Here time is measured with respect to (w.r.t.) time of detection ${T=0}$ and we set ${T_0=-\infty}$ so that ${{\cal W}=1}$ (\ref{calW}).}{\begin{flushleft}\vspace{-1em}Only constant ${\kappa(t)=\kappa_0}$ yields non-zero probability amplitude at ${t=0}$ (where it is maximum). This illustrates the two roles $\kappa(t)$ plays: determining the probability $\kappa(t)dt$ with which an excitation can enter the system at a time $t$, and the rate $\kappa(t)$ at which the excited state will decay.This latter effect drives down the probability amplitude for the distant past. For ${\kappa(T)\neq 0}$, the most likely time that a photon entered the system is \emph{now} (${t=0}$) whereas if ${\kappa(T)=0}$, there is some time of maximum likelihood determined by competition between the two effects (absorption and decay). For no order $n$ are the retrodictive probability distributions continuously differentiable; these simple polynomial couplings do not yield measurements projecting onto smooth wave packets unlike the couplings plotted in Fig. \ref{dataCouplingSimp}. Note that $\kappa(t)$ is not normalized and diverges as ${t\rightarrow -\infty}$, yet it results in well-behaved, normalized retrodictive probability amplitudes.\end{flushleft}}
	\label{rainbow}
\end{figure}

Whereas $f(t)$ is a normalized wavepacket, $\Psi(t)$ is subnormalized for finite $T_0$, since
\be\label{calW}
{\cal W}=\int_{T_0}^T dt |\Psi(t)|^2=1-\exp
\left[-\int_{T_0}^T\! dt\, \kappa(t)\right].
\ee We can interpret $\Psi(t)$ as a retrodictive probability amplitude (for simple examples, see Fig. \ref{rainbow} and Fig. \ref{rainbow2}), identifying at which times a photon likely entered the system given a detector ``click'' at $t=T$.

  \begin{figure}[h!] 
  \centering
	\includegraphics[width=.8\linewidth]{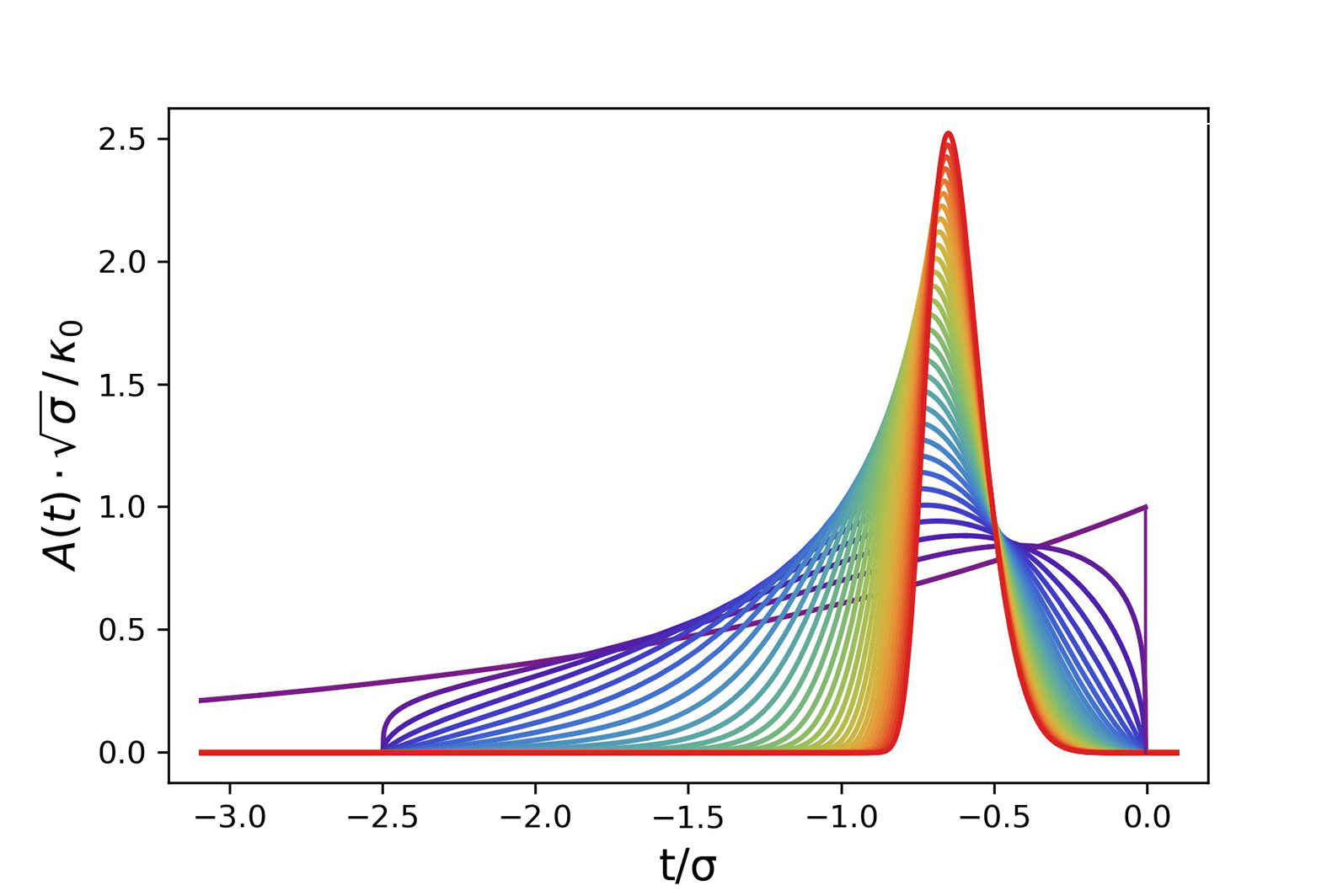} 
	\caption{Retrodictive probability amplitudes $|\Psi(t)|$ defined in (\ref{timeCSolPack}) are plotted for polynomial decays of the form ${\kappa(t) =\kappa_0 \left(\frac{t-T_0}{\sigma}\right)^n\left(\frac{T-t}{\sigma}\right)^n}$ for ${T_0\leq t\leq T}$, with the polynomial order $n$ varying from $0$ (violet, lowest peak) to $15$ (red, highest peak) denoted by the color. Time is measured w.r.t. detection time ${T=0}$.}{\begin{flushleft}\vspace{-1em}Except for ${n=0}$ (also in Fig. \ref{rainbow} and included here for reference), these decays become non-zero when the photodetector is turned on at ${T_0=-2.5\sigma}$ so that ${\cal W}$ from (\ref{calW}) is less than unity. As in Fig. \ref{rainbow}, for no order $n$ are the retrodictive probability distributions continuously differentiable (indeed, this is the source of the singularity in Ref.~\cite{Nurdin2016}, and do not project onto smooth wave packets as in Fig. \ref{dataCouplingSimp}. Also as in Fig. \ref{rainbow}, we observe a time of maximum likelihood determined by competition between the two effects (absorption and decay).\end{flushleft}}
	\label{rainbow2}
\end{figure}

We can define a normalized single-photon state 
\bea\label{simpleState}
\ket{\Psi_T} ={\cal W}^{-1/2} \int_{T_0}^{T}\! dt\, \Psi(t) \hat{a}^\dagger_{\rm in} (t) \vac\nonumber \\
\eea with the creation operator $\hat{a}^\dagger_{\rm in} (t) $ acting on the input continuum of states (it's Fourier transform $\hat{a}^\dagger_{\rm in}(\omega)$ creates a monochromatic state with frequency $\omega$). 
The arbitrary input single-photon state (which may have been created long before our detector was turned on at $T_0$ or long after the measurement ended at time $T$) is

\bea\label{inputStateDef}
\ket{f} &= \int_{-\infty}^{\infty} dt f(t) \hat{a}^\dagger_{\rm in} (t) \vac.
\eea
The commutator relation for the input field operator is $[\hat{a}_{\rm in} (t), \hat{a}^\dagger_{\rm in} (t')] = \delta(t-t')$.

The probability for an arbitrary  input photon wavepacket $f(t)$ to result in the system being found in the excited state at a time $T$ is 
 $|C_1(T)|^2=
	{\cal W}\braket{f}{\Psi_T}\braket{\Psi_T}{f} $.
The measurement does not project onto times \emph{after} we have checked if the system is in the excited state, nor onto times before the detector was turned on.

\begin{figure}[h] 
  \centering
	\includegraphics[width=.6\linewidth]{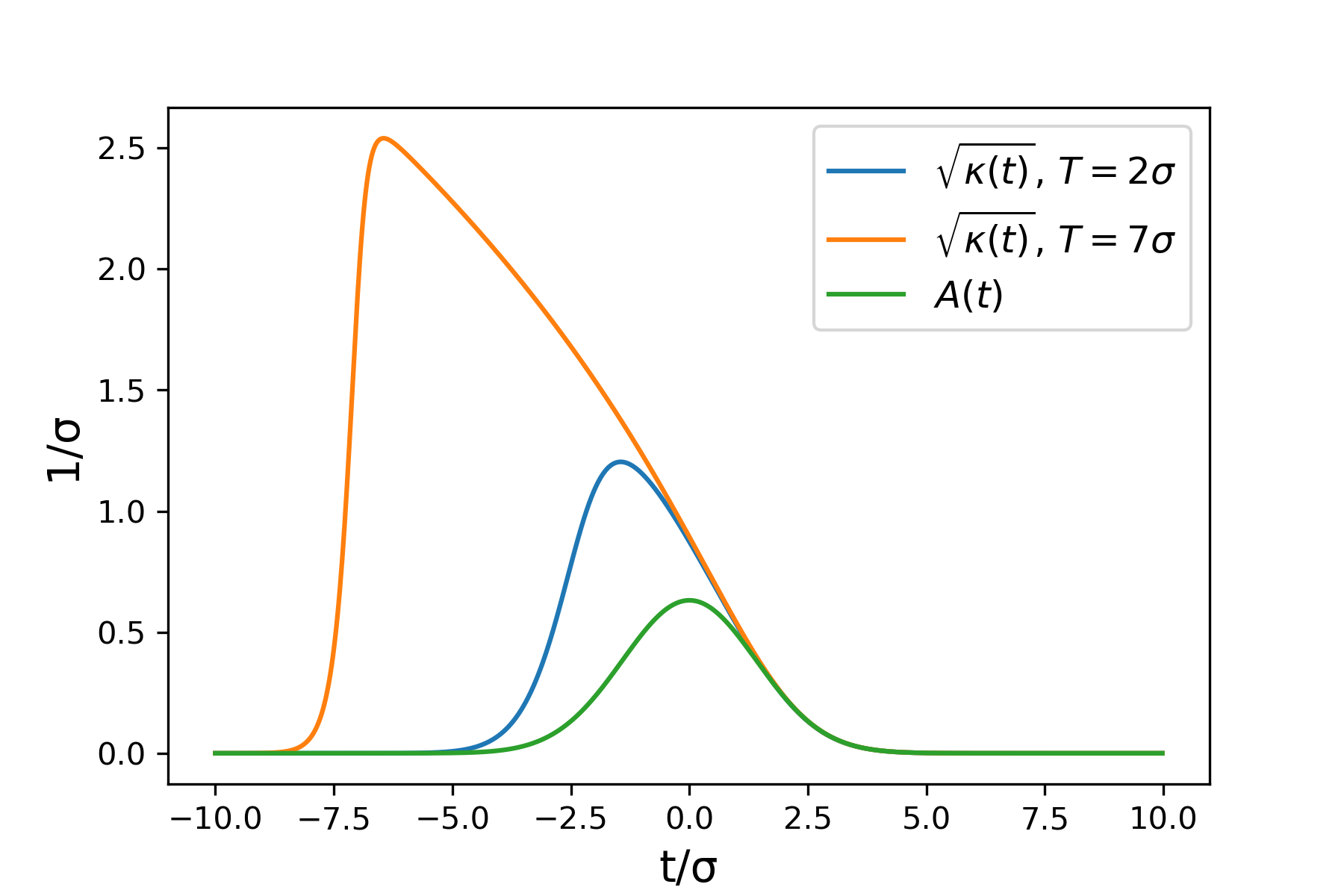} 
	\caption[]{Time-dependent couplings $\kappa(t)$ defined in (\ref{ErrfSol}) generating a minimum uncertainty (Gaussian) wavepacket are plotted for times of detection ${T=2\sigma}$ (${{\cal W} = 0.98}$) and ${T=7\sigma}$ (${{\cal W} = 1-10^{-10}}$) and detector on-time ${T_0=-\infty}$. The retrodictive probability amplitude is now a well-defined wavepacket, with amplitude $A(t)$ depicted in the rotating frame (without fast oscillations at central frequency $\omega_0$). Time is measured w.r.t. the wavepacket's central time $t_0$. For times near ${t\approx T}$ the coupling $\sqrt{\kappa(t)}$ is approximately Gaussian as an excitation absorbed at this time will not have sufficient time to decay back out. For earlier times, $\sqrt{\kappa(t)}$ is strictly larger than $A(t)$ and is skewed towards earlier times to incorporate the time it takes the system to respond (${\sim 1/\kappa(t)}$); one needs to prepare the two-level system for the photon, accomplished via a ``sharkfin'' coupling that precedes the single-photon wavepacket \cite{lukin2007}. Near ${t=-T}$, we observe the coupling $\kappa(t)$ rapidly drops to zero, which is a direct consequence of the temporal symmetry of the wavepacket about $t_0$. [From (\ref{timeCSolPack}) one can verify that, if $A(t)$ is time-symmetric around ${t=t_0}$, the coupling $\kappa(t)$ must satisfy ${\kappa(t_0-t) = e^{\int_{t_0-t}^{t_0+t} dt' \kappa(t')} \kappa(t_0+t)}$.]}
	\label{dataCouplingSimp}
\end{figure}

We rewrite this probability in terms of a POVM element containing a single element 
 \bea\label{simpleAns}
\hat{\Pi}_T = {\cal W}\ket{\Psi_T}\bra{\Psi_T}
\eea 
To the extent our detector has been open long enough, such that ${\cal W}\rightarrow 1$, our detector could act as a perfectly efficient detector for a specific single-photon wavepacket with temporal mode function $\Psi(t)$\footnote{For measurements projecting onto a Gaussian wavepacket as in (\ref{ErrfSol}) and Fig. \ref{dataCouplingSimp}, the weight (\ref{calW}) has the simple form ${\cal W} = \frac{1}{2} {\rm Erf}[\frac{t_0-T_0}{\sqrt{2}\sigma}] + \frac{1}{2} {\rm Erf}[\frac{T-t_0}{\sqrt{2}\sigma}]$, going to unity for $T,T_0\gg \sigma$.}. This wavepacket is the time reverse of the wavepacket that would be emitted by our system if it started in the state $\ket{1}$\footnote{Instead of solving (\ref{simpleC}), one can find the Green's function $G(t)$ of the time-reversed problem: at $t_0$ the two-level system is started in the excited state and at time $T$ we check whether the excitation has leaked out. Taking $t\rightarrow T-t$, one arrives at (\ref{simpleState}) with $\Psi(t)=G(T-t)$. This clarifies the role of the Green's function; propagating back in time starting from $t = T$ (when the photon is detected) back to the infinite past, which indeed is what the POVM does as well.}.

For this simple system, the POVM element is both pure (containing just one term\footnote{As an auxiliary and less-conventional figure of merit, we define purity of the POVM element ${\rm Pur}[\hat{Pi}_k] = \frac{{\rm Tr}[\hat{\Pi}_k^2]}{\left({\rm Tr}[\hat{\Pi}_k]\right)^2}\leq 1$ where the upper limit is reached only when the POVM element projects onto a single state (von Neumann measurement).}) and (almost) maximally efficient; the weight ${\cal W}$ may approach unity as close as we wish by lengthening the time the detector is on for (that is, taking $T_0$ to the distant past and making sure the projected wavepacket [approximately] ends before checking at time $T$).

Here we observe an obvious tradeoff between efficiency and photon counting rate: one cannot project onto a long single-photon wavepacket in a short time interval without cutting off the tails, lowering the overall detection efficiency\footnote{The limitation to photon counting rate imposed by efficient detection of long temporal wavepackets is avoided via signal multiplexing, see Ref.~\cite{Nehra20}.}. 

The two-level system described in Eq. (\ref{simpleC}) is a special case but an important one; the two-level system is often a very good approximation of more complicated systems near-resonance. Furthermore, the two-level system is the foundation for generating more complicated network structures, as we will discuss in more detail in the next chapter\footnote{For an arbitrary multi-level time-independent structure, we will end up with a system of equations governing discrete state evolution
\bea\label{generalC}
\dot{\vec{C}}(t)= \mathbf{M} \vec{C} (t) + \vec{S}(t)
\eea with $\mathbf{M}$ a time-independent matrix and $\vec{S}(t)$ a time-dependent
(inhomogeneous) source term describing the input photon. The solution is then always of the form
\bea\label{generalCSol}
\vec{C}(t)= e^{\mathbf{M} t} \vec{C} (t_0) + \int_{t_0}^t dt' e^{\mathbf{M} (t-t')} \vec{S}(t').
\eea 
Writing $e^{\mathbf{M} (t-t')}$ as a Green's matrix, we can identify elements that correspond to transitions to the final monitored discrete state (detector outcomes) through standard numerical techniques \cite{duffy2001}.}. In this chapter, we will focus on the simple time-dependent system (\ref{simpleC}) as it is sufficiently general to perform a measurement described by any time-independent system, and more\footnote{In particular, time time-independent systems cannot achieve Fourier-limited measurements of time and frequency. This is because networks of discrete states experience a natural spectral broadening that is Lorentzian. While Gaussian broadening can additionally occur (for instance, due to Doppler shifts in atomic distributions \cite{siegman86}) this only increases the product uncertainty further from the minimum of ${\Delta \omega \Delta t = e\pi}$ \cite{epi1975}, attained \emph{only} by pure measurements projecting onto Gaussian wavepackets.}. Indeed, (\ref{simpleC}) is general enough to project onto a \emph{completely arbitrary} single-photon wavepacket, a result we will now prove.

\lettersection{Proof} Consider a photon with complex wavepacket $\Psi^*(t)=A(t)e^{i\phi(t)}$, positive amplitude $A(t)$, and phase $\phi(t)$. Inserting this into (\ref{timeCSolPack}), we arrive at two separate expressions 

\bea\label{sepExpress}
A(t)&=& \sqrt{\kappa(t)} e^{-\int_t^T dt' \frac{\kappa(t')}{2}}\nonumber \\
 \phi(t)&=& -\int_t^T dt' \Delta(t').
\eea The second line is always solvable by $\Delta(t)=\dot{\phi}(t)$ up to a constant global phase shift provided $\phi(t)$ is everywhere differentiable (smooth). We now focus on the first line. Taking the natural logarithm we arrive at an expression

\bea\label{conditionPacket}
2{\rm Ln}[A(t)] - {\rm Ln}[\kappa(t)] = -\int_t^T dt' \kappa(t').
\eea Taking the time derivative of both sides, we arrive at a Bernoulli differential equation \cite{Zwilinger1994}

\bea\label{conditionBernouli}
\kappa^{-2}(t) \frac{d \kappa(t)}{dt} - \frac{2}{A(t)\kappa(t)} \frac{dA(t)}{dt} = -1.
\eea Provided $\frac{1}{A(t)} \frac{dA(t)}{dt}$ is continuous, this is solved by

\bea\label{BernSol}
\kappa(t) = \frac{A^2(t)}{1-\int_{t}^{T} A^2(t') dt' } 
\eea in agreement with Ref.~\cite{Gheri2004} and Ref.~\cite{Molmer2020}. Here, $\kappa(t)$ is given by the square of the electromagnetic field, divided by a correction factor accounting for the finite response time imposed by $\kappa(t)$ itself\footnote{We observe from (\ref{BernSol}) that now ${A(t)=0 \iff \kappa(t)=0}$ for smooth wavepackets, whereas for a general retrodictive distribution (\ref{timeCSolPack}) we find ${\kappa(t)=0 \implies A(t)=0}$; to generate a smooth wavepacket, $\kappa(t)$ must go to $0$ in the distant past.}. From (\ref{BernSol}), we observe that the only condition imposed on $A(t)$ is that $A^2(t)$ has an antiderivative (indefinite integral). We simply require $A^2(t)$ be continuous, which in turn requires $A(t)$ to be continuous. Thus, any wavepacket with smooth phase profile $\phi(t)$ and smooth amplitude $A(t)$ is projected onto by \emph{some} physically realizable single-photon detection scheme. \qed

\lettersection{Special Case: Fourier-Limited Measurement} A Fourier-limited simultaneous measurement of time and frequency is achieved with a Gaussian time-frequency distribution. We want a temporal wavepacket $\Psi^*(t)$ that is the complex square root a Gaussian distribution 

\bea\label{gaussianWavePacket}
\Psi^*(t)= \frac{1}{(2\pi\sigma^2)^{\frac{1}{4}}} e^{-\frac{(t-t_0)^2}{4\sigma^2}} e^{i\frac{\omega_0}{2}t}
\eea where $\sigma$ is the temporal half-width, and $t_0$ and $\omega_0$ are the central time and frequency of the Gaussian distribution. We find that this wavepacket is projected onto by a time-dependent system with constant resonance $\Delta(t) = \omega_0$ and a time-dependent coupling 

\bea\label{ErrfSol}
\kappa(t) = \frac{e^{-\frac{(t-t_0)^2}{2\sigma^2}}}{\sqrt{2\pi\sigma^2}(1+\frac{1}{2}{\rm Erf}[\frac{t_0-T}{\sqrt{2}\sigma}] - \frac{1}{2}{\rm Erf}[\frac{t_0-t}{\sqrt{2}\sigma}])}
\eea as in Fig. \ref{dataCouplingSimp}. Note that the coupling $\kappa(t)$ is $T$-dependent even though the projected state (\ref{gaussianWavePacket}) is $T$-independent, in agreement with the general case (\ref{BernSol}). 

%
%
%

\section{The Three-Stage Model of Photodetection}

The model of a SPD as an isolated two-level system is highly idealized. In a more realistic system, photodetection is an extended process wherein a photon is transmitted into the detector, interacting with the system and triggering a macroscopic change of the photodetector state (amplification) which can then be measured classically. Many theories of single-photon detection have been developed over the past century, \cite{glauber1963,kelley1964,scully1969,yurke1984,MandelWolf95,ueda1999,schuster2005,helmer2009,clerk2010,young2018,dowling2018,leonard2019} and indeed there are numerous implementations of SPD technology \cite{Allen39,mcintyer81,marsili2013,frog,wollman2017uv}. 

  \begin{figure}[h!] 
  \centering
	\includegraphics[width=.8\linewidth]{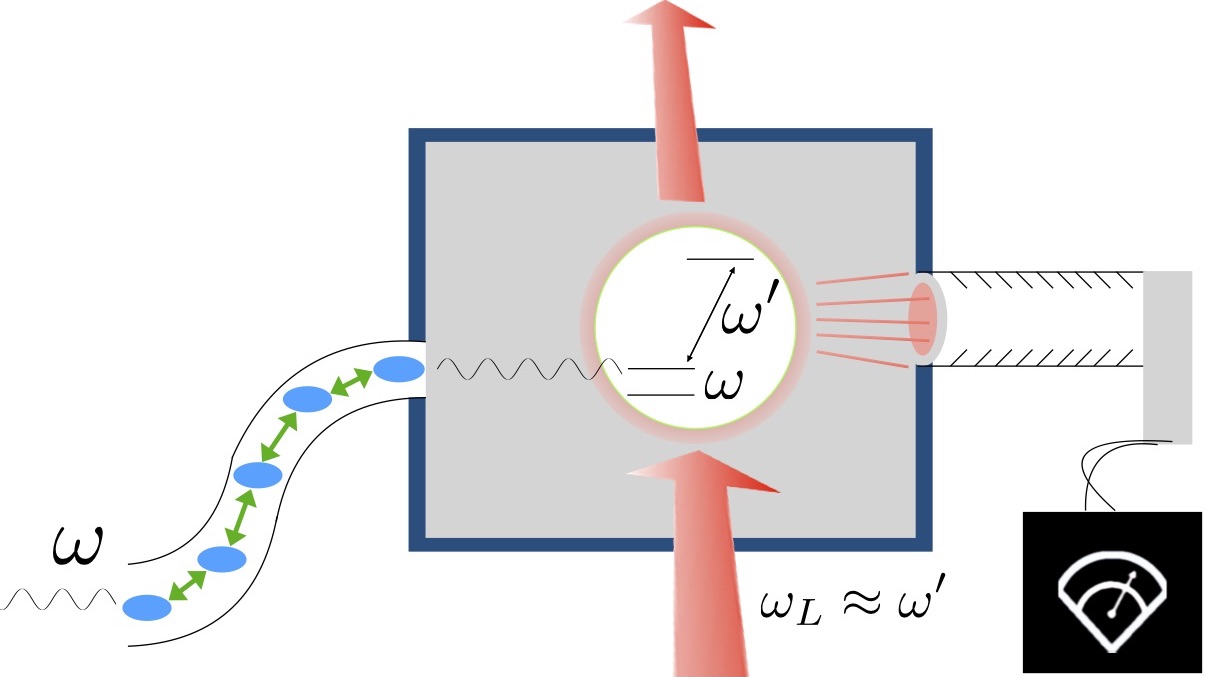} 
	\caption[]{The three stages of photodetection in a physical implementation: transmission, amplification, and measurement. An input photon with frequency $\omega$ is transmitted irreversibly via network of coherently coupled discrete states into contact with the photodetection mechanism, here modeled as a two level system. Once the amplification mechanism is triggered by a photon, a macroscopic optical signal with frequency ${\omega'\approx \omega_L}$ (the laser frequency) is irreversibly created via amplification, in our model implemented with a three-level electron shelving scheme in a single mode cavity; absorption of the single photon at frequency $\omega$ results in many flourescence excitations at frequency $\omega'$. This is then converted to a classical electrical readout through a photomultiplier tube (which could have low efficiency) in a third irreversible step. Both initial transmission into the system and absorption by the two-level system filter the light, changing which quantum state is projected onto by the final measurement. }
	\label{detectbox2}
\end{figure}

Across all systems, we identify these three stages of transmission, amplification, and measurement as universal in Fig. \ref{detectbox2}. In the next three chapters, we will study the transmission and amplification stages in detail before deriving a POVM that incorporates all three stages quantum mechanically and includes fluctuations of system parameters. The time-dependent two-level system from the previous section enabling arbitrary wavepacket projection will be incorporated into the three-stage model as the trigger for the amplification mechanism. We will assume in this analysis that the system is left on for a sufficient time such that the subnormalization of $\Psi(t)$ is minimal and ${\cal W}\approx 1$ (\ref{calW}).

\chapter{Transmission}

We now study in detail the topic of the ``transmission'' stage of photo detection, wherein a photonic signal is transmitted into the photodetector, potentially changing form (transduction) and being absorbed by the device. To accomplish this, the photon has to interact with one or more charged particles, its excitation energy will be converted into other forms of energy which will (in the next stage) lead to a macroscopic signal (amplification), and then a ``click'' (measurement). A simple example of this is a fiber optical cable: here, a discrete fiber mode acts as intermediary between two continua. More complicated forms of energy transport are also described by quantum networks of coupled coupled discrete quantum states and structured continua (e.g. band gaps) provide generic models for that first part of the detection process, as we will discuss in this chapter. The input to the network is a single continuum (the continuum of single-photon states), the output is again a single continuum describing the next (irreversible) step. The process of a single photon entering the network, its energy propagating through that network, and finally exiting into another output continuum of modes can be described by a single dimensionless complex transmission amplitude, $T(\omega)$. Along with $T(\omega)$, we calculate a complex reflection amplitude $R(\omega)$ which characterizes light reflected off of the photodetector. In this chapter, we discuss how to obtain from $T(\omega)$ the photo detection efficiency, how to find sets of parameters that maximize this efficiency, as well as expressions for other input-independent quantities such as the frequency-dependent group delay and spectral bandwidth of the transmission portion of a single-photon detector.  
We then study a variety of networks, discuss how to engineer different transmission functions $T(\omega)$ amenable to photo detection, and discuss implications for single-photon detection technology.

The underlying basics of photodetection theory was developed in the early 1960s \cite{Glauber, MandelWolf,kelley1964}, with the quantum nature of light being taken into account, and with later additions to the theory also incorporating the backaction of the detector on the detected quantum field \cite{scully1969,yurke1984,ueda1999,schuster2005,clerk2010}. More recent additions to the theory have analyzed more deeply the amplification process by itself \cite{proppamp} and its relation to the absorption and transduction part of the process \cite{young2018,helmer2009}. In particular, it turns out that for an ideal detector one should decouple the two processes, by having an irreversible step in between the two, such that the amplification part does not interfere negatively with the absorption (and possibly, transduction) stage, here called transmission \cite{young2018,clerk2010}. This decoupling will be assumed in this analysis, too.

In order to develop a useful fully quantum-mechanical theory
we cannot be {\em completely} general; or, rather, if we are completely general, then the only statements on fundamental limits we can make are likely going to be merely examples of Heisenberg's uncertainty relations. So we will make three restrictive but---we think---reasonable assumptions about our quantum theory of photo detection.

\begin{center}
  \begin{figure}[h] 
 \centering
	\includegraphics[width=.8\textwidth]{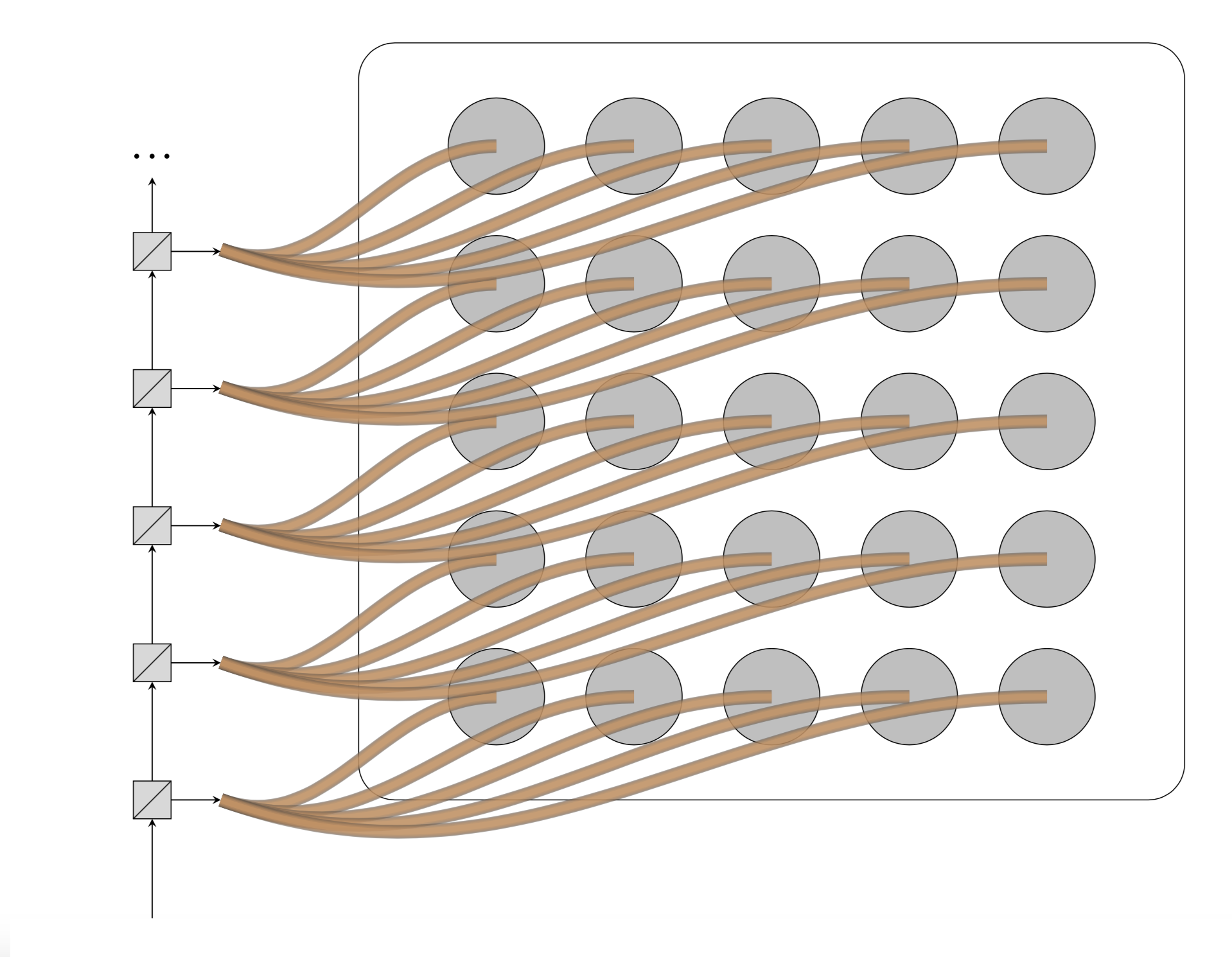} 
	\caption{An array of single-photon detecting (SPD) pixels. A series of beamsplitters with low reflectivity ensure that at most one photon is incident on each single-photon detector.}
	\label{array}
\end{figure}
\end{center}

First, we focus on {\em single}-photon detection.
The main reason is that number-resolved photo detection is possible using arrays of SPDs where each ``pixel'' receives at most one photon as in Fig. \ref{array} (also see \cite{photonnumber2017,Wollman19}, or \cite{migdall2003} for the time-reversed process of creating a single photon on demand).
So we focus on an individual pixel here. (See Ref.~\cite{combes2017} for a modeling framework for systems with multiple inputs and Refs.~\cite{fan2010,caneva2015,xu2015} for non-linear S-matrix treatments of few-photon transport.)  

Second, although a general state of a single photon is a function of four quantum numbers, one related to the spectral degree of freedom, two related to the two transverse spatial degrees of freedom, and one related to the polarization or helicity degree of freedom, we will restrict ourselves to the spectral (or, in the Fourier transformed-picture, the temporal) degree of freedom.
That is, the input state can be defined in terms of frequency-dependent creation operators $\had(\omega)$ acting on the vacuum.
The reason is that the other three degrees of freedom can, in principle, if not in practice, be sorted before detection. For example, if one wishes to distinguish between horizontally and vertically polarized photons, one may use a polarizing beam splitter and put two detectors behind each of the two output ports. Similarly, efficient sorting of photons by their orbital angular momentum quantum number \cite{osullivan2012} and spatial mode are also possible \cite{Bouchard2018,Fontaine2019}. It is easier to consider sorting as part of the pre-detection process, rather than a task for the detector itself \footnote{To describe sorting as well within this framework, we simply write more transmission functions, e.g. $T_j(\omega)$ for all the different input continua $j$, each leading to their own output continuum. Even more complicatedly, we could consider multiple outputs $i$ for a given input $j$ and write $T_{i|j}(\omega)$. But even in this case we can focus on a particular $i$ and $j$ as we do this in this thesis.}. On the other hand, the spectral response of a detector cannot be eliminated; the time-frequency degree of freedom is intrinsic to the resonance-structure of the photo detecting device. 

Third, we are going to assume that the transmission stage of each pixel's operation is passive. That is, apart from being turned on at some point, and being turned off at some later point, it operates in a time-independent manner with time-independent decay rates, couplings, and resonance structure. Thus, an incoming photon will interact with a time-independent quantum system. As we will see, active filtering is not needed for perfect detection provided the photodetector has no internal losses (couplings to additional continua or side channels). Furthermore, including a time-dependent amplification trigger allows for arbitrary wavepacket detection, a result we will show in Chapter V.

We can now describe the interaction of a single photon with an arbitrary quantum system as follows. The system may be naturally decomposed into
subsystems, each of which may have discrete and/or continuous  energy eigenstates. (For example, the photon may be absorbed by a molecule or atom or quantum dot or any structure with a discrete transition that is almost resonant with the incoming photon.)
The continua will in general be structured (for example, containing bands and band gaps in between) \cite{Tellinghuisen1975,Odegard2002,Nygaard2008}, but structured continua can be equivalently described as structureless (flat) continua coupled to (fictitious) discrete states \cite{pseudomodes2,pseudomodes3,pseudomodes4,pseudomodes1}, enabling a Markovian description of the system independent of an input photon's bandwidth. Indeed, it is well known that a non-Markovian open system can always be made Markovian by expanding the Hilbert space (the converse of the Stinespring dilation \cite{stinespring}). And so an arbitrary quantum system may be described by a network of discrete states (some physical, some fictitious), coupled to flat continua. The latter coupling makes the time evolution irreversible. Of course, an actual detector is indeed irreversible.
In particular, the amplification process (converting the microscopic input signal into a classical macroscopic output signal) is intrinsically irreversible. 

 \begin{figure}[h] 
 \centering
	\includegraphics[width=.6\textwidth]{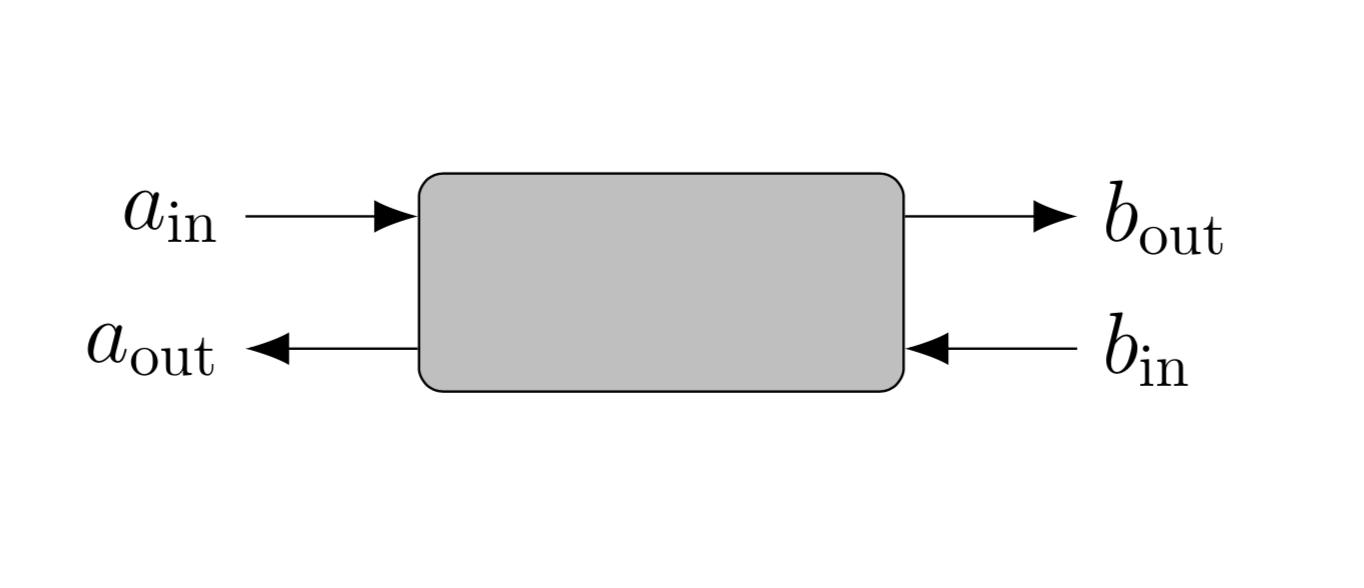} 
	\caption{Input and output fields coupled to a two-port black box quantum network.  The input and output operators are continuous mode (as functions of frequency $\omega$) annihilation operators, satisfying canonical commutation relations \cite{input1985}. Here the mode $\ain(\omega)$ carries the single-photon input state, the reflected mode is described by $\aout(\omega)$, and the output mode $\bout(\omega)$ contains the excitation energy if the input energy successfully traversed the network (and may serve as input to the next part of the photo detection process, e.g., amplification, see Ref.~\cite{proppamp}). $\bin(\omega)$ may contain thermal excitations, but is never occupied by the photon we wish to detect.}
	\label{2ports}
\end{figure}

As we will show, this first part of the process can then be fully described in terms of a complex transmission amplitude $T(\omega)$, which is the probability amplitude for the component of the input signal at frequency $\omega$ to enter the next (amplification) process. (By then the energy $\hbar\omega$ will have been converted into a different type of energy, i.e., a different type of excitation, but that plays no particular role here.) It is straightforward to calculate the transmission function through a network (the equations are linear!). The point here is that from $T(\omega)$ we can determine three input-independent quantities of interest to the photo detection process. (That one complex function $T(\omega)$ is sufficient is due to our considering only the time/frequency degree of freedom.) 

This chapter is organized as follows. After defining our quantities of interest in section II (short), we start with analyzing the simplest quantum network to illustrate how we calculate $T(\omega)$ and how our three quantities of interest behave in section III (also short). We then present a systematic survey of more complicated networks in the much longer section IV; for each network class, we analytically calculate $T(\omega)$ and discuss the behavior of the three quantities of interest. Then in section V, we briefly discuss extensions to arbitrary quantum networks including the effects of couplings to additional continua/side channels before summarizing our findings in the conclusions.

\section{Quantities of Interest}

Here, we define three quantities of interest derivable from $T(\omega)$, and discuss their applicability to photo detecting systems.

First, $|T(\omega)|^2$ itself gives an upper bound on the probability for the frequency component $\omega$ to be detected. (If there were no losses downstream in the photodetection process, it would equal the probability of detection of monochromatic light at that frequency.) 
We are thus particularly interested in identifying quantum systems for which there is at least one frequency $\omega_i$ for which $|T(\omega_i)|=1$.

Second,
an upper bound to the total detectable frequency range of input light is given by the spectral bandwidth of the quantum network (not to be confused with the channel bandwidth in \cite{vanenk2017} or the range of a single frequency band \cite{Sundararajan2008}). We define this quantity 
\be\label{bandwidthdef}
\tilde{\Gamma}=\frac{1}{\pi}\int_0^\infty d\omega |T(\omega)|^2.
\ee 
(The $\frac{1}{\pi}$ factor gives agreement with a classical Lorentzian filter; the transmission function is $T(\omega)=\frac{\Gamma}{\Gamma-i(\omega-\omega_0)}$ for a filter with damping factor $\Gamma$ and resonant frequency $\omega_0$ and we find $\tilde{\Gamma}=\Gamma$, see Eq. (\ref{GammaInt}) below.) The inverse of this quantity is also a measure of the time the photon spends in the detector. (Indeed, $\tilde{\Gamma}^{-1}$ is a lower bound on timing jitter from integrated detection event when quasi-monochromatic photon states are detected with high efficiency, see footnote\footnote{Timing jitter for a photo detection has three contributing components \cite{vanenk2017}. The first comes from the temporal spread of the mode onto which the measurement projects \cite{spectralPOVM} and is intrinsic to the resonance structure of the photodetector. The second comes from integrating a continuously monitored continua to form discrete detection event, which is necessary for an accurate information-theoretic characterization of photo detection and influences photo detection efficiency. It is this part of the jitter that $\tilde{\Gamma}^{-1}$ is a lower bound for; it sets the timescale for integration where monochromatic states are detected with high efficiency. The third contribution comes from input signals with long temporal wave-packets and, since it is input-dependent, will be ignored in this analysis since here we assume no priors about the single-photon input.}.)

Third,
we can also define a delay (latency) using the polar decomposition of $T(\omega)$ 
\be
T(\omega)=|T(\omega)|\exp(i\phi(\omega))
\ee
and using the standard definition of group delay
as
\be \label{groupdelaydef}
\tau_g(\omega)=-\frac{d\phi(\omega)}{d\omega}.
\ee 

We can see how this group delay directly relates to an experimentally-measured latency by considering a temporal wavepacket $\psi_{\rm in}(t)$ and Fourier transform $\tilde{\psi}_{\rm in}(\omega)$. We can then write the output single-photon state 

\bea\label{filtered state}
\ket{\psi'} &=&\int_0^\infty d\omega \tilde{\psi}_{\rm in}(\omega) R(\omega) \had_{\rm out} (\omega) \vac \nonumber \\
&+& \int_0^\infty d\omega \tilde{\psi}_{\rm in}(\omega) T(\omega) \hbd_{\rm out} (\omega) \vac 
\eea where the first and second terms correspond to the reflected and transmitted parts of the single-photon state, respectively  (for details, see Ref.~\cite{spectralPOVM}) and we have introduced a reflection coefficient $R(\omega)$ satisfying $|T(\omega)|^2 + |R(\omega)|^2 =1$ and $R^*(\omega) T(\omega) + R(\omega) T^*(\omega) = 0$ at every frequency $\omega$. From (\ref{filtered state}) we note that, after interacting with the network, the transmitted wavepacket will have the form $\psi_{\rm out,T}(t) = FT^{-1} [\tilde{\psi}_{\rm in}(\omega) T(\omega)]$. (Note that the wavepacket described by $\psi_{\rm out,T}(t)$ will be sub-normalized by construction; it only corresponds to the portion of the full single photon state $\psi'(t)\equiv \braket{t}{\psi'}$ that is transmitted!) For a long input pulse with central frequency $\omega'$, we find $\psi_{\rm out,T}(t)\approx |T(\omega')| \psi_{\rm in} (t-\tau) e^{i\omega'\tau'}$ with $\tau'$ the difference between the group delay $\tau_g(\omega')$ defined above and the (here irrelevant) phase delay. 

The effect of the group delay on an arbitrary input photon state is to selectively delay and reshape the transmitted wavepacket. Of course, if an input photon has a wide spread of frequencies, a differential group delay $\tau_g(\omega)=-\frac{d\phi(\omega)}{d\omega}$ may increase (or decrease) the temporal spread of the wavepacket, as it will affect each frequency differently. This increase (or decrease) in the arrival times of different frequencies is manifestly input-dependent and will not play a role in the input-independent temporal uncertainty or jitter. We can, however, define an additional quantity characterizing the input-independent group delay-induced dispersion 

\be\label{maxdisp}
\mathcal{T}_g =\int_{0}^\infty d\omega \left |\frac{d\tau_g(\omega)}{d\omega}\right | |T(\omega)|^2
\ee whose definition agrees with our physical intuition that a constant group delay over the transmission window will not contribute to dispersion, nor will frequencies that are not transmitted regardless of how large $\frac{d\tau_g(\omega)}{d\omega}$ may be. We find that, for a flat transmission function that is unity over some spectral range $\omega_0\pm\delta\omega$ and zero everywhere else and a monotonic group delay $\tau_g(\omega)$, (\ref{maxdisp}) gives the difference in group delay between the minimum and maximumly transmitted frequencies: $\mathcal{T}_g=|\tau_g(\omega_0-\delta\omega) - \tau_g(\omega_0+\delta\omega)|$. This is the maximum dispersion possible for any input to this system. 

Finding key conditions that change the transmission function $T(\omega)$ and frequency-dependent group delay $\tau_g(\omega)$ are important for the design of coupled-resonator optical waveguide (CROW) networks \cite{yariv1999} for delay-lines \cite{poon2004} and spectral filtering \cite{madsen2000}, where the transmission efficiency, frequency-dependent group delay, and spectral bandwidth will all affect performance. 
These are the three quantities we focus on in the rest of the chapter. 

Knowing $T(\omega)$ for a specific photodetector also allows one to construct a simplified positive-operator valued measure (POVM) by assuming any excitation in the output continuum results in a click (for details, see the last section of this chapter). The simple POVM element corresponding to a click after the photodetector has been left on for a very long time (in particular, long compared to the bandwidth $\tau\gg \tilde{\Gamma}^{-1}$) has the particularly simple form

\bea\label{povmlongtime}
\hat{  \Pi}=\int_{0}^{\infty}\,d\omega\,|T(\omega)|^2 \ket{\omega}\bra{\omega}.
\eea

$\hat{ \Pi}$ is defined such that
the probability of a photon in an arbitrary input state $\hat{ \rho}$ being detected is given by the Born rule $\textnormal{Pr}=\textnormal{Tr}\left( \hat{ \Pi}\hat{ \rho}\right)$. For example, any photon state $\hat{ \rho}=\sum_i \lambda_i\,\ket{\omega_i '}\bra{\omega_i '}$ where $|T(\omega_i ')|^2 = 1$ and $\sum_i \lambda_i = 1$ will be detected with unit probability. (The states the photodetector can detect perfectly include both pure states [when only one $\lambda_i$ is non-zero] and mixed states comprised entirely of frequencies where $|T(\omega_i')|^2 = 1$.) Of course, no photon is truly monochromatic (or discretely polychromatic), but it can be effectively so if the wave-packet envelope is long compared to the inverse spectral bandwidth $\tilde{\Gamma}^{-1}$. 

%

\vspace{-1em}

\section{Simple example}
 \begin{figure}[h] 
 \centering
	\includegraphics[width=.8\textwidth]{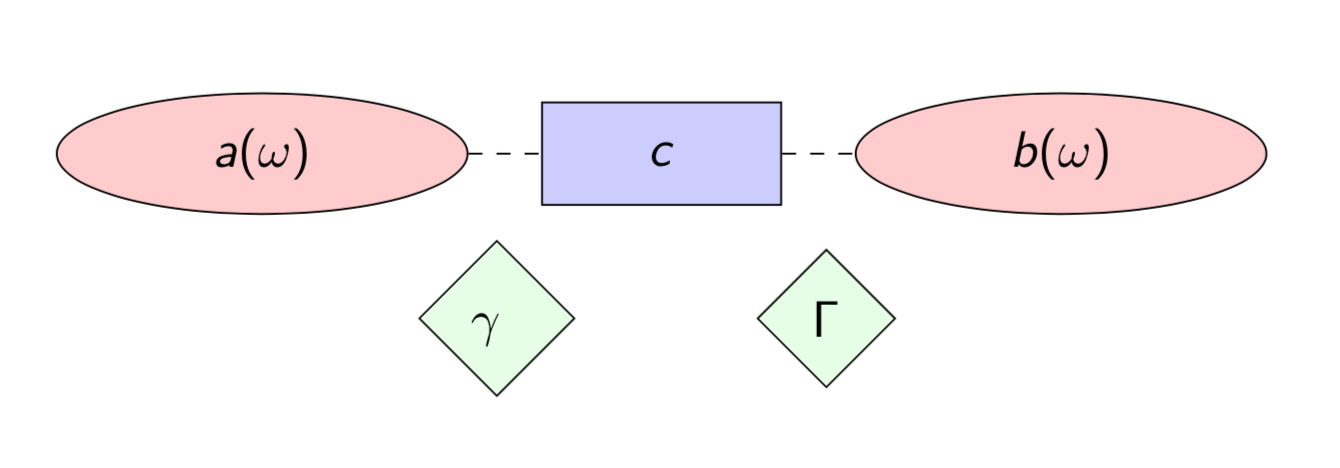} 
	\caption{A simple network comprised of a single discrete state described by an operator $c$ (the bosonic replacement for the two-level system fermionic lowering operator $\sigma^-$) and decaying to left (input, external) and right (output, internal) continua $a$ and $b$ at rates $\gamma$ and $\Gamma$, respectively. That is, here the black box of Fig. \ref{2ports} contains just one two-level system. This exactly describes filtering of light by a Fabry-Perot cavity.} 
	\label{simpleschem}
\end{figure}

The simplest quantum network consists of a single two-level system with a ground state $\ket{g}$ and an excited state $\ket{e}$, with the incoming photon coupling these states (Fig. \ref{simpleschem}). The two-level system is described by fermionic raising and lowering operators $\sigma^+ = \ket{e}\bra{g}$ and $\sigma^- = \ket{g}\bra{e}$ (this is also the simplest model of a photodetector, see \cite{Glauber}). Physically, this excited state could be any discrete state of a single absorber, e.g. the p-state of an atom (that is, an atomic excited state with exactly one quanta of angular momentum) which then decays to a monitored flat continuum. (Indeed, single absorbers such as atoms \cite{Cirac1997,vanEnk2000,Tey2008,Pinotsi2008,Wang2011,haroche2013}, single molecules \cite{Wrigge2007}, NV-centers \cite{iowaPOVM}, and quantum emitters \cite{Zumofen2008,Chen2011} are able to efficiently couple to and absorb single photons.)

In the Heisenberg picture, the evolution of these raising and lowering operators $\sigma^+$ and $\sigma^-$ will determine whether a photon makes it from one side of the network to the other. By focusing our analysis on cases where at most a single photon is in the network, we can use the equivalence between the two-level system and the simple harmonic oscillator to simplify our problem ab initio: we replace the fermionic raising and lowering operators with bosonic creation and annihilation operators $c^\dagger$ and $c$\footnote{Since both operators and their expectation values (mode amplitudes) will satisfy the same systems of equations, we will omit hats throughout this chapter.}. 

We follow standard input-output theory here \cite{input1985}, separating the full evolution of both continua $a$ and $b$ into input and output modes (Fig. \ref{2ports}). By formally solving the Heisenberg evolution equations for the two input continuum mode annihilation operators $\bin$ and $\ain$, we can write the effective system Hamiltonian that governs the evolution of the system operators $c$ and $c^\dagger$

\bea\label{Hamiltonian1}
&\,\\
H&=\hbar\omega_0 c^\dagger c -  \sqrt{\gamma}(c^\dagger \ain + c\adin) - \sqrt{\Gamma} (c^\dagger \bin + c\bdin)\nonumber
\eea where we have identified $\omega_0$ as the resonance frequency, and $\gamma$ and $\Gamma$ as the left and right side couplings to two continua $a$ and $b$ respectively\footnote{In assuming these couplings to be frequency independent, we are invoking a modified version of the first Markov approximation. Formally, we define ${\gamma_i=2\pi\,\kappa_i^2(\omega_i)}$ where $\omega_i$ is the resonance frequency of the $i$th discrete state and $\kappa_i(\omega)$ is the coupling between the $i$th discrete state and the left continuum at the discrete state frequency. This implies that, in an experiment, the decays $\gamma_i$ and resonances $\omega_i$ cannot be varied independently, which is well known in the context of the Thomas-Reiche-Kuhn sum rule for electric dipole transitions \cite{kuhn1992}.}. In the Heisenberg picture, the time evolution of the discrete state annihilation operator is given by
\bea\label{1state}
\dot{c}=-i\omega_0 c -\frac{\gamma+\Gamma}{2} c - \sqrt{\gamma} \ain - \sqrt{\Gamma}\bin.
\eea

The input mode operators $\ain$ and $\bin$ and output mode operators $\aout$ and $\bout$ are determined by open quantum system evolution of the discrete state operator $c$ in (\ref{1state}) and the two boundary conditions 
\bea\label{1statebound}
\aout-\ain= -\sqrt{\gamma} c\nonumber \\
\bout-\bin= -\sqrt{\Gamma} c.
\eea

It is easiest to solve the equations by taking the Fourier transform. Unitarity implies the existence of a transfer matrix relating in and out fields in the spectral domain

\bea\label{transfer}
\begin{bmatrix}
    \aout (\omega)       \\
    \bout (\omega)      \\
\end{bmatrix} = \begin{bmatrix}
    R(\omega)   & T(\omega)   \\
    T(\omega) & R(\omega)       \\
\end{bmatrix} \,\begin{bmatrix}
    \ain (\omega)       \\
    \bin (\omega)      \\
\end{bmatrix} 
\eea 
where $|T(\omega)|^2 + |R(\omega)|^2 = 1$ (resulting from our assumption there are no internal losses) \cite{vogelswelsch}. Defining a detuning $\Delta=\omega-\omega_0$, we can easily solve (\ref{1state}) in terms of the Fourier transform of the discrete state annihilation operator 

\bea\label{1statefreq}
c(\omega)= \frac{- \sqrt{\gamma} \ain(\omega) - \sqrt{\Gamma}\bin(\omega)}{\frac{\gamma+\Gamma}{2} -i\Delta}
\eea yielding a transmission function

\bea\label{T1}
T(\omega)= \frac{\sqrt{\gamma\Gamma}}{\frac{\gamma+\Gamma}{2} -i\Delta}.
\eea
We can see from (\ref{T1}) that perfect transmission ($|T(\omega)|^2=1$) occurs only when $\gamma=\Gamma$ and $\Delta=0$. These are the well-known conditions of balanced mirrors and on-resonance required for perfect transmission through a Fabry-Perot cavity \cite{siegman86}. 

We can also calculate the frequency dependent group delay

\bea\label{T1}
\tau_{g} (\omega)=\frac{\frac{\Gamma+\gamma}{2}}{\left(\frac{\Gamma+\gamma}{2}\right)^2 + \Delta^2}.
\eea  Like $T(\omega)$, the group delay is also a Lorentzian with width $\frac{\Gamma+\gamma}{2}$, and that frequencies close to resonance spend the most time in the network with a maximum group delay of $\frac{2}{\Gamma+\gamma}$ on resonance. We similarly find the input-independent group delay-induced dispersion (\ref{maxdisp}) to be $\mathcal{T}_g = \frac{8\gamma\Gamma}{(\gamma+\Gamma)^3}$.

We can also use $T(\omega)$ to calculate a spectral bandwidth (not to be confused with the channel bandwidth discussed in \cite{vanenk2017})

\bea \label{GammaInt}
 \tilde{\Gamma} &= \frac{1}{\pi}\int_{0}^\infty \,d\omega\,|T(\omega)|^2\nonumber\\
 &= \frac{2\Gamma\gamma}{\Gamma+\gamma}
\eea which is a measure of the number of frequencies that can be efficiently detected. For this simple case, we note that $\tau_g (\omega)=  \tilde{\Gamma}^{-1} |T(\omega)|^2$ and thus
$\int\, d\omega \tau_g(\omega)=\pi$. While this expression will not be true for a general network, the shift in phase by $\pi$ as a resonance frequency is crossed is a universal feature of networks, as we shall see shortly

\section{Quantum Networks}

We now set up the general problem of an arbitrary network of discrete states connecting two continua. The Hamiltonian is a straightforward generalization of (\ref{Hamiltonian1})

\bea\label{HamiltonianGen}
\hfill H=\sum_i\hbar\omega_i c^\dagger_i c_i &-& \sum_{ij} g_{ij} (c_ic^\dagger_j + c^\dagger_ic_j)\hfill\phantom{211}\\
- \sum_i \sqrt{\gamma_i}(c^\dagger_i \ain+ c_i\adin) &-& \sum_i \sqrt{\Gamma_i} (c^\dagger_i \bin + c_i \bdin)\nonumber \eea where we've now defined a real coherent coupling between discrete states $g_{ij}$ (we define $g_{ii}=0$ for each state). Some states may not be coupled to one (or both) continuum, in which case either $\sqrt{\gamma_i}$ or $\sqrt{\Gamma_i}$ (or both) will be zero. 

We can similarly generalize the operator evolution in (\ref{1state}) for an arbitrary network; moving to the spectral domain, we write the spectral dependence of the discrete state operators 

\bea\label{quantlangspect}
-i\Delta_i c_i (\omega)= &- \sum\limits_j\,\left(\frac{\sqrt{\gamma_i\,\gamma_j} + \sqrt{\Gamma_i\,\Gamma_j}}{2} + i g_{i\,j}\right) c_j(\omega) -\sqrt{\gamma_i}\,a_{\rm in}(\omega)  -\sqrt{\Gamma_i}\,b_{\rm in}(\omega).
\eea

Similarly to (\ref{1statebound}), we can write boundary conditions for the two continua with an arbitrary network 

\bea\label{Nstatebound}
\aout(\omega)-\ain(\omega)= -\sum\limits_{i}\sqrt{\gamma_i} c_i(\omega)\nonumber \\
\bout(\omega)-\bin(\omega)= -\sum\limits_i \sqrt{\Gamma_i} c_i(\omega).
\eea

To move from (\ref{Nstatebound}) to the transfer matrix (\ref{transfer}) involves solving $N$ systems of $N$ coupled first-order differential equations\footnote{Luckily, there is some redundancy. Once one has solved for one $c_i$ in terms of the input and output fields, one can permute the labels to generate the remaining solutions.}, as each discrete state amplitude $c_i(\omega)$ depends on every other amplitude. Three methods for solving these equations are as follows.

The first is to take the weak-coupling limit $2\,g_{ij}\ll\sqrt{\gamma_i\,\gamma_j} + \sqrt{\Gamma_i\,\Gamma_j}\,\forall i,j$ and truncate the solutions after some power in $g_{ij}$, making (\ref{RGen}) the $0$th order approximate solution with higher order corrections. However, this method fails if even a single discrete state decouples from both continua, as is the case for many networks of interest.

The second is to use numerical techniques to diagonalize the systems of equations and find the transmission function numerically \cite{datta2005}. However, this rapidly gets harder with large systems, and masks the analytic conditions for perfect transmission we are interested in identifying. 

The third method is to use a variety of mathematical techniques to solve these $N$ systems of $N$ coupled first-order differential equations analytically, which is the route taken in this analysis. These tricks include making use of correlations between discrete states, permuting labels, and using generalized continued fraction formulae. Here, we will use these techniques to describe three large classes of systems that can be solved exactly with arbitrary couplings, decays, and resonant frequencies. These are parallel networks, series networks, and hybrid networks.

\underline{Parallel}: each discrete state is directly coupled to both continua ($\gamma_i\neq0$ and $\Gamma_i\neq 0 \,\forall i$) but not to each other. Thus there are multiple parallel paths to the same final state and hence we'll get interference. Physical photo detection platforms described by parallel networks include (i) single atoms with multiple p-states that then decay directly to a continuum (similarly for trapped ions/atoms due to Stark and Zeeman effects, see  Ref.~\cite{Higginbottom2016} for this in generating single photons), (ii) quantum dots \cite{Livache2019}, (iii) structured continua with multiple pseudomodes \cite{Hughes2018}.

\underline{Series}: each of the two continua are coupled to their own discrete state, which are in turn coherently coupled by a chain of intermediate single discrete states. There is just one path from the input continuum to the output continuum. This model describes (i) an atom in a p-state that first decays to a d-state or metastable s-state before decaying to a flat continuum (and analogously for a molecule \cite{young2018}), (ii) coherently coupled frequency filtering in front of a photo detecting platform (for instance, using an antireflective coating \cite{Rosfjord2006} or optical cavities \cite{Dilley2012}), (iii) atomic chains coupled through their p-states (which could be mediated by, for example, a fiber mode \cite{Song2018}).

\underline{Hybrid}: a combination of the above cases. For example, two parallel paths of three steps each or, more generally, layered structured continua (a photon must pass through one before the other). A physical system that can be modeled with a hybrid network is a photosynthetic light-harvesting (i.e. Fenna–Matthews–Olson) complex with multiple paths for coherent transport \cite{Scholes2017,Valleau2017,Chan2018}.

In all cases, we will solve  the systems of equations for $R(\omega)$ directly and make use of the identities $|T(\omega)|^2 = 1-|R(\omega)|^2$ and $T^2(\omega) = -R^2(\omega) \frac{|T(\omega)|^2}{|R(\omega)|^2} $when we calculate the transmission efficiency, spectral bandwidth, and group delay.

\vspace{-1em}
\subsection{Parallel Networks}

 \begin{figure}[h] 
 \centering
	\includegraphics[width=.8\textwidth]{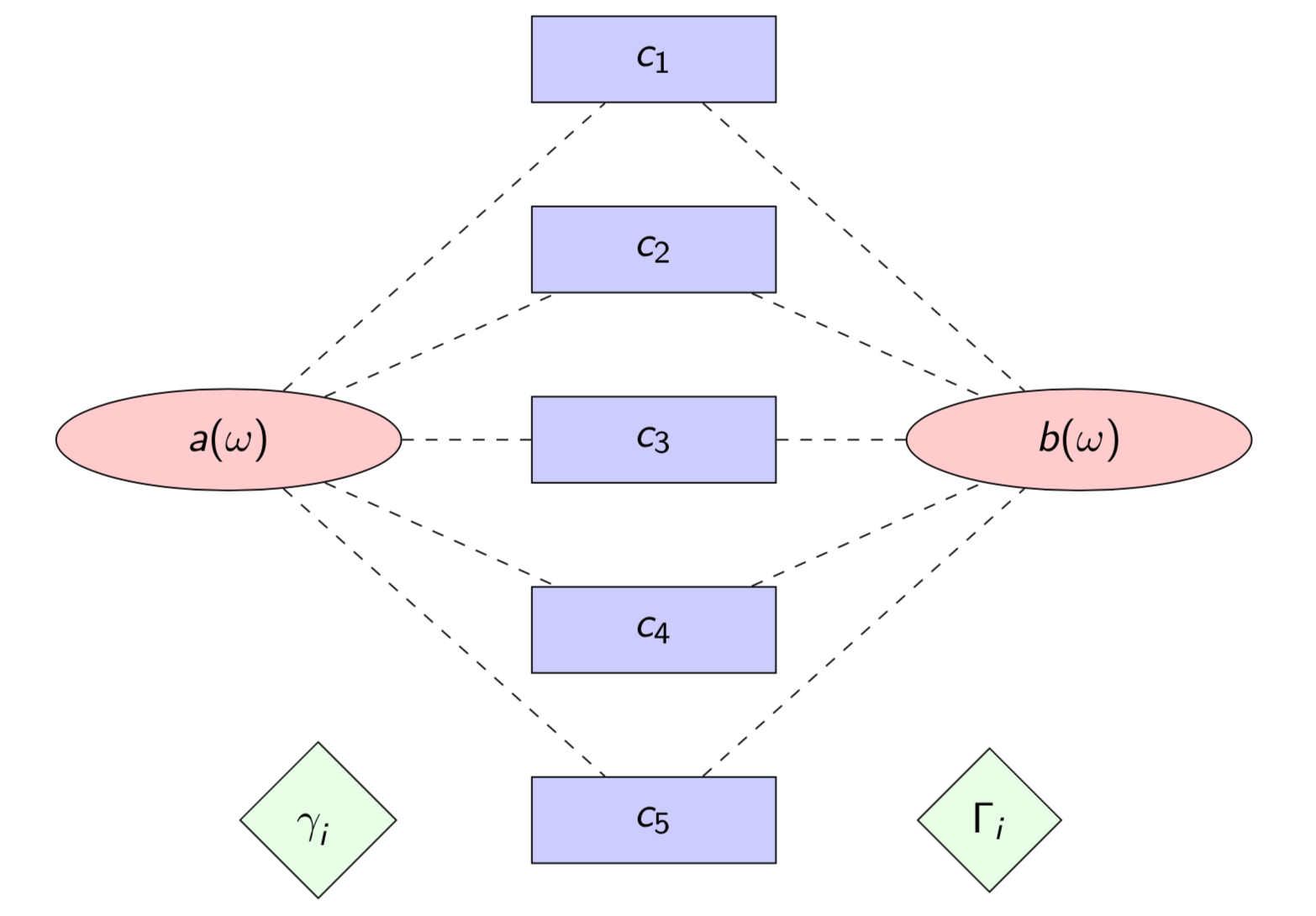} 
	\caption{A parallel network of ${N=5}$ decoupled discrete states, each described by an operator $c_i $ and decaying to left (input, external) and right (output, internal) continua $a$ and $b$ at rates $\gamma_i$ and $\Gamma_i$, respectively.}
	\label{pschem}
\end{figure}

We now consider a parallel quantum network where each discrete state is directly coupled to both continua (Fig. \ref{pschem}). If the discrete states are directly coupled to each other, it is helpful to first diagonalize the system in the basis where they are decoupled. We then consider the new modes to be the discrete states a photon can occupy, each with a new resonance frequency and two decay rates to the two continua. (In the strong coupling limit, Rabi splitting comes into effect and these dressed states are the more physical description anyways \cite{steckquoptics}.) This decoupled manifold of discrete states, each decaying at a rate $\gamma_i$ to the input continuum and a rate $\Gamma_i$ to the output continuum (Fig. \ref{pschem}), is the next simplest case to analyze, and provides a simple model of a photodetector with a simple band structure.

We find the frequency dependence of the discrete state operators

\bea\label{quantlangspectpar}
-i\Delta_i c_i (\omega)= &- \sum\limits_j\,\frac{\sqrt{\gamma_i\,\gamma_j} + \sqrt{\Gamma_i\,\Gamma_j}}{2} c_j(\omega) \nonumber\\
&-\sqrt{\gamma_i}\,a_{\rm in}(\omega)  -\sqrt{\Gamma_i}\,b_{\rm in}(\omega).
\eea

A salient feature of Eq. \ref{quantlangspectpar} is the decay terms produce purely virtual coupling between discrete states coupled to the same continua. These \emph{cannot} be thought of as coupling mediated by continua, as the continua are flat (Markovian) and perfectly dissipative. If an excitation exits the discrete states and enters a continuum, there is no possibility of its ever returning to another discrete; energy transfer from one discrete state is impossible! Thus, the coupling is purely virtual, and we must instead consider it to be a purely information-theoretic phenomena\footnote{Interestingly, this coupling does persist for higher numbers of excitations, an effect we intend to explore in future work.}.

We use (\ref{transfer}) to find $R(\omega)$ by considering an input on only one side of the network (thus setting the expectation value of the other input field operator to zero). This yields an expression $a_{out} (\omega) =R(\omega) a_{in}(\omega)$ (or $b_{out} (\omega) =R(\omega) b_{in}(\omega)$), from which we reconstruct $T(\omega)$. We analytically find the general form of $R(\omega)$ 

\bea\label{RGen}
R(\omega)=\frac{\prod\limits_i \left(\frac{\Gamma_i - \gamma_i}{2} - i\Delta_i\right) - X_-^{(N)}}{\prod\limits_i \left(\frac{\Gamma_i + \gamma_i}{2} - i\Delta_i\right) - X_+^{(N)}}
\eea where $X_\pm^{(N)}$ is a polynomial of order $N-2$ in the detunings $\Delta_i$. (We find $X_\pm^{(1)} = 0$.)  For $N=2$, we can see that $X_\pm^{(2)}$ is symmetric (anti-symmetric) between $\Gamma_i$ and $\gamma_i$

\bea\label{X2}
X_\pm^{(2)} = \left(\frac{\sqrt{\Gamma_1\Gamma_2} \pm \sqrt{\gamma_1\gamma_2}}{2}\right)^2.
\eea

This (anti-)symmetry is also present in higher-$N$ coefficients. From (\ref{RGen}) we can determine a key feature of parallel quantum networks: if some subset of the discrete states have \emph{balanced} decay rates such that $\gamma_i=\Gamma_i$, for large spacings between discrete states compared to the other decay rates $|\omega_i-\omega_j|\gg \gamma_j,\Gamma_j$, we find $R(\omega_i)=0$; input monochromatic photons with frequencies on resonance with those discrete states are transmitted perfectly through the network\footnote{For ${N=2}$, the requirement for large spacing such that ${R(\omega_i)=0}$ is less stringent, we only need it much larger than difference in decay rates ${|\omega_i-\omega_j|\gg (\sqrt{\Gamma_j}-\sqrt{\gamma_j})^2}$.}. 

In general, finding the specific form of $X_\pm^N$ is a numerical task, and we will focus on a simpler case where we can utilize another salient feature of parallel networks: the purely virtual coupling present in (\ref{quantlangspectpar}). Before we consider a network that is uniformly coupled (all decays are the same), we can consider a network with couplings that are inhomogeneous but uniformly unbalanced such that $\Gamma_i = k\,\gamma_i \,\forall i$. We can then write (\ref{quantlangspectpar}) in a simplified form

 \begin{figure}[h] 
 \centering
	\includegraphics[width=.8\textwidth]{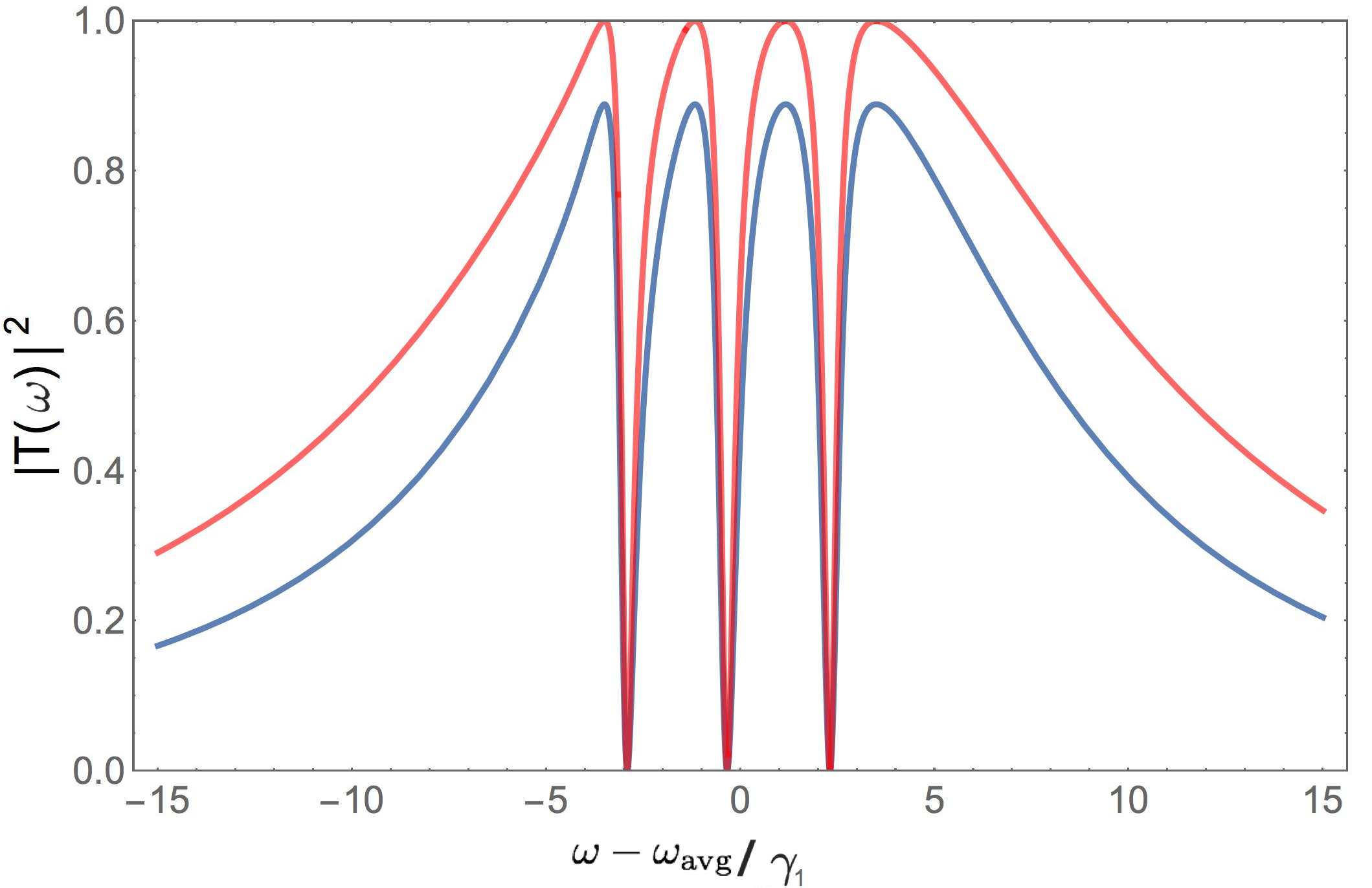} 
	\caption{Transmission probability for parallel networks with ${N=4}$ equally spaced discrete states, each with a decay rate to the input continuum ${\gamma_i=\gamma_1 \,(\frac{7}{5})^{i-1}}$. The decays to the monitored output continuum are uniformly unbalanced (${\frac{\Gamma_i}{\gamma_i}=k\,\forall i}$), with ${k=\frac{1}{2}}$ for the blue (lower) curve and ${k=1}$ for the red (upper) curve. Frequency is measured w.r.t. the average resonance frequency. The four resonance frequencies $\omega_i$ have maximum transmission probability ${|T(\omega_i)|^2=\frac{4k}{(k+1)^2}}$. The three frequencies of perfect reflection (i.e, ${T(\omega_i)=0}$) correspond to solutions of ${\sum_i\frac{\gamma_i}{\Delta_i}=0}$.}
	\label{parralelk}
\end{figure}

    \begin{figure}[h]
    \centering
            \includegraphics[width=.8\textwidth]{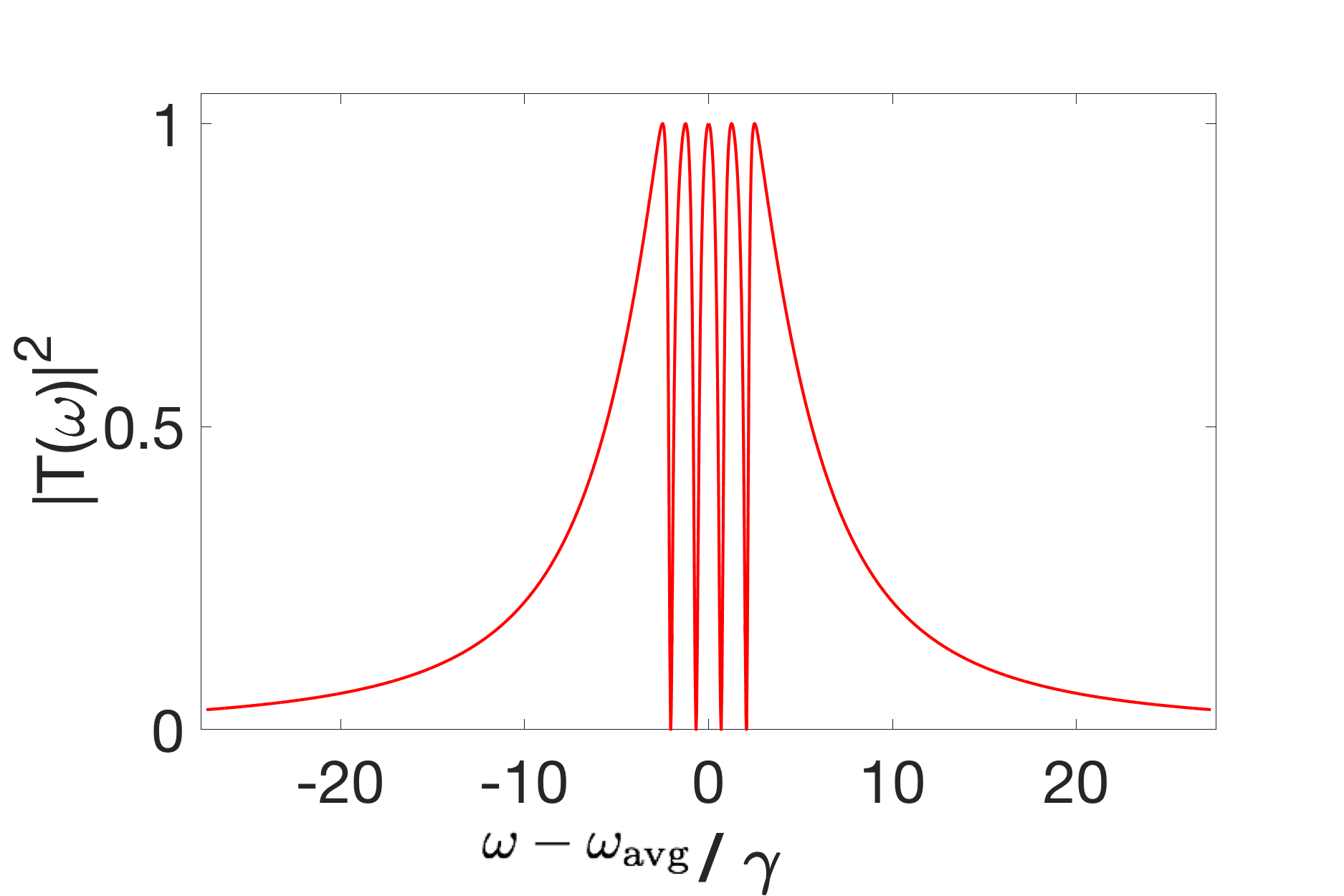}
         \caption{Transmission probability for a parallel network with ${N=5}$ equally spaced discrete states with balanced decay rates to both continua (${\gamma=\Gamma}$). Frequency is measured w.r.t. the average resonance frequency. We observe the five resonance frequencies $\omega_i$ each correspond to a perfectly transmitted frequency. The four frequencies of perfect refection correspond to solutions of ${\sum_i\frac{1}{\Delta_i}=0}$. }  
            \label{parralelhomoT}
        \end{figure}

    \begin{figure}[h]
    \centering
            \includegraphics[width=.8\textwidth]{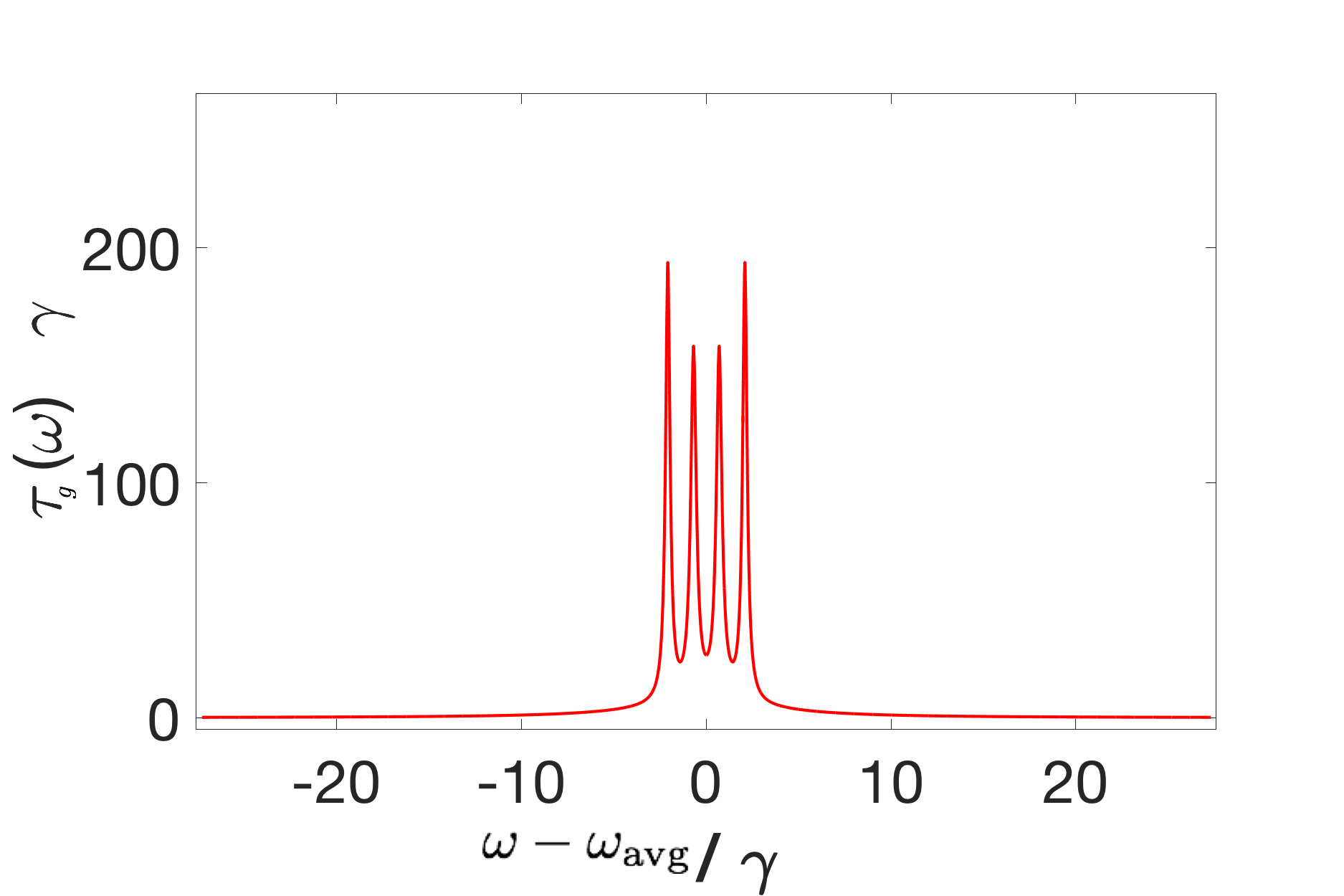}
                     \caption{Group delay (in units of $\frac{1}{\gamma}$) for a parallel network with ${N=5}$ equally spaced discrete states with balanced decay rates to both continua (${\gamma=\Gamma}$). The group delay is always largest for the highest and lowest frequency resonances (except in the large spacing limit, where they are of equal magnitude).}  
                        \label{parralelhomoG}
        \end{figure}

    \begin{figure}[h]
    \centering
             \includegraphics[width=.8\textwidth]{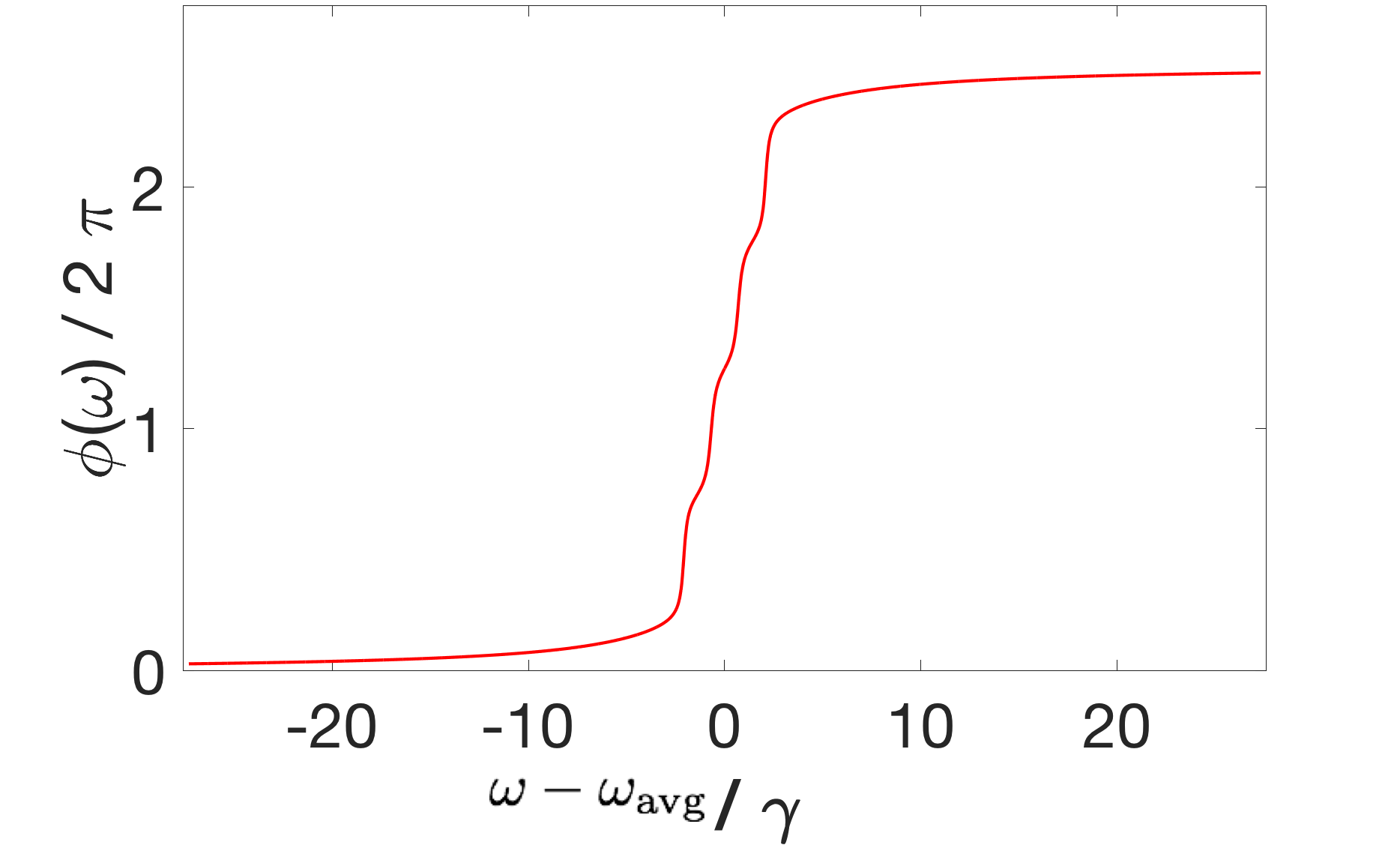}
         \caption{Transmission function phase (in units of $2\pi$) for a parallel network with ${N=5}$ equally spaced discrete states with balanced decay rates to both continua (${\gamma=\Gamma}$). The total change in phase of the transmission function is directly proportional to the number of discrete states. This is due to the phase shift in $\pi$ that occurs when crossing a resonance frequency, as was discussed for the simple example previously.}  
                     \label{parralelhomoP}
    \end{figure}

\bea\label{quantlangspectcorrk}
\,&\,\\
&\frac{i\Delta_i}{\sqrt{\gamma_i}}c_i (\omega) &= \sum\limits_j\frac{\sqrt{\gamma_j}(1+k)}{2} c_j(\omega) +a_{\rm in}(\omega)  +\sqrt{k}\,b_{\rm in}(\omega)\nonumber
\eea so that we can identify strong correlations between discrete state amplitudes
\bea\label{quantlangspectcorrk2}
&\frac{i\Delta_i}{\sqrt{\gamma_i}}c_i (\omega) = \frac{i\Delta_{i'}}{\sqrt{\gamma_{i'}}}c_{i'} (\omega).
\eea

For non-degenerate states, Eq. \ref{quantlangspectcorrk2} means that a photon that is resonant with one state will only excite that state. (For infinitely-narrow discrete states, these correlations are satisfied trivially of course; $c_{i'}(\omega)$ is only non-zero at $\omega_{i'}$.) Eq. \ref{quantlangspectcorrk2} also indicates that the relative phase of discrete state amplitudes is frequency dependent; when $\omega_i < \omega < \omega_{i'}$, there is a relative phase of $\pi$ between $c_i$ and $c_{i'}$. (Outside of photo detection, this purely virtual coupling provides a possible alternative explanation for the destructive interference present in atoms along a fiber \cite{atomwire1, atomwire2} and multi-mode Fabry-Perot cavities \cite{multimodefabry}.) We can see destructive interference directly from the form of the reflection coefficient

\bea\label{reflectk}
R(\omega)=\frac{i-(k-1)\sum\limits_i \frac{\gamma_i}{2(\Delta_i)}}{i-(k+1)\sum\limits_i \frac{\gamma_i}{2(\Delta_i)}}
\eea from which we can determine $|T(\omega)|^2$ and the other quantities of interest. 

We can further specialize to the case of homogenous coupling where $\gamma_i=\gamma$ and $\Gamma_i=\Gamma$. This case is of interest for several reasons: most generally, this assumption directly follows from the first Markov approximation if the spacing between discrete states is small compared to the decays. It also simplifies the form of the correlations between discrete states $\Delta_i c_i(\omega)=\Delta_{i'} c_{i'}(\omega)$\footnote{We now can perform a small sanity check. Considering the two continuum fields together, we define total input and output flux operators ${J_{in} = \sqrt{\gamma} a_{in} + \sqrt{\Gamma} b_{in}}$ and ${J_{out} = \sqrt{\gamma} a_{out} + \sqrt{\Gamma} b_{out}}$. Solving for the total outputs directly from (\ref{Nstatebound}), we find ${J_{out} =\frac{i+ \frac{\gamma+\Gamma}{2}\,\sum_i \frac{1}{\Delta_i}}{i - \frac{\gamma+\Gamma}{2}\,\sum_i \frac{1}{\Delta_i}}\,J_{in}}$. We can immediately see that ${|J_{out}|=|J_{in}|}$; photon flux is preserved through the system at every frequency, as it must be since there are no side channels present.}, as well as the form of the reflection coefficient 

\bea\label{reflecthomo}
R(\omega)=\frac{i-\frac{\Gamma-\gamma}{2}\sum\limits_i \frac{1}{\Delta_i}}{i-\frac{\Gamma+\gamma}{2}\sum\limits_i \frac{1}{\Delta_i}}.
\eea

\lettersection{Perfect transmission} From (\ref{reflectk}) and (\ref{reflecthomo}) we can see that in both cases there are $N$ frequencies of maximum transmission corresponding to each resonant frequency $\Delta_i=0$ with transmission probability $|T(\omega_i)|^2=\frac{4k}{(k+1)^2}$ and $|T(\omega_i)|^2=\frac{4\gamma\Gamma}{(\gamma+\Gamma)^2}$, respectively.  We also see $N-1$ frequencies of destructive interference corresponding to the $N-1$ solutions of $\sum\limits_{i}\frac{\gamma_i}{\Delta_i} =0$ (Fig. \ref{parralelk}) determined solely by the decays and resonances (and notably not by $k$). We similarly observe for the case of a network with homogenous decay rates $N-1$ frequencies of destructive interference corresponding to the solutions of $\sum\limits_{i}\frac{1}{\Delta_i} =0$ (Fig. \ref{parralelhomoT}). One might think that, in principle, a resonant frequency could coincide with a frequency of perfect reflection when $N>2$, in which case they can annihilate. However, this only occurs when two discrete states are energetically degenerate. Since the discrete states would also couple to the same $1$D continuum, this is forbidden by unitarity. To avoid this, we see a resonant frequency that is perfectly transmitted and a frequency of destructive interference move closer together as the spacing between discrete states is decreased. When the degeneracy becomes exact, a discrete state is forced to decouple from the system as the frequencies of perfect reflection and transmission annihilate. 

In general, the condition for perfect transmission through a parallel network is that all the couplings be balanced ($\gamma=\Gamma$ or in the inhomogeneous uniformly balanced case, $k=1$). As we saw in the case for a completely arbitrary parallel network, we see that perfect transmission at some discrete state frequency $\omega_i$ not only requires balanced coupling $\gamma_i=\Gamma_i$, but also that all the other discrete states either be far away in frequency compared to the their decay rates ($|\omega_i-\omega_j|\gg\gamma_j,\Gamma_j$), or also be balanced ($\gamma_j=\Gamma_j$), or a mix of the two.

\lettersection{Spectral Bandwidth} Once we have the reflection coefficient, we can calculate the three quantities of interest. For all three cases we've discussed, we find that the spectral bandwidth is purely additive $\tilde{\Gamma} = \sum_i \frac{2\gamma_i\Gamma_i}{\gamma_i+\Gamma_i}$ and is completely independent of the spacing between discrete states\footnote{The uncertainty in frequency, as defined in \cite{vanenk2017} and calculated entropically from the spectral POVM \cite{spectralPOVM}, is in this case directly proportional to the bandwidth (and hence also independent of discrete state spacing).}.

\lettersection{Group Delay} We note that, unlike the simple model, the sharp peaks in the group delay (Fig. \ref{parralelhomoG}) correspond to frequencies of destructive interference and are greatest for the outermost frequencies of destructive interference despite all the frequencies of note in (Fig. \ref{parralelhomoT}) being completely destructive or constructive. We also observe that the relationship between the three quantities of interest discussed for the simple model is not present: here $\tau_g(\omega_i) > \tilde{\Gamma}^{-1} |T(\omega_i)|^2$ for each resonant frequency. We find that the phase of $T(\omega)$ increases by $\pi$ with each resonant frequency (\ref{parralelhomoP}). This provides a novel application for single-photon interferometry for resolving tightly-structured resonance structures and explains why, whereas the spectral bandwidth $\tilde{\Gamma}$ is independent of discrete state spacing, the group delay increases with close spacing; the same change in phase is occurring in a smaller spectral range so the magnitude of the group delay $|\tau_g(\omega)|=|\frac{d\phi(\omega)}{d\omega}|$ increases.

\subsection{Series Networks}

\begin{figure}[h!] 
\centering
	\includegraphics[width=.8\textwidth]{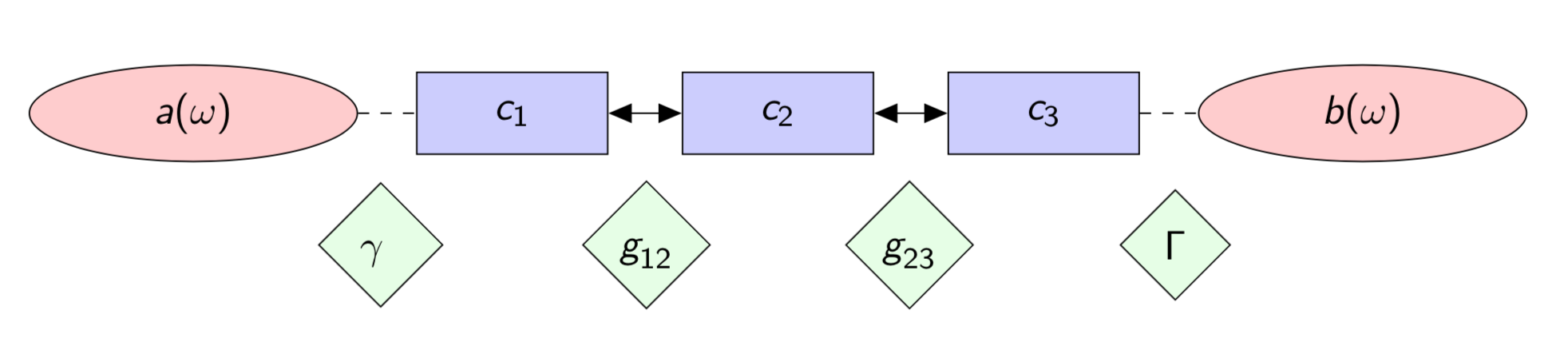} 
	\caption{A series network of three coherently coupled discrete states, each described by an operator $c_i$ and coupled to each other at rates $g_{ij}$. The first and last states decay to left (input, external) and right (output, internal) continua $a$ and $b$ at rates $\gamma$ and $\Gamma$, respectively.}
	\label{seriesschem}
\end{figure}

        \begin{figure}[h]
        \centering
            \includegraphics[width=.8\textwidth]{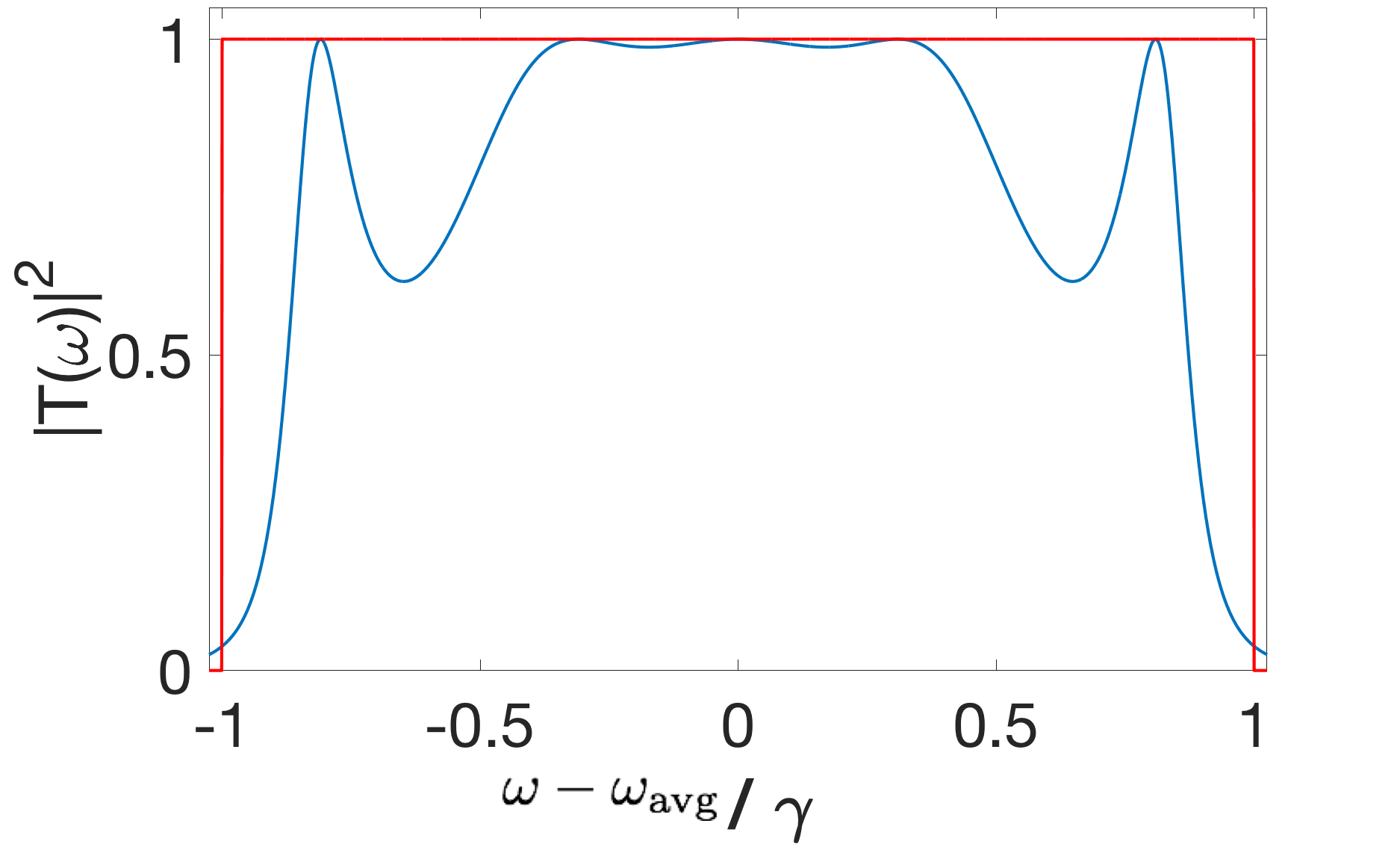}
        \caption{\small Transmission probability for a quantum network with ${N=5}$ discrete states in series with no relative detuning and balanced decay rates ${\gamma=\Gamma}$ for the special case of homogenous critical coupling ${g_{ij}\rightarrow g=\frac{\sqrt{\gamma\,\Gamma}}{2}}$. For comparison with the actual transmission probability (blue curve), we also plot a fictitious transmission probability (red bounding box) that is unity for ${-2g\leq \omega \leq 2g}$ and zero elsewhere. Frequencies are measured w.r.t. resonance.  Meeting both the balanced decay and critical coupling conditions ensures that the on-resonance transmission is both unity and maximally broadened, but are not necessary conditions for perfect transmission to occur at some frequency.} 
            \label{CriticalOddBalanced}
        \end{figure}

        \begin{figure}[h]
        \centering
            \includegraphics[width=.8\textwidth]{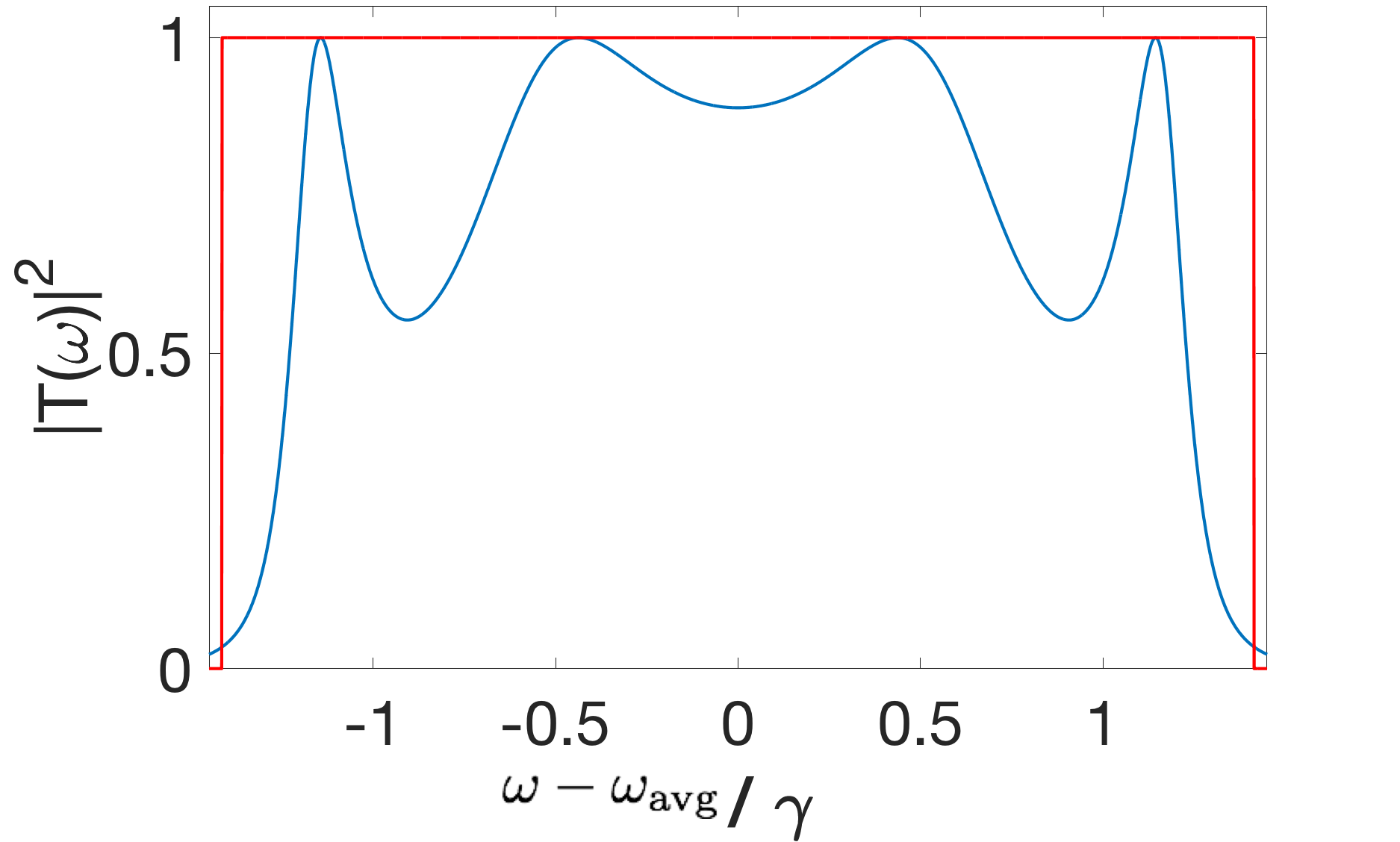}
                  \caption{\small Transmission probability for a quantum network with ${N=5}$ discrete states in series with no relative detuning and unbalanced decay rates ${2\gamma=\Gamma}$ for the special case of homogenous critical coupling ${g_{ij}\rightarrow g=\frac{\sqrt{\gamma\,\Gamma}}{2}}$. Again, we observe perfect transmission for certain frequencies.} 
            \label{CriticalOddUnbalanced}
        \end{figure}

        \begin{figure}[h]
        \centering
            \includegraphics[width=.8\textwidth]{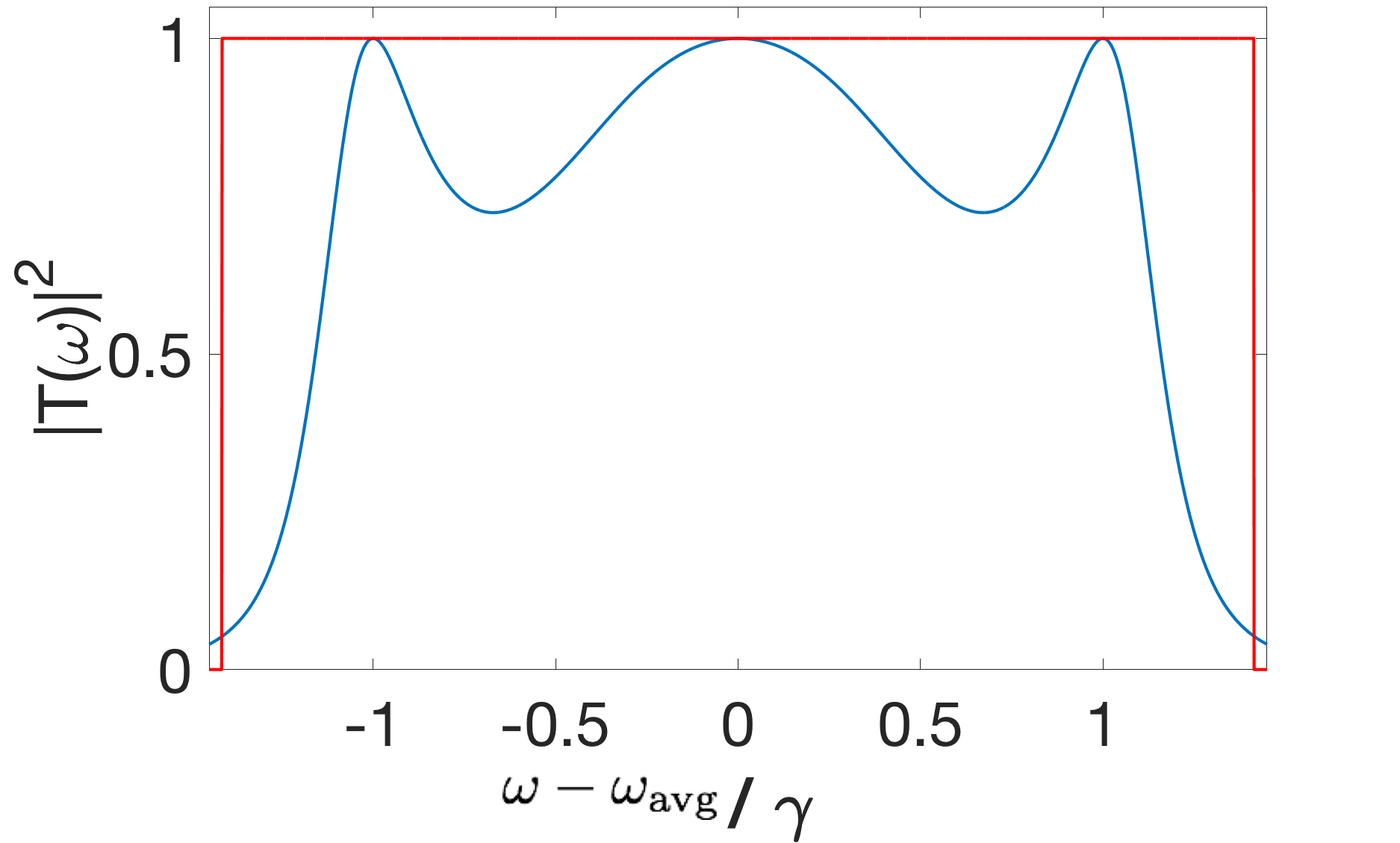}
                  \caption{\small Transmission probability for a quantum network with ${N=4}$ discrete states in series with no relative detuning and unbalanced decay rates ${2\gamma=\Gamma}$, again for the special case of homogenous critical coupling ${g_{ij}\rightarrow g=\frac{\sqrt{\gamma\,\Gamma}}{2}}$ ensuring perfect transmission at some frequencies} 
            \label{CriticalEvenUnbalanced}           
        \end{figure}
        
        \begin{figure}[h]
        \centering
            \includegraphics[width=.8\textwidth]{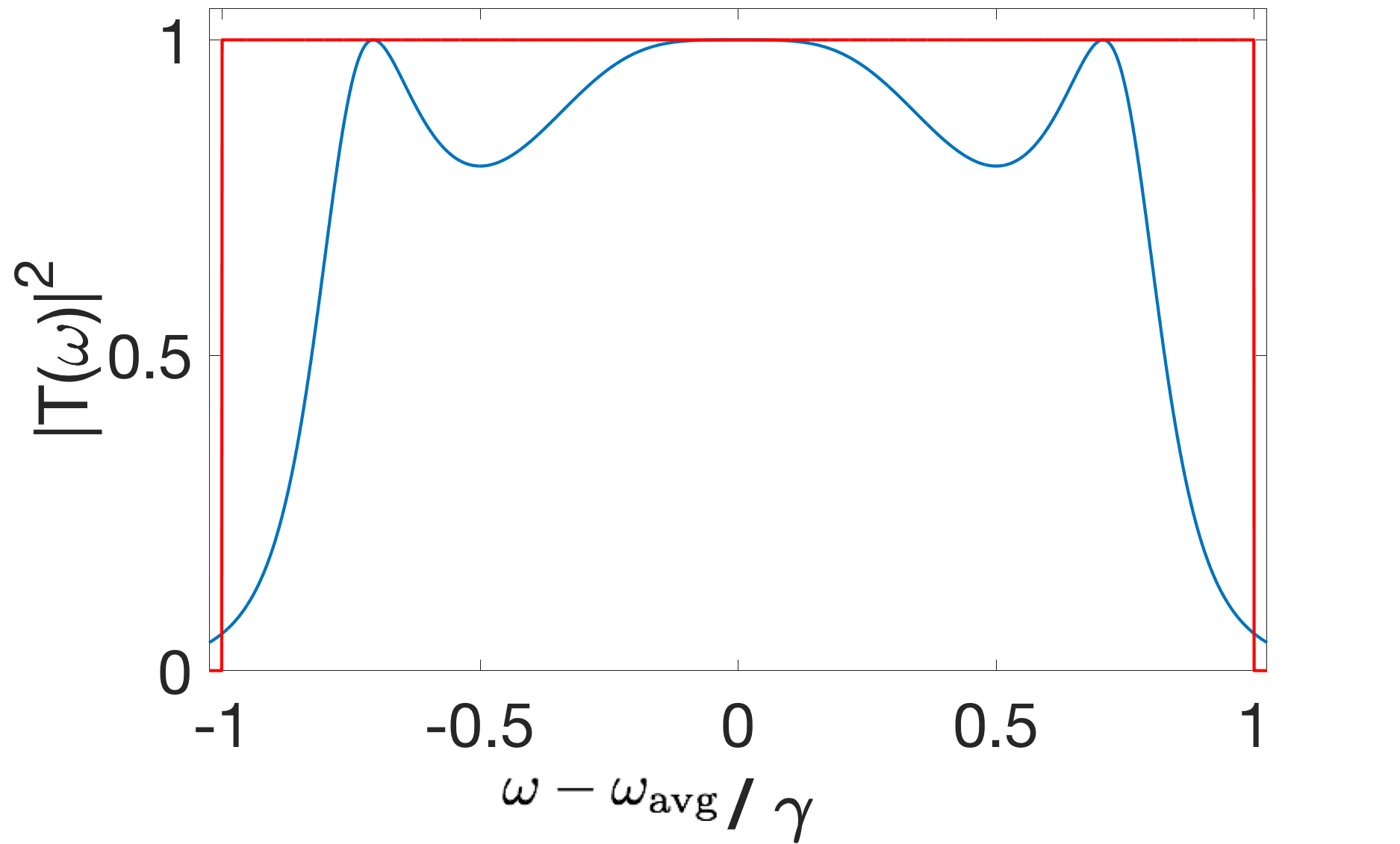}
 \caption{\small Transmission probability for a quantum network with ${N=4}$ discrete states in series with no relative detuning and balanced decay rates ${\gamma=\Gamma}$. Here critical coupling ${g_{ij}\rightarrow g=\frac{\sqrt{\gamma\,\Gamma}}{2}}$ ensures both perfect transmission and maximal broadening on-resonance.} 
            \label{CriticalEvenBalanced}
        \end{figure}

In general, we cannot diagonalize a completely arbitrary network in terms of a single set of parallel states; there may be a causal relationship embedded in the network structure; for example, for an atom in an s-state coupled via a photon to one or many p-states which subsequently decay to multiple d-states, we would have to diagonalize the p-states and d-states separately. A class of such systems are series quantum networks (Fig. \ref{seriesschem}), where only one state is coupled to each continuum, with the other states forming a chain between the two outer ones ($\gamma_{i>1} = 0$, $\Gamma_{i<N} = 0$, and $g_{ij} = 0$ for $j\neq i\pm1$). This provides a natural (and especially simple) model for energy transport and repeated spectral filtering (a series of Fabry-Perot cavities).  Furthermore, we can analytically determine the transmission function for arbitrary series networks, as we will now proceed to do.

\begin{figure}[h]
\centering
            \includegraphics[width=.8\textwidth]{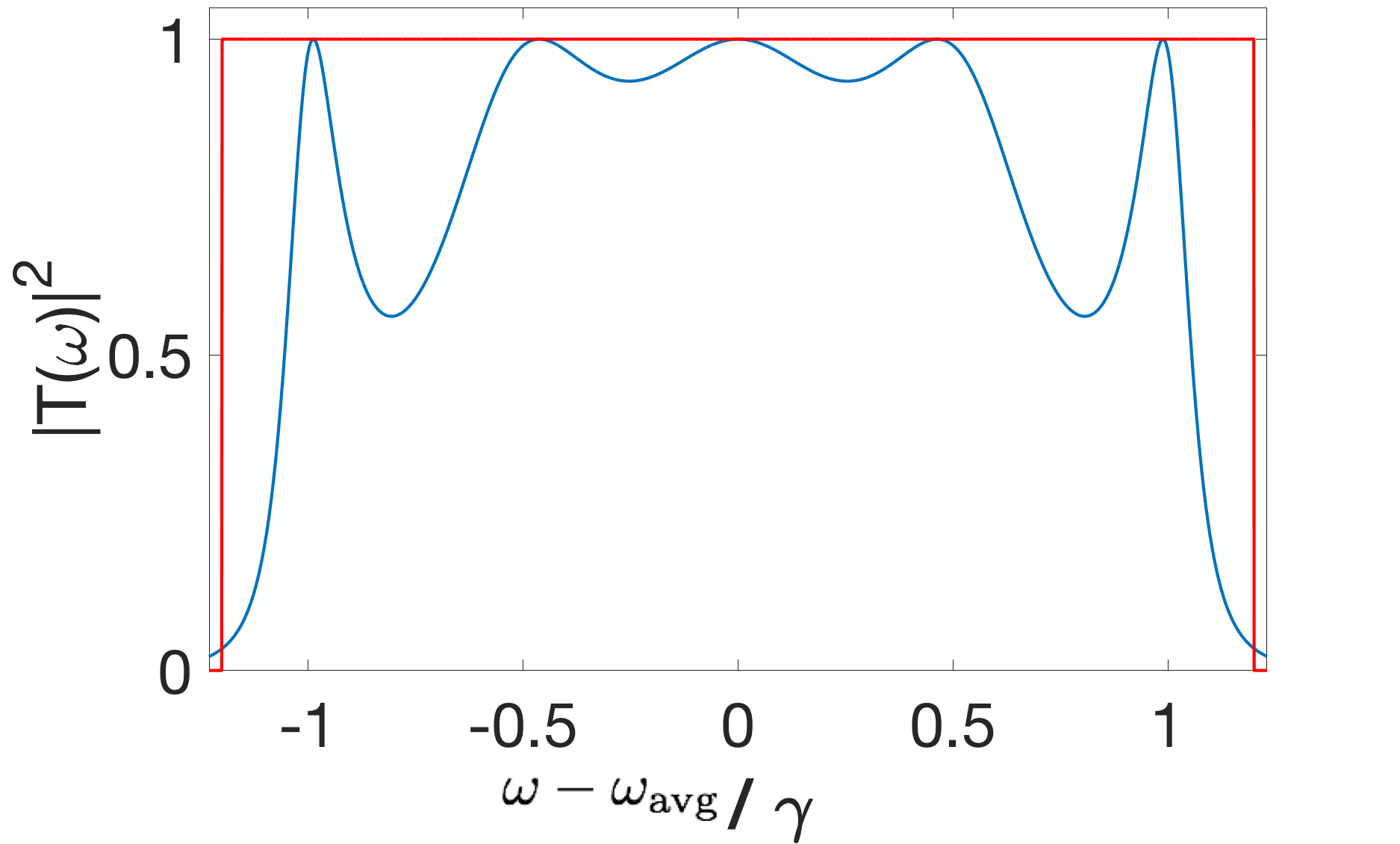}
            \caption{Transmission probability for networks with ${N=5}$ discrete states in series with no relative detuning for the special case of homogenous balanced decays ${\gamma=\Gamma}$ (blue curve) and over-coupling ${g=\frac{6}{5}\frac{\sqrt{\gamma\Gamma}}{2}}$. A balanced over-coupled network will always have more peaks of perfect transmission than an under-coupled or critically-coupled one, though on-resonance transmission may not be a local maxima.}    
            \label{OverOddBalanced}
        \end{figure}
        
\begin{figure}[h]
\centering
            \includegraphics[width=.8\textwidth]{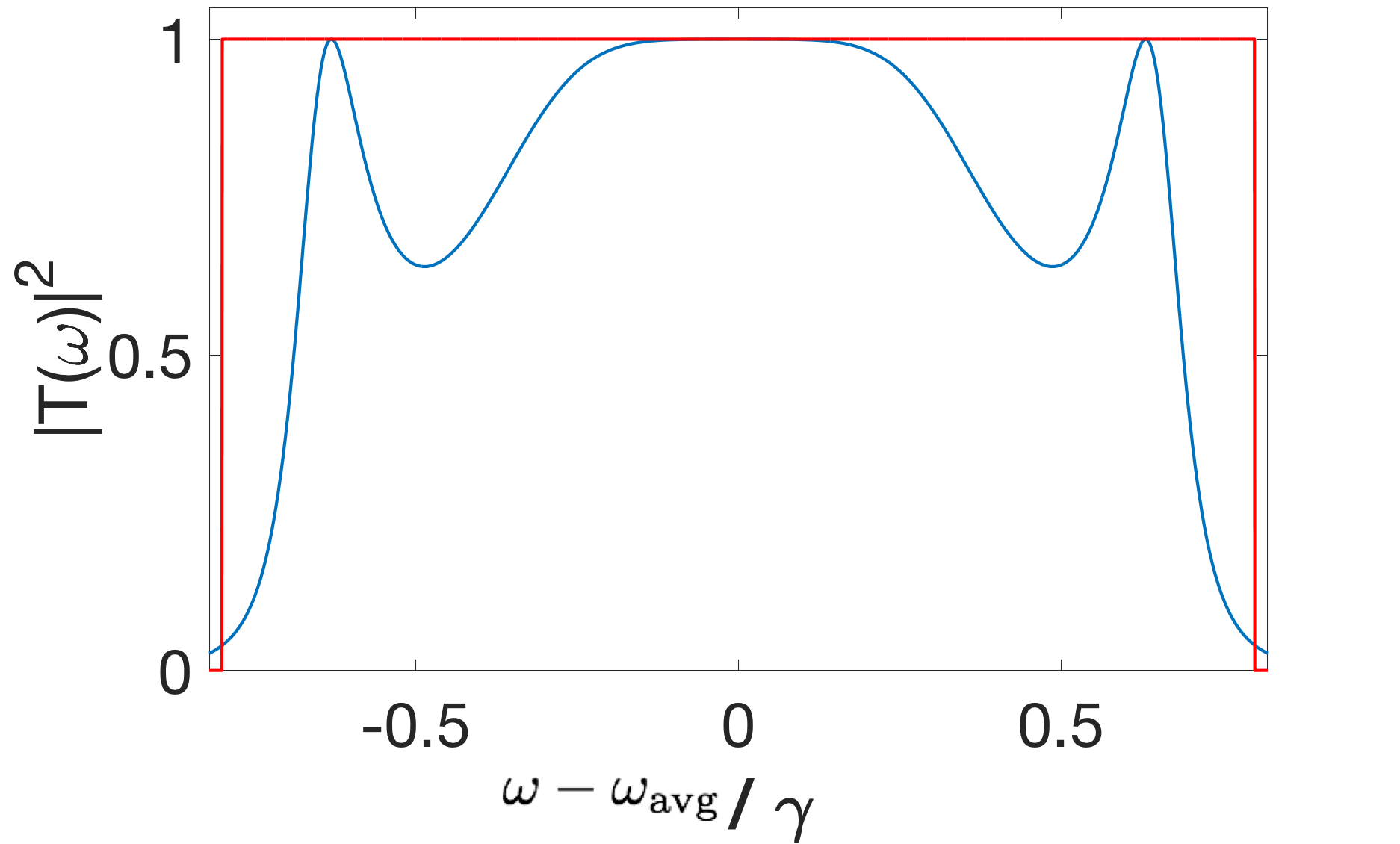}
            \caption{Transmission probability for networks with $N=5$ discrete states in series with no relative detuning for the special case of homogenous balanced decays ${\gamma=\Gamma}$ (blue curve) and under-coupling ${g=\frac{4}{5}\frac{\sqrt{\gamma\Gamma}}{2}}$. }       
            \label{UnderOddBalanced}
        \end{figure}

\begin{figure}[h]
\centering
            \includegraphics[width=.8\textwidth]{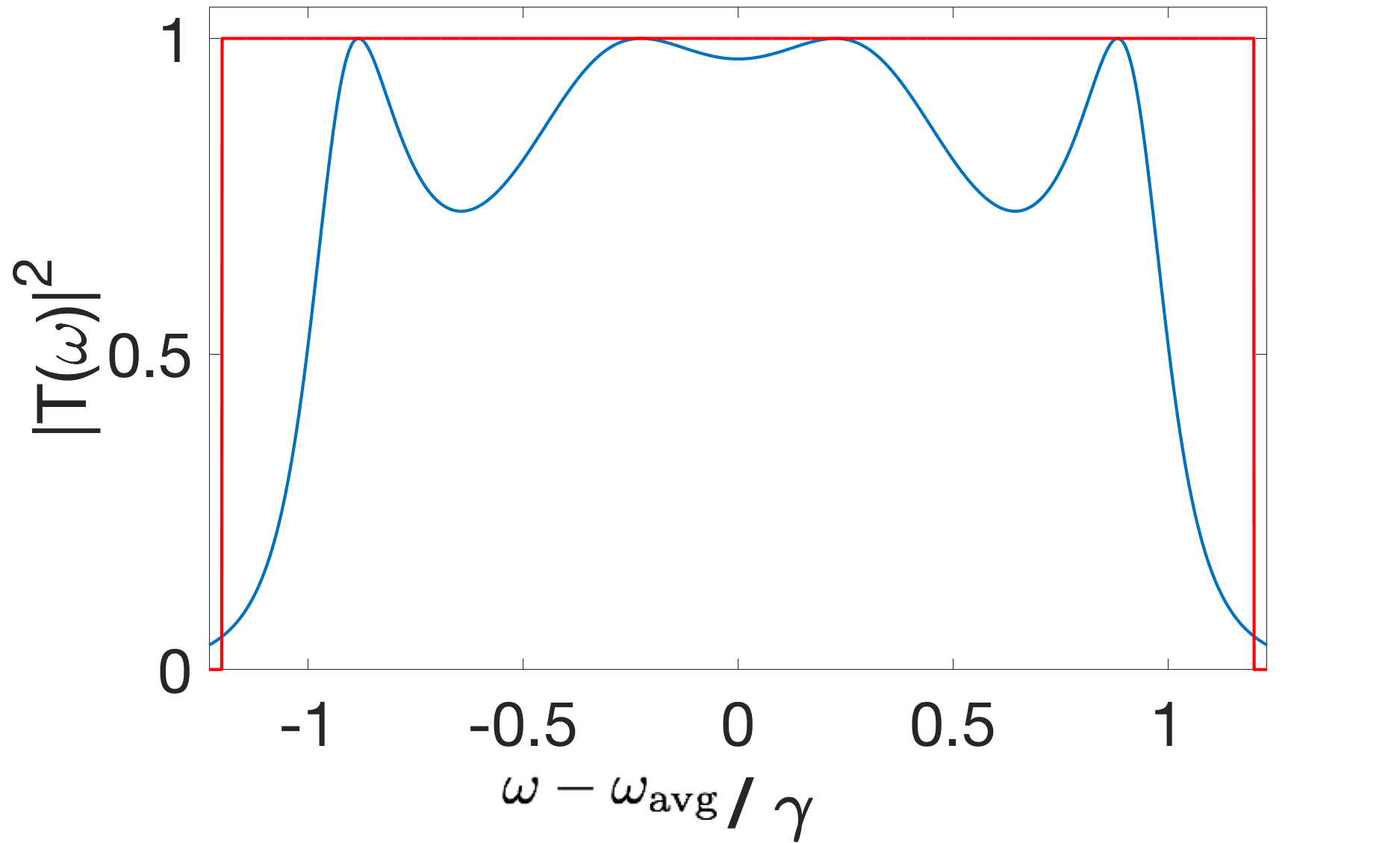}
                \caption{Transmission probability for networks with ${N=4}$ discrete states in series with no relative detuning for the special case of homogenous balanced decays ${\gamma=\Gamma}$ (blue curve) and over-coupling ${g=\frac{6}{5}\frac{\sqrt{\gamma\Gamma}}{2}}$. Again, balanced over-coupled network will always have more peaks of perfect transmission than an under-coupled or critically-coupled one, though on-resonance transmission may not be a local maxima.}     
            \label{OverEvenBalanced}
        \end{figure}

\begin{figure}[h]
\centering
            \includegraphics[width=.8\textwidth]{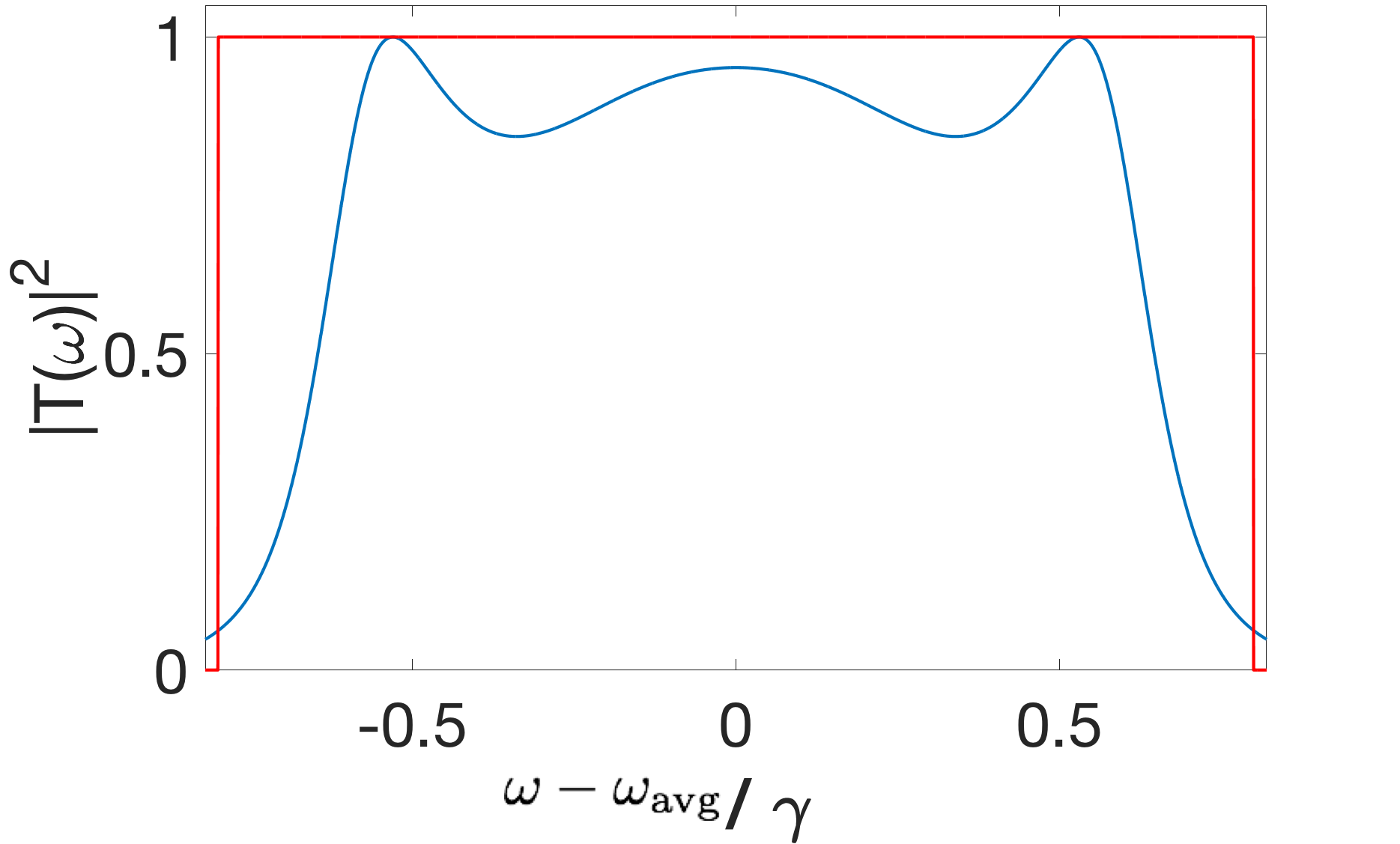}
             \caption{Transmission probability for networks with ${N=4}$ discrete states in series with no relative detuning for the special case of homogenous balanced decays ${\gamma=\Gamma}$ (blue curve) and under-coupling ${g=\frac{4}{5}\frac{\sqrt{\gamma\Gamma}}{2}}$. }     
            \label{UnderEvenBalanced}
        \end{figure}
        

Since each state is only coupled to the two adjacent states, the system of $N$ equations of the form of (\ref{quantlangspect}) describing evolution of the system is solvable in a stepladder-type approach. Setting the expectation value of the second input continuum to zero, we solve for the $N$th state in terms of the $N-1$th. As we step up the ladder, we find that the expression for the reflection coefficient has the form of a generalized continued fraction

\bea\label{refseries}
R(\omega) &= 1-\cfrac{\gamma}{\frac{\gamma}{2}-i\Delta_1+\cfrac{g_{12}^2}{-i\Delta_2+\cfrac{g_{23}^2}{\dots+\cfrac{g_{N-1 N}^2}{\frac{\Gamma}{2}-i\Delta_N}}}}
\eea (we have dropped the subscripts on $\gamma_1$ and $\Gamma_N$). This equation looks difficult to analyze, but can be greatly simplified using the Wallis-Euler recursion relations for continued fractions. We define two function $A_{n}$ and $B_{n}$ given by the following recurrence relations 

\bea\label{WallisEuler}
&B_{-1}=0, B_0=1, A_{-1}=1, A_0=b_0\nonumber \\ 
&A_n=b_n\,A_{n-1}+a_n\,A_{n-2}\,(n\geq1)\nonumber \\
&B_n=b_n\,B_{n-1}+a_n\,B_{n-2}\,(n\geq1)
\eea with coefficients $a_n$ and $b_n$ given in Tbl. \ref{table}. The reflection coefficient is then given $R(\omega)=\frac{A_N}{B_N}$ for $N$ discrete states. (For $n<N$, the functions $A_n$ and $B_n$ have no clear physical meaning.) From this, we easily solve for conditions where $R(\omega)=0$ (perfect transmission).

\begin{table}
\centering
\begin{tabular}{| c | c | c|}
\hline
$n $&$ a_n $&$ b_n $\\
\hline\hline
$0 $&$ 0 $&$ 1 $\\
\hline
$1$&$-\gamma $&$ \frac{\gamma}{2} -i\,\Delta_1 $\\
\hline
$2 $&$ g_{12}^2 $&$ -i\,\Delta_2 $\\
\hline
$\vdots$ & $\vdots$ & $\vdots$ \\
\hline
$N-1 $&$ g_{N-2 N-1}^2 $&$ -i\,\Delta_{N-1}$\\
\hline
$N  $&$ g_{N-1 N}^2 $&$ \frac{\Gamma}{2}-i\,\Delta_{N}$\\
\hline
$>N  $&$0 $&$0$\\
\hline
\end{tabular}
\caption{Wallis-Euler coefficients for ${N>1}$ discrete states in series. For a series network with an arbitrarily high number of discrete states $N$, these can be used to generate transmission functions (that correctly encode the causal structure that is inherent to the network outside of the strong-coupling limit) using the Wallis-Euler recursion relations (\ref{WallisEuler}).}\label{table}
\end{table}

\begin{figure}[h] 
\centering
	\includegraphics[width=.8\textwidth]{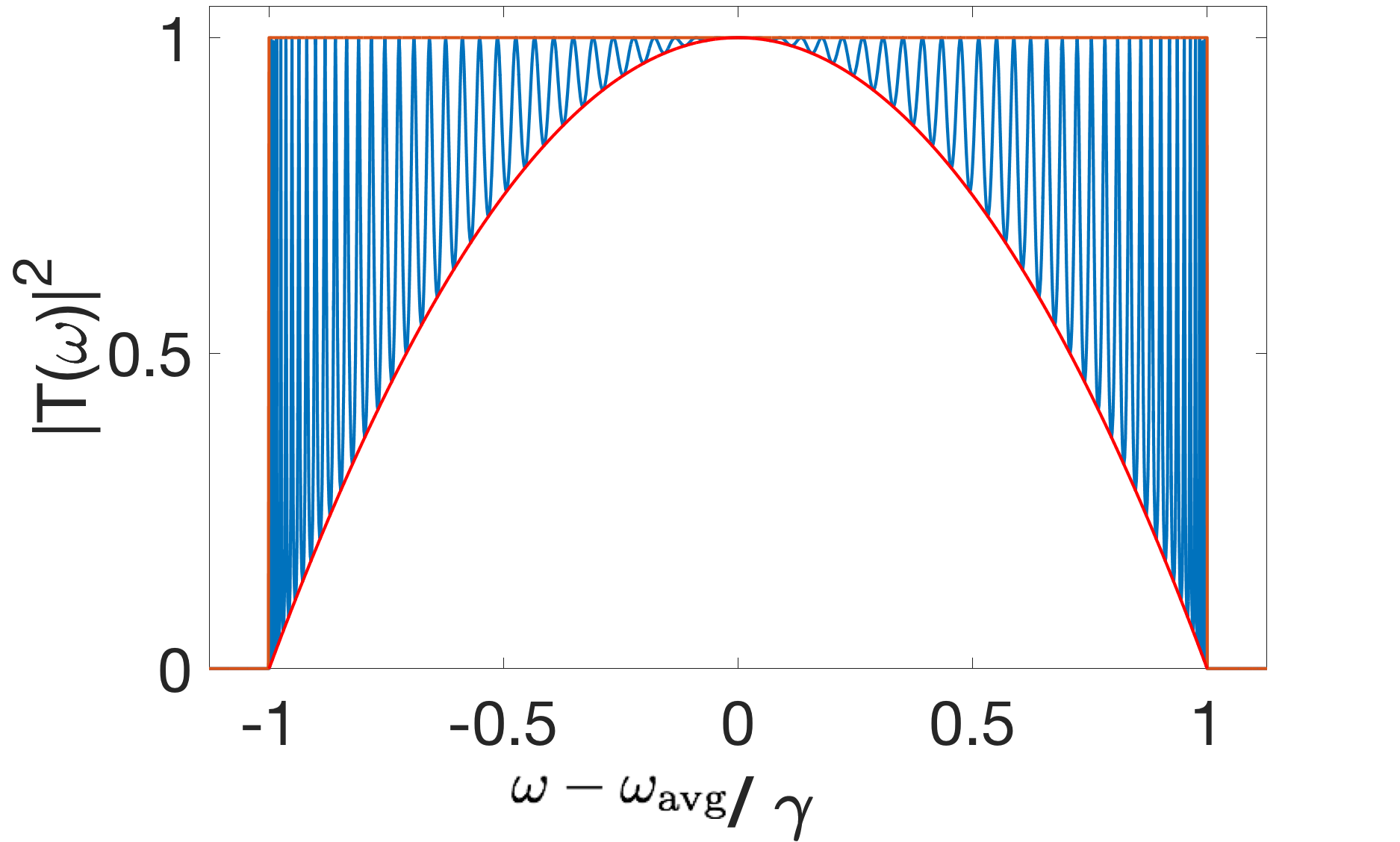} 
	\caption{Transmission probability for a series network with ${N=70}$ identical discrete states (${\omega_i=\omega_j}$) with both balanced decays (${\gamma=\Gamma}$) and uniform critical coupling (${g=\frac{\sqrt{\gamma\Gamma}}{2}}$) conditions met. This results in a maximally-broadened on-resonance transmission. In the large-$N$ limit, the amplitude of this ideal (perfect and broadened) transmission function $|T(\omega)|$ is bounded below by a circle (parabola for $|T(\omega)|^2$) and above by a square (both red) with widths $2g$ and maxima of unity.}
	\label{CircleAndSquare}
\end{figure}

\lettersection{Perfect transmission} We begin by considering the unphysical but illuminating case of an infinite series of identical discrete states ($g_{i\,i+1}= g \,\forall  i$ and $\Delta_i = \Delta \,\forall i$) analytically. We find that the limit $\lim_{N\rightarrow\infty} R(\omega)$ only converges on-resonance for the special ``critical'' case \be
g=\frac{\sqrt{\gamma\,\Gamma}}{2}.
\ee 

Furthermore, the limit only converges to zero on resonance (perfect transmission) when we also have $\gamma=\Gamma$. This first condition corresponds to a series of discrete states coupled through their decays (e.g. Fabry-Perot cavities coupled through their evanescent fields), and the second condition corresponds to the same requirement of balanced decays we saw for parallel quantum networks.


While infinite series networks of discrete states are not realistic, these two conditions play different but important roles in all finite series networks where the discrete states are identical (though introducing relative detuning will change the critical values of $g$ and $\Gamma/\gamma$, as we will see). That there are two conditions can be explained thus; since $R(\omega)$ is in general complex, the condition $R(\omega)=0$ gives two constraint equations on the real and imaginary parts of $R(\omega)$. When $N$ is even [odd], the real part is an order $N$ [$N-1$] polynomial in the detunings $\Delta_i$, while the imaginary part is order $N-1$ [$N$]. In general, this means there are conditions for a minimum of $N-1$ frequencies of perfect transmission and we may find $N$ frequencies of perfect transmission only if the lower-order equation is satisfied trivially for all frequencies. When considering finite series networks of identical discrete states, these same two conditions appear in the constraint equations for perfect transmission\footnote{For ${N=1}$, the lower order equation is just the requirement that ${\gamma=\Gamma}$ and the higher order equation simply requires ${\Delta_1=0}$ (on-resonance). For ${N=2}$, the lower order equation is ${\gamma\Delta_2=\Gamma\Delta_1}$ and the higher order equation is ${g^2=\Delta_1 \Delta_2+\frac{\gamma\Gamma}{4}}$. The lower order equation is only frequency independent when ${\gamma=\Gamma}$ and ${\Delta_2=\Delta_1}$. When the discrete states are degenerate, we find that ${g\geq \frac{\sqrt{\gamma\,\Gamma}}{2}}$ is the requirement for perfect transmission at one (${g= \frac{\sqrt{\gamma\,\Gamma}}{2}}$) or two (${g> \frac{\sqrt{\gamma\,\Gamma}}{2}}$) frequencies.}. 

We now explore in detail the effects of these two conditions on series networks of identical discrete states with uniform coupling; first, consider fixing the coupling $g$ to be critical ($g=\frac{\sqrt{\gamma\Gamma}}{2}$). For odd $N$, we find that when the decays are balanced ($\gamma=\Gamma$), this ensures $N$ frequencies of perfect transmission  (Fig. \ref{CriticalOddBalanced}) with the resonance frequency at a local maxima of unity, and when the decays are unbalanced ($\gamma\neq\Gamma$), $N-1$ frequencies of perfect transmission (Fig. \ref{CriticalOddUnbalanced}) with the resonant frequency at a local minima. For even $N$, letting $g=\frac{\sqrt{\gamma\Gamma}}{2}$ always results in $N-1$ frequencies of perfect transmission (Fig. \ref{CriticalEvenUnbalanced}) with the on-resonant frequency at a local maxima. Here, having balanced decays broadens the the on-resonance maxima (Fig. \ref{CriticalEvenBalanced}), which will be desirable for detection of non-monochromatic photons (wavepackets). 

Now we instead consider balanced decays ($\gamma=\Gamma$) and observe a switch in the on-resonance behavior; whereas above we found that for even $N$ the critical coupling condition was sufficient for on-resonance transmission to be at a local maxima, we now find that, for odd $N$ that the balanced decay condition results in $N$ peaks of unity transmission (Fig. \ref{OverOddBalanced}) with on-resonance transmission at a local maxima. Here three peaks become degenerate to give $N-2$ frequencies of perfect transmission when $g\lesssim \frac{\sqrt{\gamma\Gamma}}{2}$ (Fig. \ref{UnderOddBalanced}). (This inequality rapidly becomes exact [$g< \frac{\sqrt{\gamma\Gamma}}{2}$] with increasing $N$.) Similarly, for even $N$ we find the behavior of $T(\omega)$ depends strongly on the coupling, flipping between $N$ frequencies of perfect transmission with on-resonance transmission at a local minima for $g> \frac{\sqrt{\gamma\Gamma}}{2}$ (Fig. \ref{OverEvenBalanced}) and $N-2$ frequencies of perfect transmission with on-resonance transmission at a non-unity local maxima for $g< \frac{\sqrt{\gamma\Gamma}}{2}$ (Fig. \ref{UnderEvenBalanced}).

For both even and odd $N$, the width is increasingly determined by $g$ with increasing $N$, with the half-width asymptotically approaching $2g$. In the large-$N$ limit, we observe that the transmission function is asymptotically bounded between a circle and a square when both conditions for perfect transmission are met (Fig. \ref{CircleAndSquare}). We also observe that increasing $g$ past $\frac{\sqrt{\gamma\,\Gamma}}{2}$ while maintaining $\gamma=\Gamma$ induces Rabi splitting, with the $N$ or $N-1$ frequencies of perfect transmission spreading outwards (for odd and even $N$, respectively). For the special case of $N=2$, the presence of a second coupled discrete state resulting in a splitting and shift of the resonant frequency is in agreement with the effect seen in Ref.~\cite{Rosfjord2006} for an antireflective coating; perfect transmission is still possible provided $\gamma=\Gamma$ but the frequency that is perfectly transmitted is split in to two. 

\begin{figure}[h] 
\centering
	\includegraphics[width=.8\textwidth]{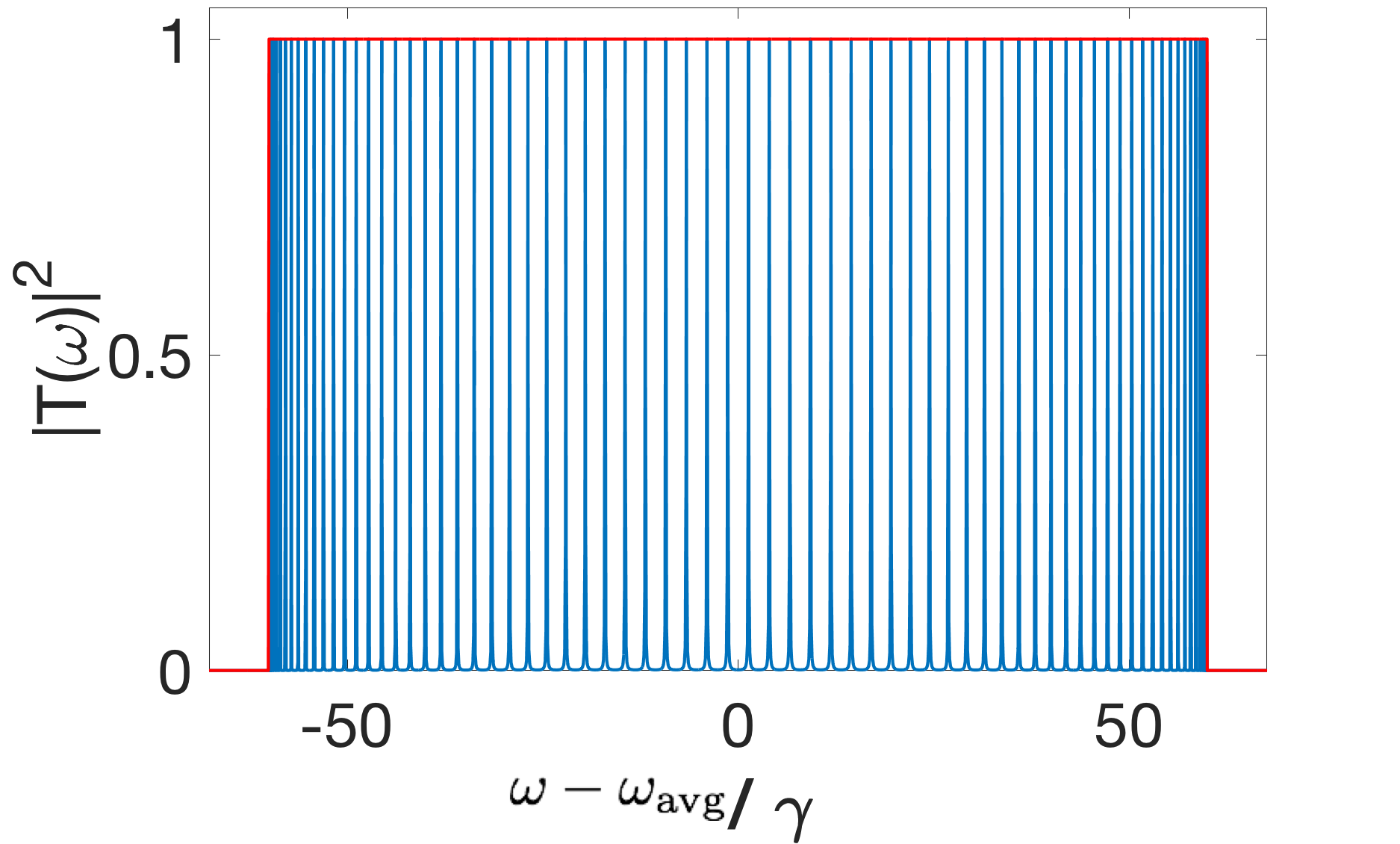} 
	\caption{Transmission probability for a series network with ${N=70}$ identical discrete states with balanced decays (${\gamma=\Gamma}$) yielding perfect transmission in the strong-and-uniform coupling limit (${g\gg\frac{\sqrt{\gamma\Gamma}}{2}}$). The result is a frequency-comb structure with $70$ frequencies of unity transmission between $-2g$ and $2g$, separated by $69$ regions of near-perfect reflection.} 
	\label{FrqeuencyComb}
\end{figure}

 In the strong coupling (high-$g$) limit, a frequency-comb structure emerges (Fig. \ref{FrqeuencyComb}), with dips approaching perfect reflection. This is due to the asymptotic irrelevance of the causal ordering (that is, the order in which an excitation must traverse discrete states to pass from one continuum to the other): the network becomes approximately diagonalizable as parallel modes with purely virtual coupling and the $N-1$ frequencies of perfect reflection from (\ref{reflectk}) manifest.

\begin{figure}[h] 
\centering
	\includegraphics[width=.7\textwidth]{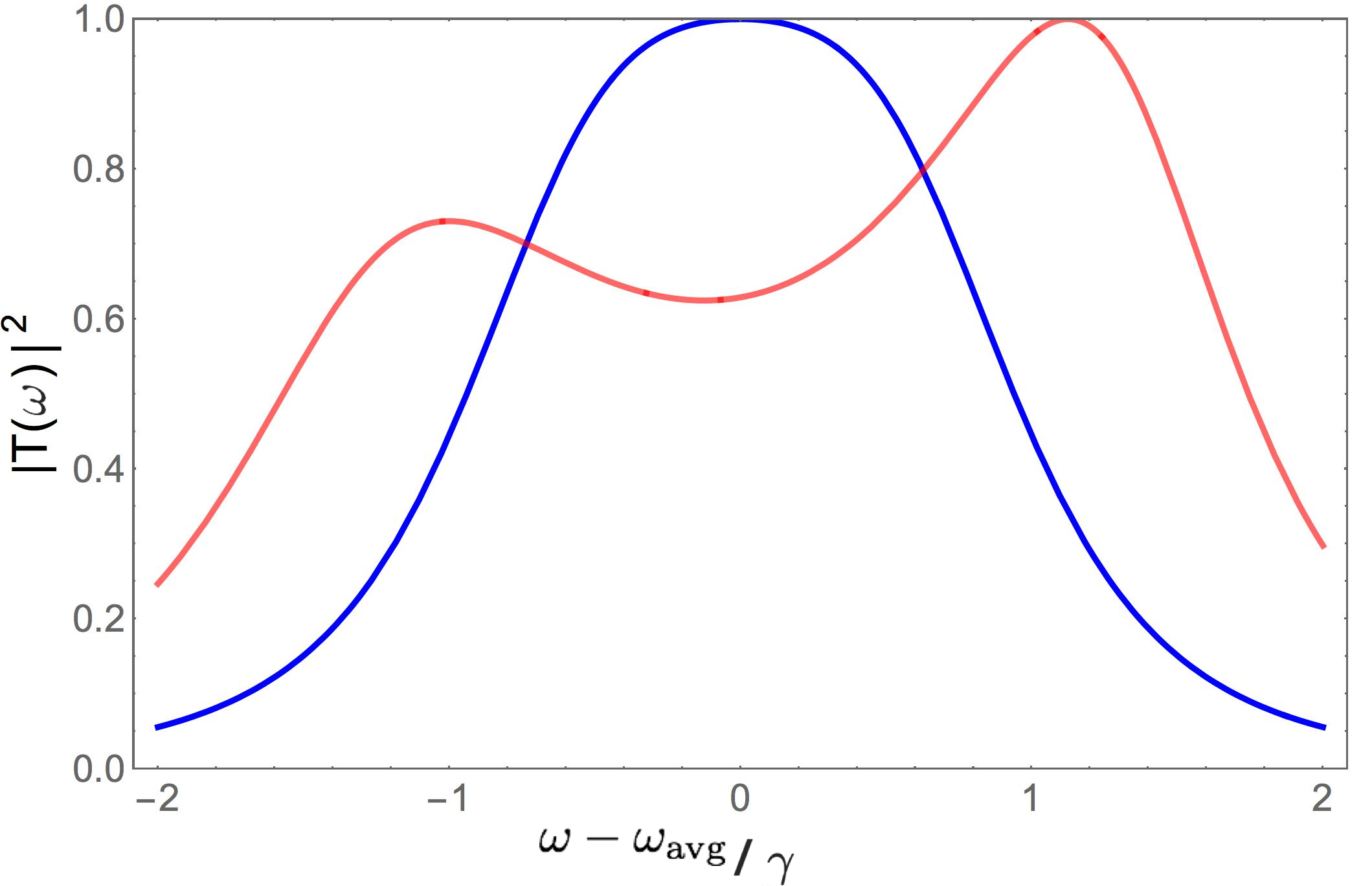} 
	\caption{Transmission probabilities for series networks with ${N=2}$ discrete states with no relative detuning (blue with a single maxima), and with non-zero relative detuning ${\frac{\omega_2-\omega_1}{\gamma}=\frac{3}{4}}$ (red with two maxima). In both cases, the decays are not balanced (${\frac{\Gamma}{\gamma}=2}$) but the couplings are chosen such that perfect transmission at some frequency is achieved nonetheless.}
	\label{detunedstates}
\end{figure}

We now consider the effects of introducing relative detunings between discrete states as, generally, a quantum network will not be comprised of identical states. Still there are sufficient degrees of freedom in (\ref{refseries}) such that, by tuning the parameters, perfect transmission at \emph{some} frequencies is always possible. This is even true when $\gamma\neq\Gamma$ (for $N>1$): considering only two discrete states in series, we find that perfect transmission occurs at a frequency $\omega=\frac{\omega_1+\omega_2}{2} + \frac{\Gamma\omega_1-\gamma\omega_2}{\Gamma-\gamma}$ when $g_{12}=\sqrt{\frac{\gamma\Gamma}{4} + \left(\frac{\Gamma\omega_1-\gamma\omega_2}{\Gamma-\gamma}\right)^2 - \left(\frac{\omega_1-\omega_2}{2}\right)^2}$ (Fig. \ref{detunedstates}). For two detuned discrete states in series with balanced decays ($\gamma=\Gamma$), perfect transmission is impossible as the critical value of $g$ is infinite; the transmission efficiency asymptotically becomes perfect in the strong-coupling limit. However, this a special case and is not true for larger numbers ($N>2$) of discrete states. Critically, given a series network of $N$ discrete states with arbitrary relative detunings, we can \emph{always} find at least one set of parameters (couplings and decay rates) such that $N-1$ frequencies are perfectly transmitted.

 \lettersection{Spectral Bandwidth} Once we have the form of the transmission function, we can calculate the spectral bandwidth for these systems. The spectral bandwidth decreases with additional discrete states and is strictly bounded above by the single discrete state bandwidth $\tilde{\Gamma}\leq\frac{2\gamma\Gamma}{\gamma+\Gamma}$. For discrete states without detuning, equality is reached in the strong-coupling limit of $g\gg\frac{\sqrt{\gamma\Gamma}}{2}$ but independently of whether $\gamma=\Gamma$ (Fig. \ref{tildeGammaN1}). Introducing detuning between discrete states lowers the bandwidth, but equality with the upper limit still occurs for sufficiently strong coupling (Fig. \ref{N1RelativeBandwidthDetuning}); the strong coupling is able to better mask the discrepancy between discrete state frequencies as the dressed states become the more physical description and as the system more strongly resembles a parallel network. (In this limit, each dressed state is coupled to the continua at reduced decays so that the total bandwidth is still bounded by $\tilde{\Gamma}\leq\frac{2\gamma\Gamma}{\gamma+\Gamma}$ with $\gamma$ and $\Gamma$ the two original decay rates for the series network.)

 \lettersection{Group Delay} Lastly, we consider the group-delay for series networks, which increases with both $g$ and $N$ but for different reasons. As $g$ increases, the peaks of the transmission function sharpen so that the phase changes more rapidly. This results in an increased group delay (Fig. \ref{2discretestatesSeriesTrans}-d). As $N$ increase, we observe (as we did for parallel networks) that the peaks in the group delay are not of uniform magnitude, even when the transmission function itself is rather flat (Fig. \ref{20discretestatesSeriesTrans}). Instead, the frequencies of maximum delay are those closest to $\pm 2g$ (Fig. \ref{20discretestatesSeriesTau}) and increase the most with $N$. This is because the oscillations in the transmission function are most dense here, resulting in a rapid change in transmission function phase and thus a larger group delay. On the contrary, the group delay is relatively flat in the center of the transmission window (Fig. \ref{20discretestatesSeriesTau}) so that a spectrally-narrow (compared to $g$) input photon pulse will not be significantly dispersed. (That is, the maximum dispersion-induced jitter $\mathcal{T}_g$ from (\ref{maxdisp}) will be very small.)

We can also consider the effect of detunings on the frequency-dependent group delay. We observe the group delay can be negative for series networks with detuning (Fig. \ref{DetunedN20Tau}), with an asymmetric structure that depends on the ordering of the detunings (except in the strongly coupled limit) \footnote{Note that a negative group delay does not imply superluminal signaling. For a detailed discussion of negative group delays and their interpretation, see Ref.~\cite{solli2002}}. Again, the magnitude of the group delay depends on the spacing of discrete states, increasing when the resonant frequencies are closely spaced.

\afterpage{

\begin{figure}[h!] 
\centering
	\includegraphics[width=.7\textwidth]{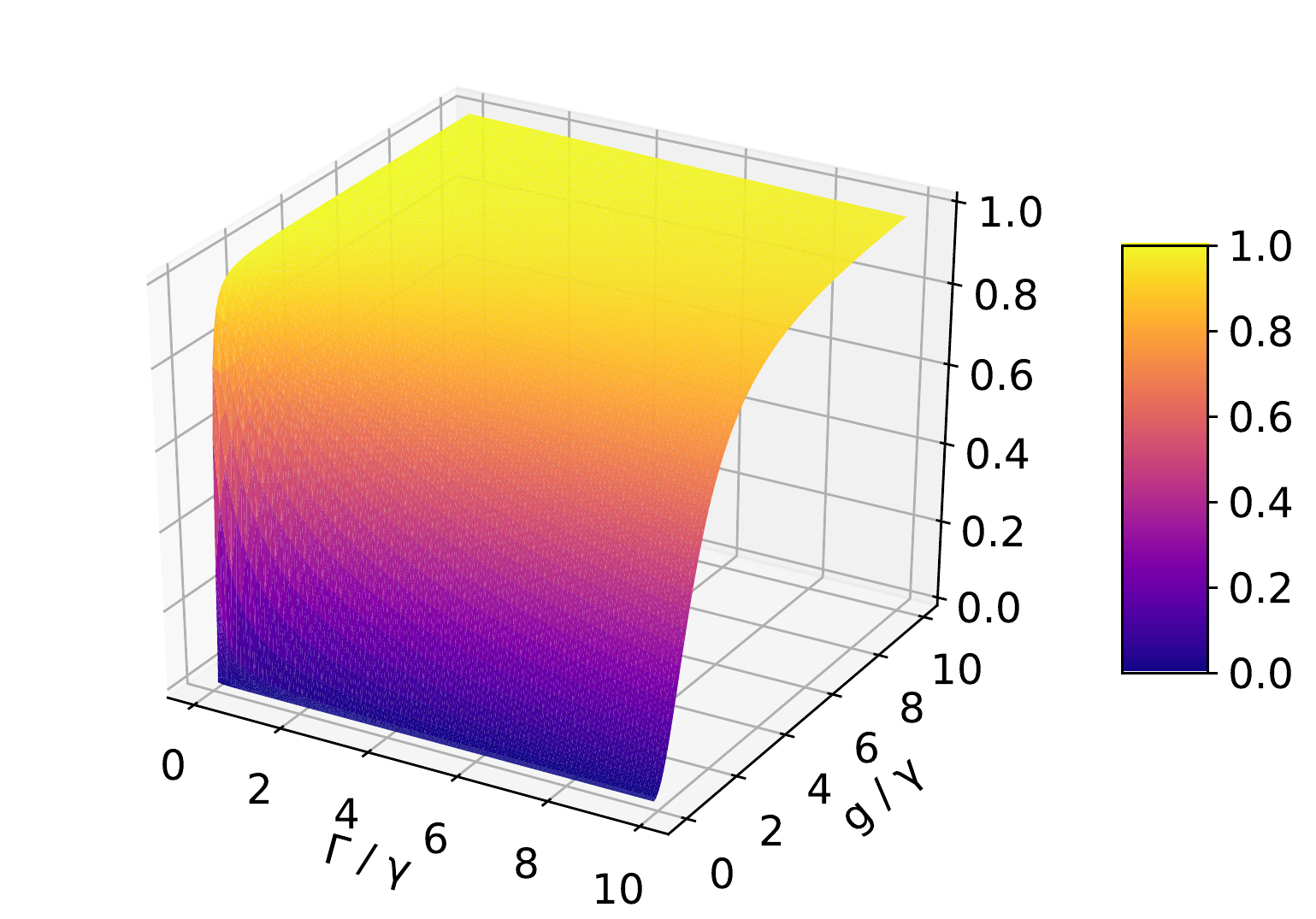} 
	\caption{Normalized spectral bandwidth $\tilde{\Gamma}/\tilde{\Gamma}_{\rm max}$ for a series network with ${N=2}$ discrete states and no relative detuning. Here the maximum bandwidth, given by that of a single discrete state ${\tilde{\Gamma}_{\rm max}=\frac{2\gamma\Gamma}{\gamma+\Gamma}}$, is reached in the strong coupling limit. The balanced decay condition does not affect bandwidth. An increase in decay rates simply increases the scale of the strong coupling limit regime (${g\gg\frac{\sqrt{\gamma\Gamma}}{2}}$).}
	\label{tildeGammaN1}
\end{figure}

\begin{figure}[h!] 
\centering
	\includegraphics[width=.7\textwidth]{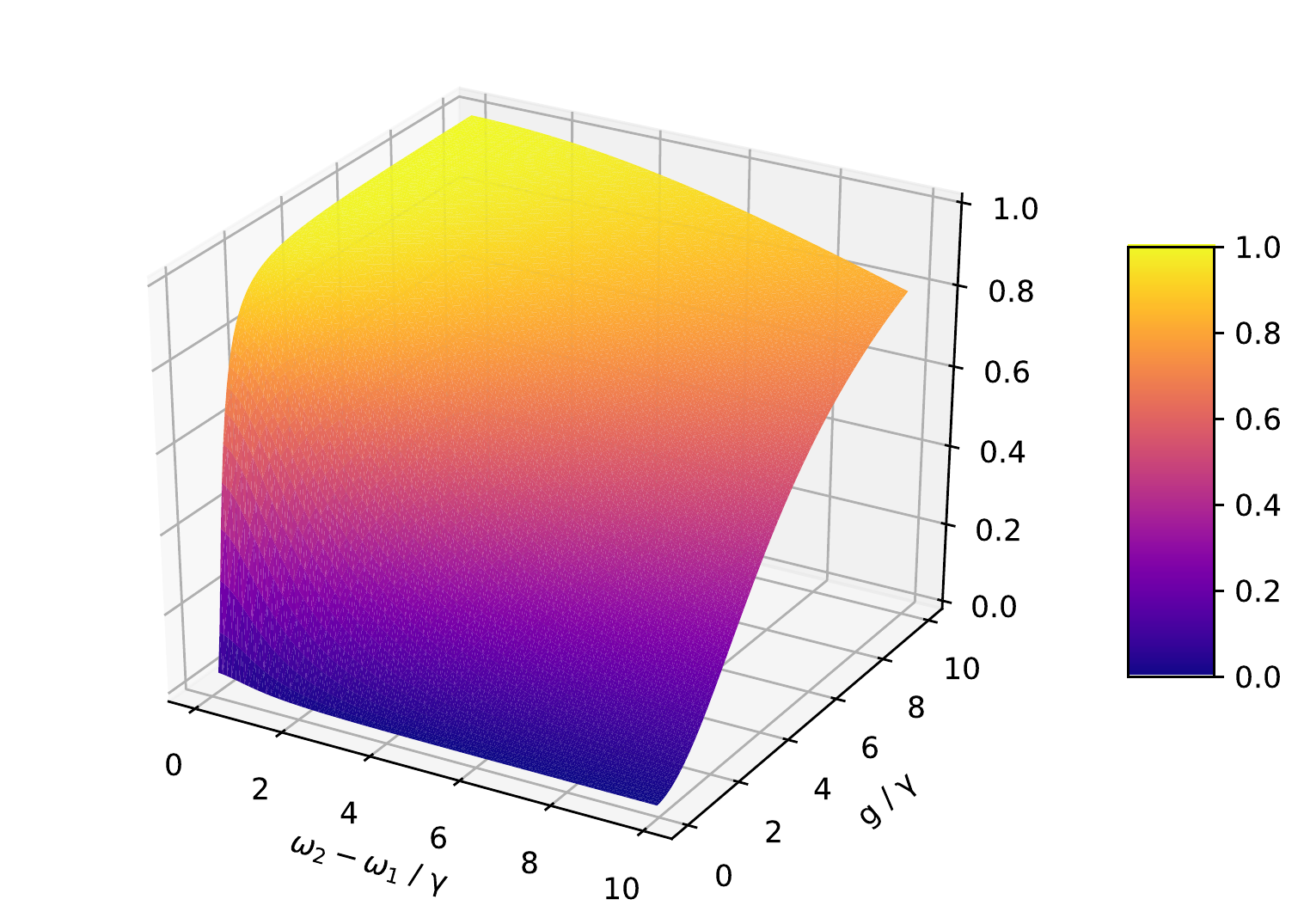} 
	\caption{Normalized spectral bandwidth $\tilde{\Gamma}/\tilde{\Gamma}_{\rm max}$ for a series network with ${N=2}$ discrete states with relative detuning ${\omega_2-\omega_1}$. We still observe ${\tilde{\Gamma}/\tilde{\Gamma}_{\rm max} = 1}$, but at a higher coupling strength $g$ for greater detunings. }
	\label{N1RelativeBandwidthDetuning}
\end{figure}

 \begin{figure}[h!]
 \centering
            \includegraphics[width=.7\textwidth]{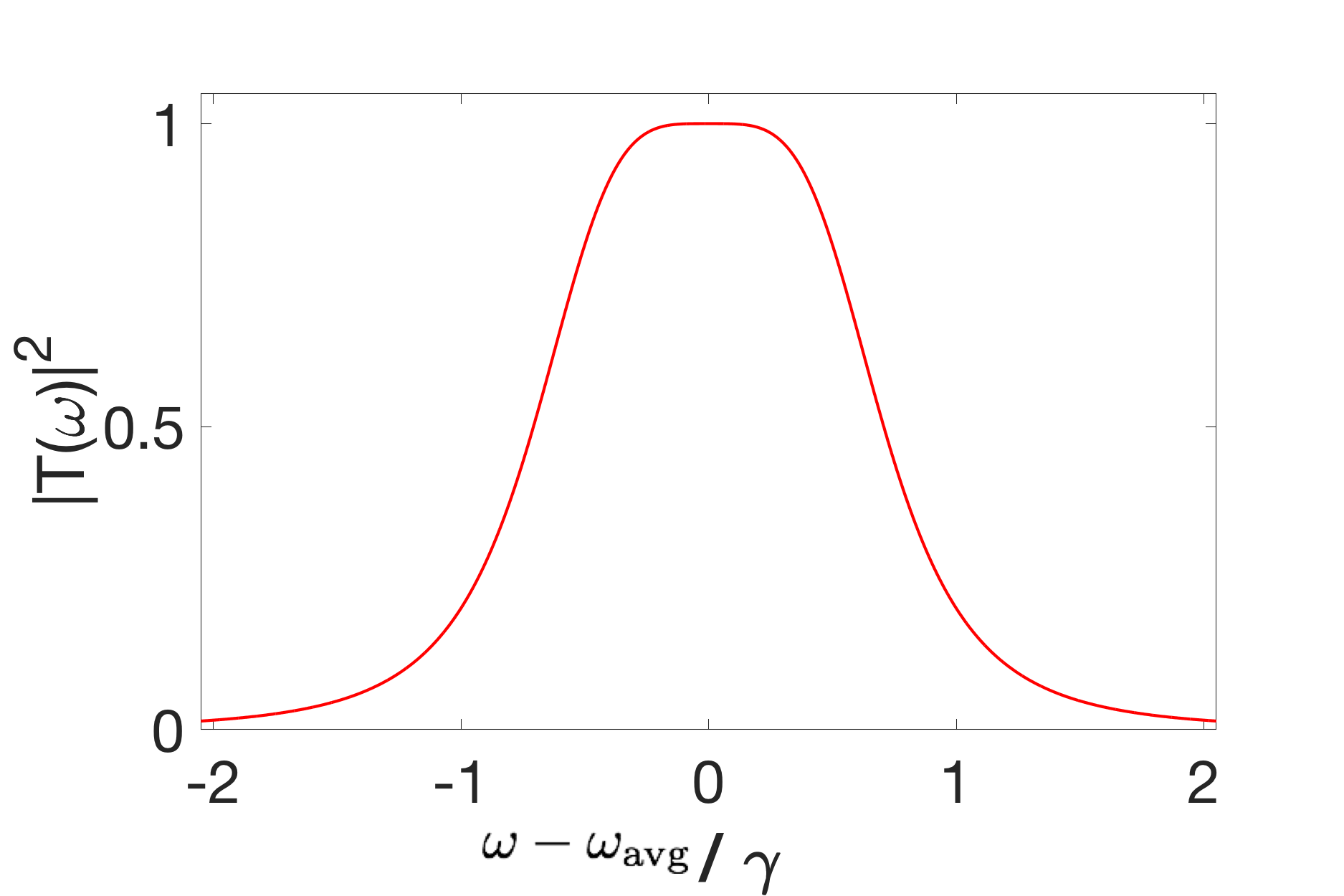}
            \caption{Transmission probability for a critically coupled ${g= \frac{\sqrt{\gamma\Gamma}}{2}}$ series network of $N=2$ discrete states with balanced decay rates (${\gamma=\Gamma}$) and no relative detuning.}    
            \label{2discretestatesSeriesTrans}
        \end{figure}

   \begin{figure}[h!]
   \centering
            \includegraphics[width=.7\textwidth]{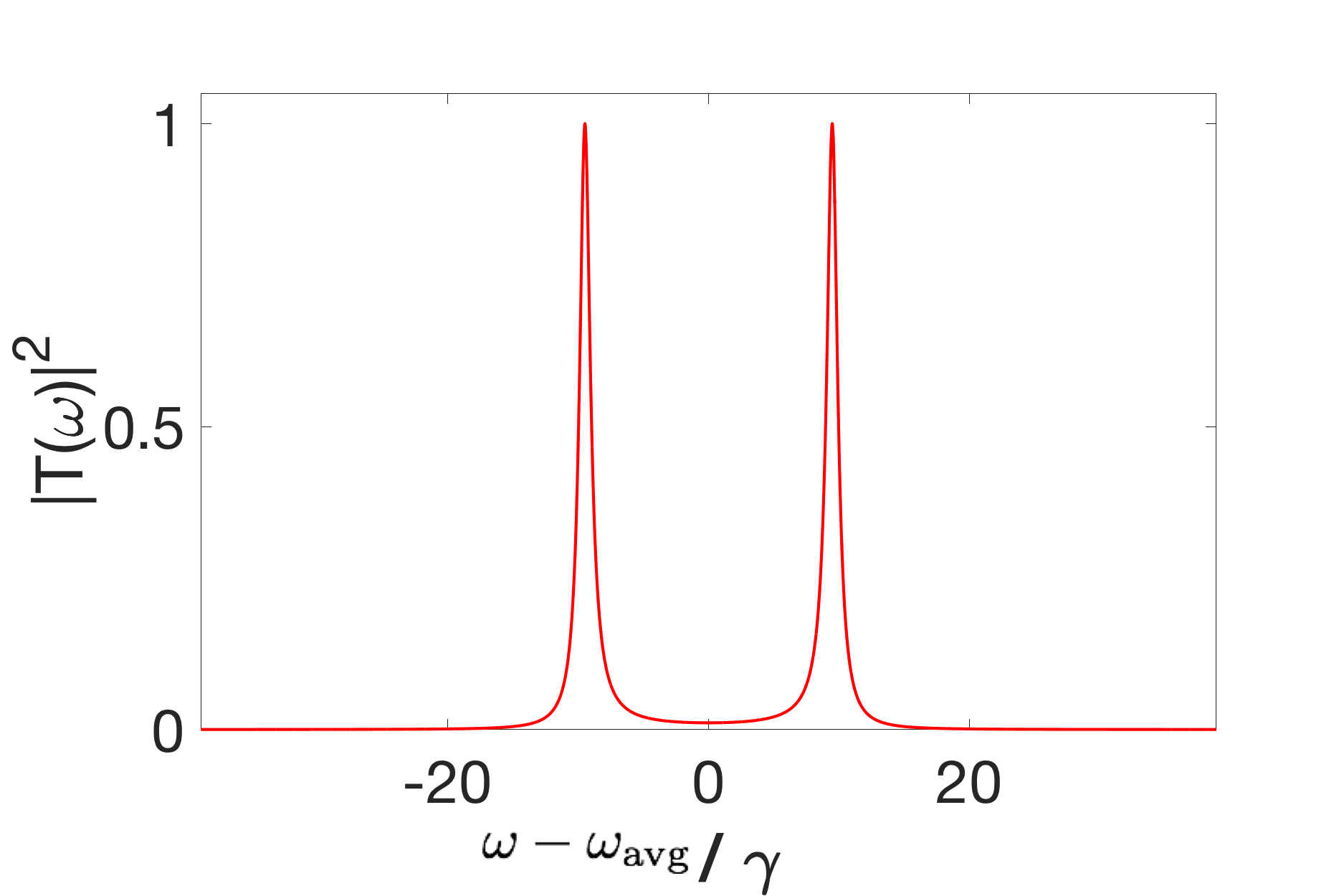}
            \caption{Transmission probability for an over-coupled ${g=18 \frac{\sqrt{\gamma\Gamma}}{2}}$ series network of $N=2$ discrete states with balanced decay rates (${\gamma=\Gamma}$) and no relative detuning.}    
            \label{2discretestatesSeriesTransOver}
        \end{figure}

 \begin{figure}[h!]
 \centering
            \includegraphics[width=.7\textwidth]{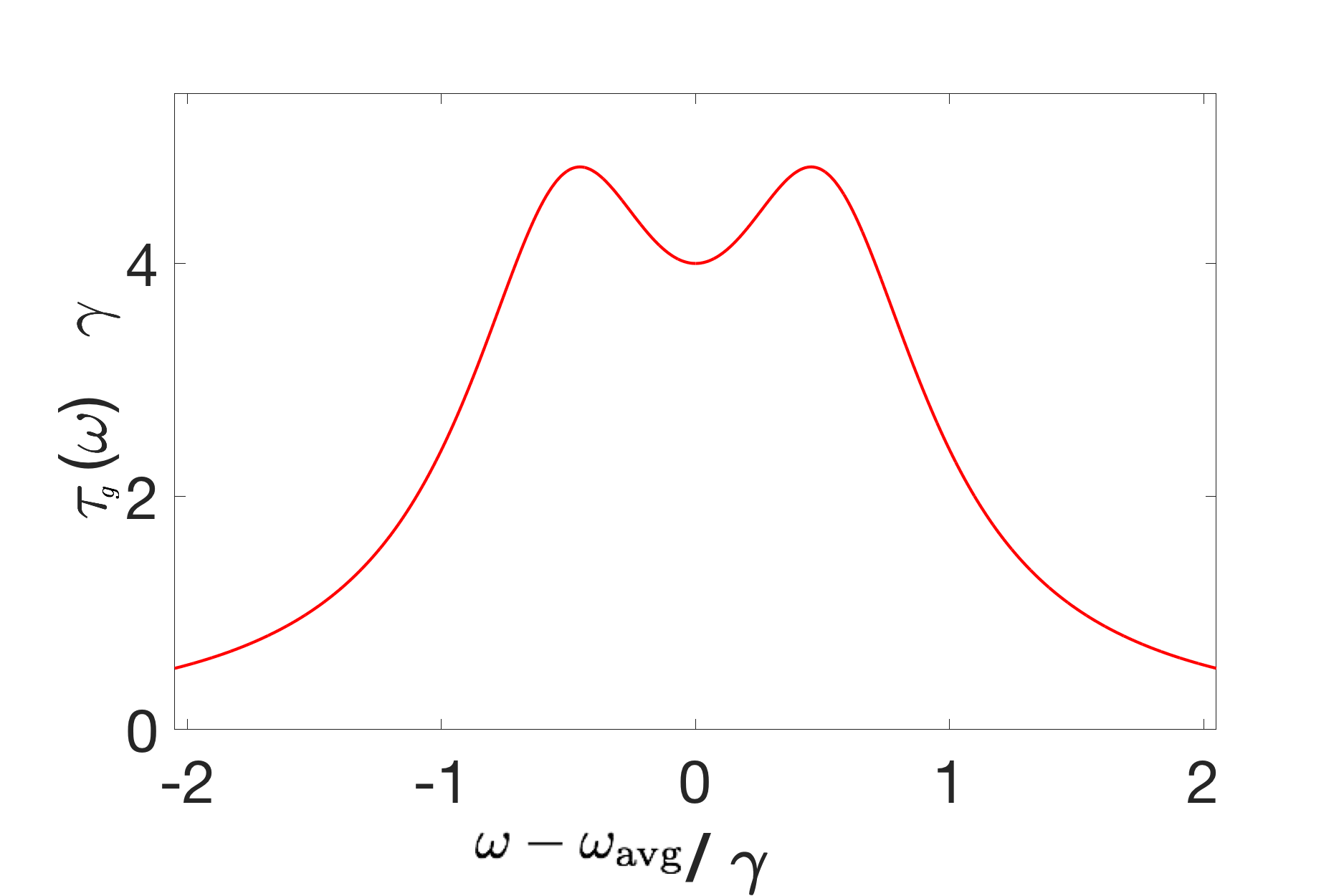}
            \caption{Group delay corresponding to a critically coupled ${g= \frac{\sqrt{\gamma\Gamma}}{2}}$ series network of ${N=2}$ discrete states with balanced decay rates (${\gamma=\Gamma}$) and no relative detuning, the transmission probability of which is given in Fig. \ref{2discretestatesSeriesTrans}. A small change in the sharpness of the transmission function's peaks can have a large effect on the magnitude of the group delay (which we measure in units of $\frac{1}{\gamma}$).}      
            \label{2discretestatesSeriesTau}
        \end{figure}

 \begin{figure}[h!]
 \centering
            \includegraphics[width=.7\textwidth]{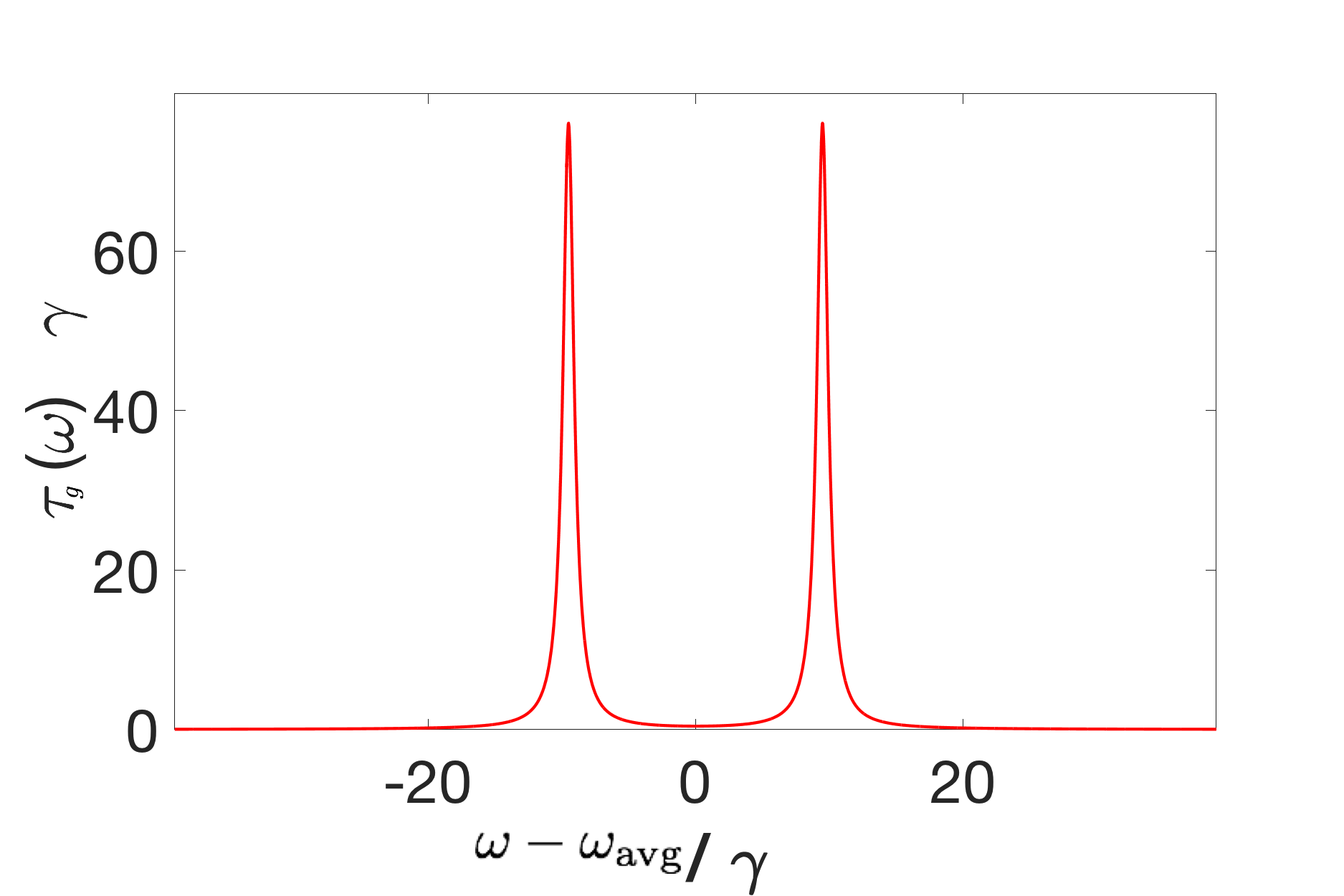}
 \caption{Group delay corresponding to an over-coupled ${g=18 \frac{\sqrt{\gamma\Gamma}}{2}}$ series network of ${N=2}$ discrete states with balanced decay rates (${\gamma=\Gamma}$) and no relative detuning, the transmission probability of which is given in Fig. \ref{2discretestatesSeriesTransOver}. Again, small change in the sharpness of the transmission function's peaks can have a large effect on the magnitude of the group delay.}     
            \label{2discretestatesSeriesTauOver}
        \end{figure}


\begin{figure}[h!]
\centering
            \includegraphics[width=.7\textwidth]{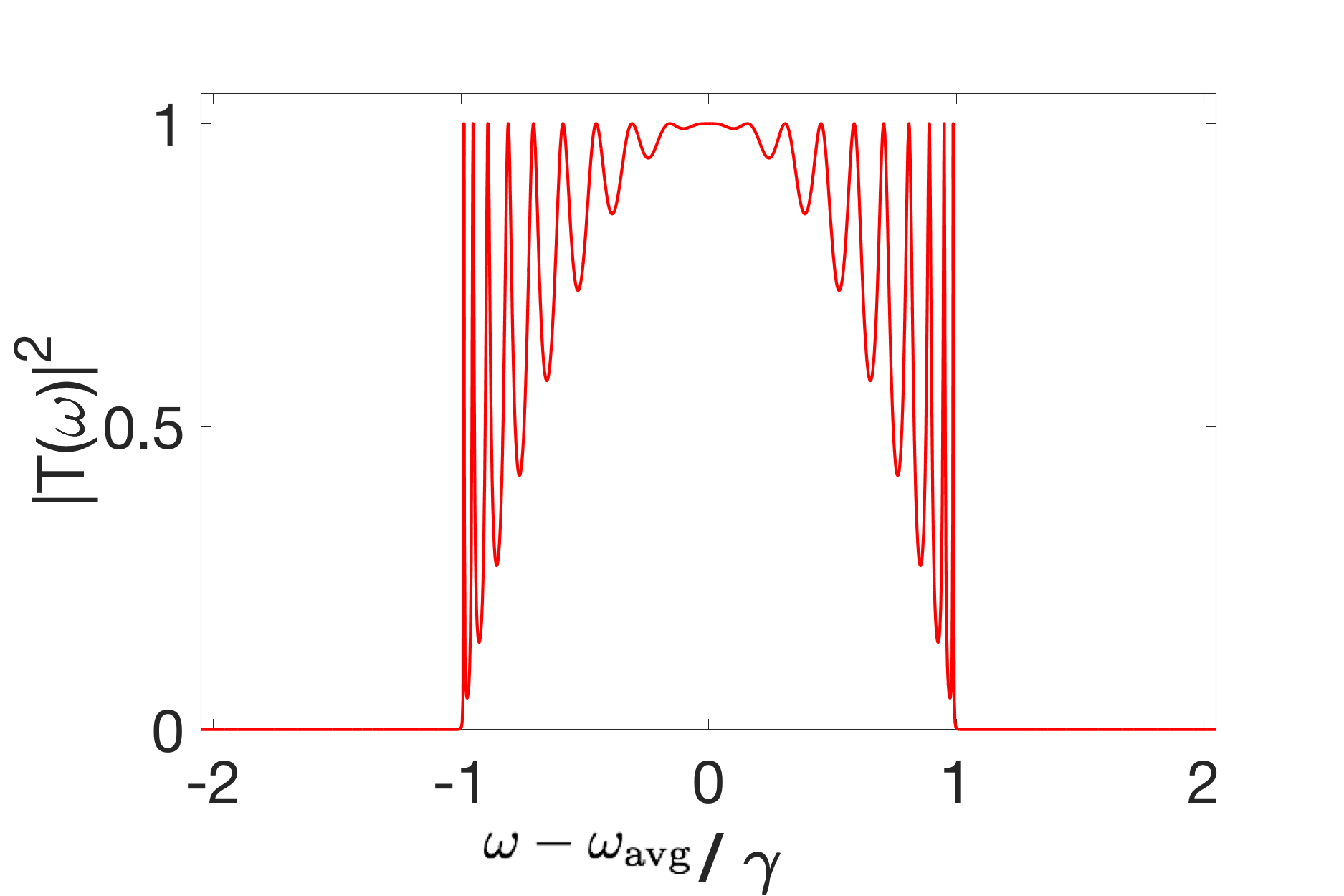}
            \caption{Transmission probability for a series networks with ${N=20}$ discrete states, balanced decay rates, critical coupling ${g=\frac{\sqrt{\gamma\Gamma}}{2}}$, and no relative detuning.}   
            \label{20discretestatesSeriesTrans}
        \end{figure}

\begin{figure}[h!]
\centering
            \includegraphics[width=.7\textwidth]{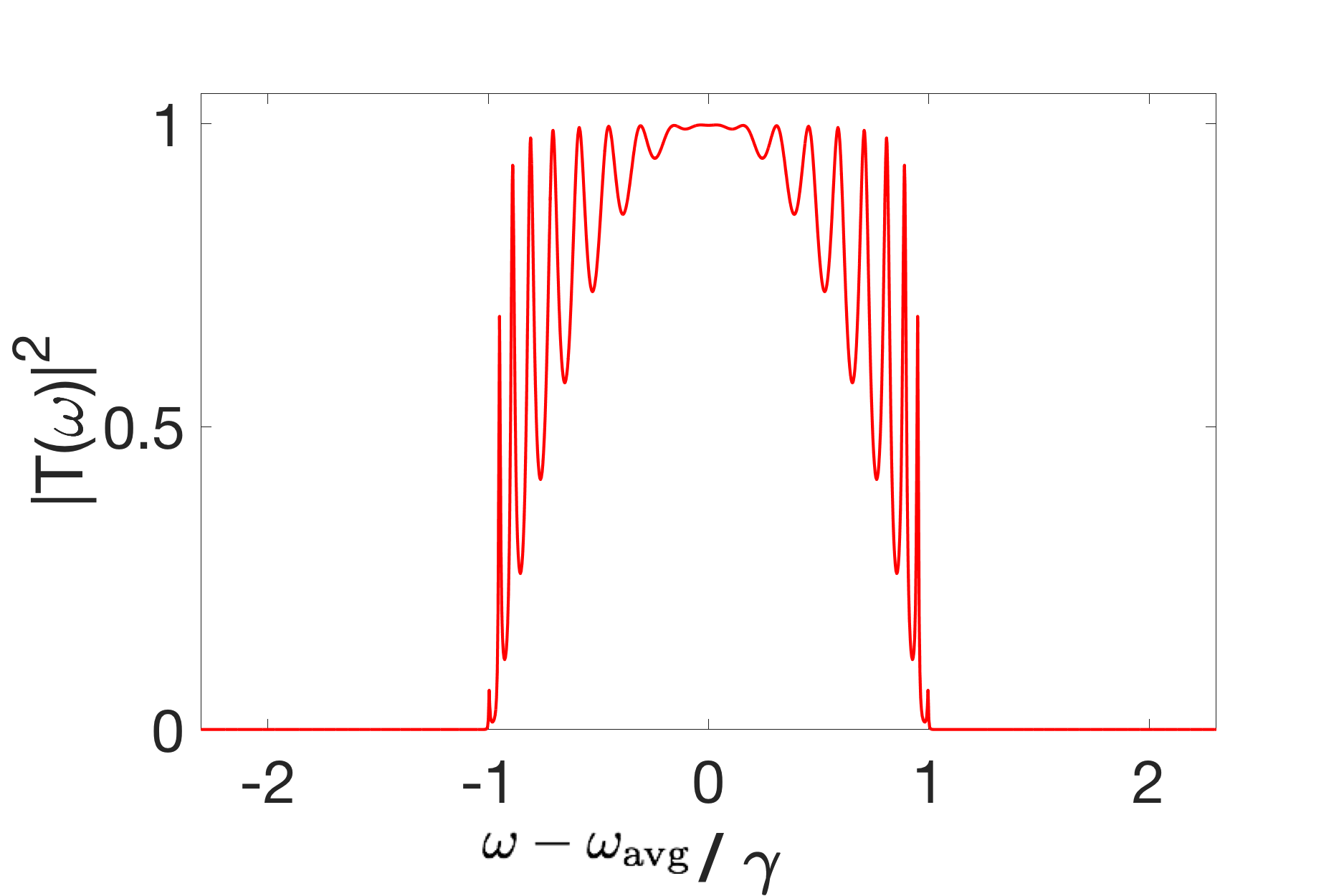}
            \caption{Transmission probability for a series networks with ${N=20}$ discrete states, balanced decay rates, critical coupling, and relative detuning  ${\omega_i - \omega_{i+1} = \frac{1}{50}\gamma}$.}   
            \label{DetunedN20Trans}
        \end{figure}

\begin{figure}[h!]
\centering
            \includegraphics[width=.7\textwidth]{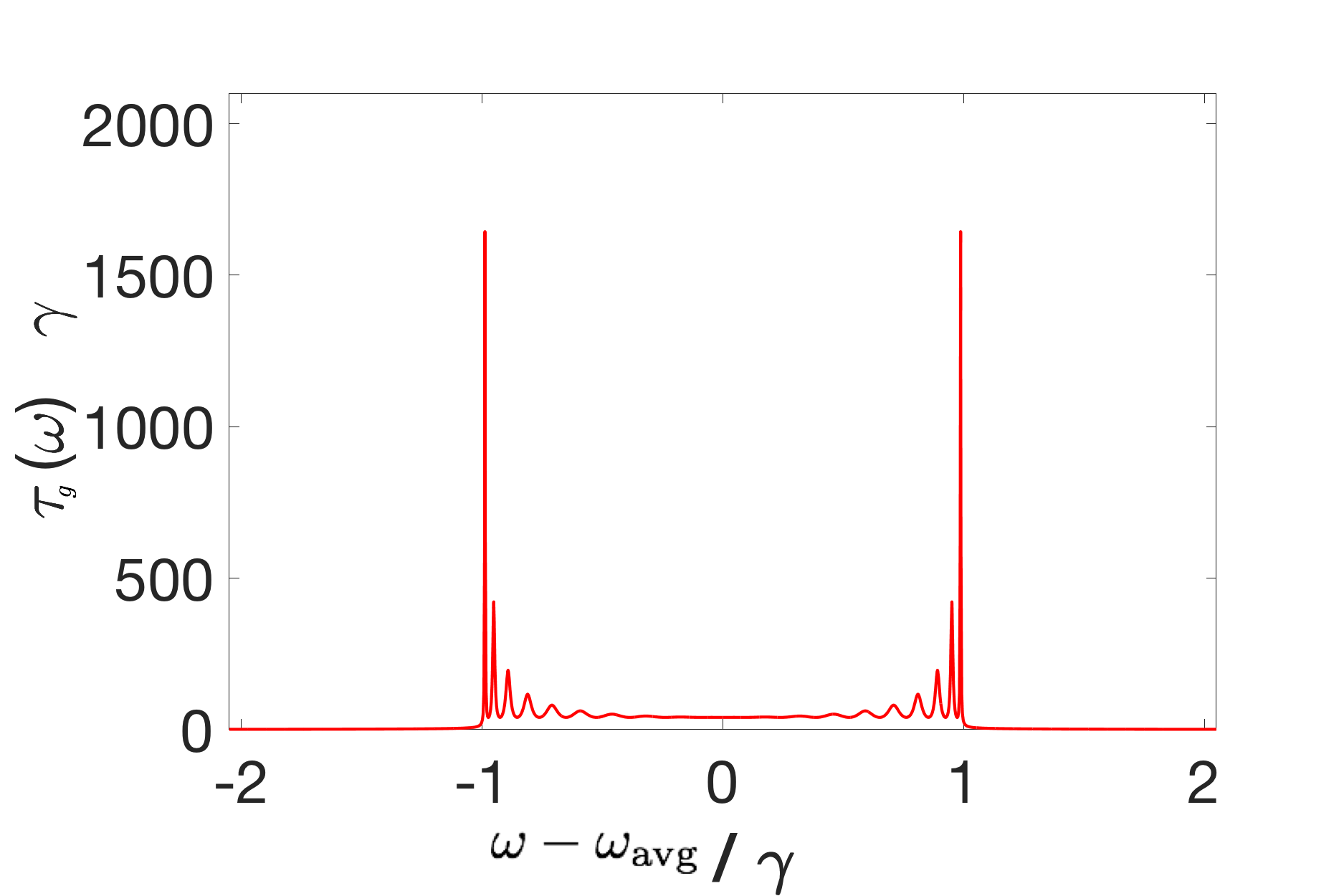}
            \caption{Group delay for a series networks with ${N=20}$ discrete states, balanced decay rates, critical coupling, and no relative detuning, the transmission probability of which is given in Fig. \ref{20discretestatesSeriesTrans}. Again, the group delay is of the largest near the edge of the transmission function (${\pm2g}$ in the high $N$ limit).}    
            \label{20discretestatesSeriesTau}
        \end{figure}

\begin{figure}[h!]
\centering
            \includegraphics[width=.7\textwidth]{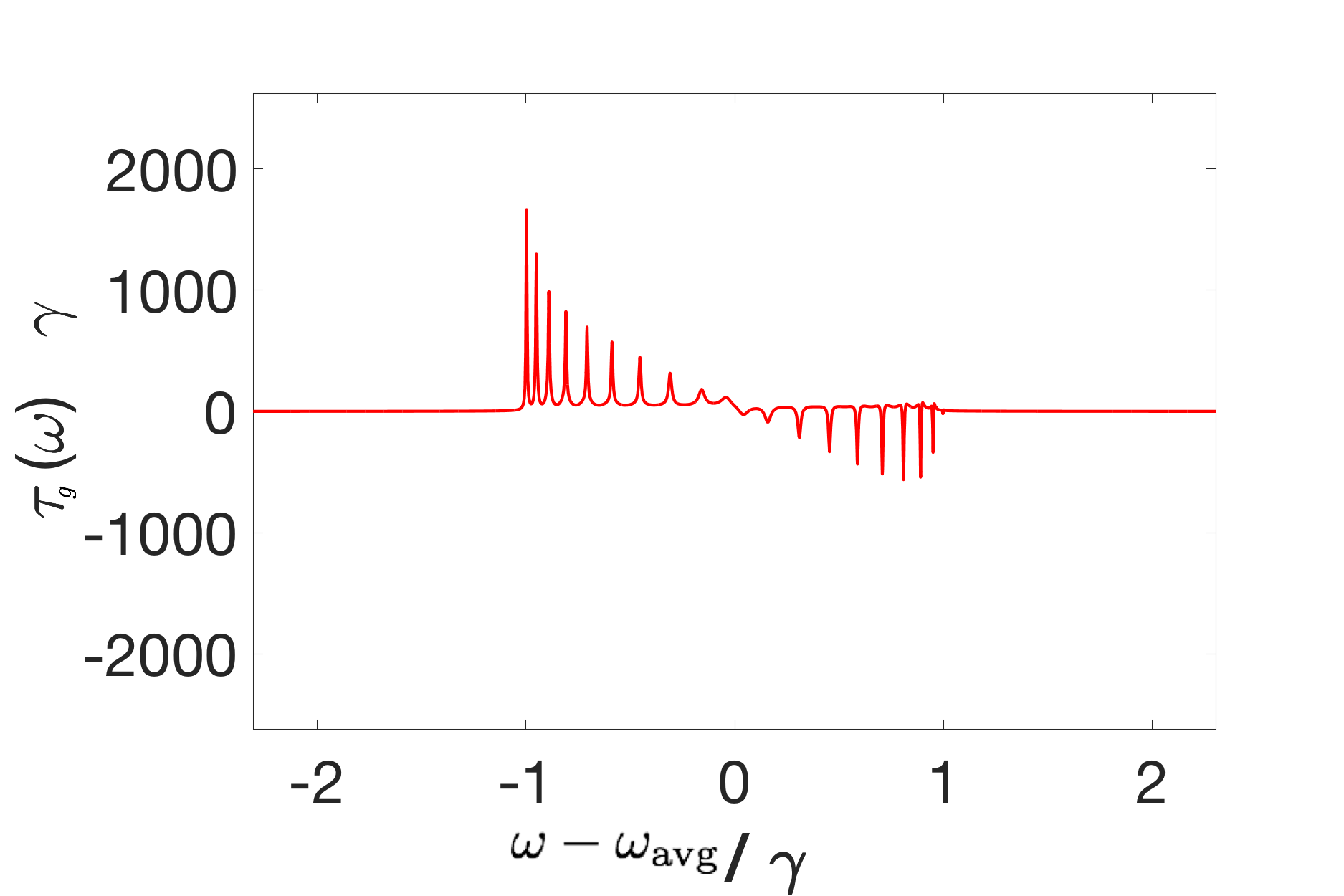}
 \caption{Group delay for a series networks with ${N=20}$ discrete states, balanced decay rates, critical coupling, and no relative detuning, the transmission probability of which is given in Fig. \ref{20discretestatesSeriesTrans}. Again, the magnitude of the group delay is of the largest magnitude near the edge of the transmission function (${\pm2g}$ in the high $N$ limit), and we observe that a relative detuning between manifolds may result in a negative group delay.}       
            \label{DetunedN20Tau}
        \end{figure}

        \clearpage
}

\newpage

\subsection{Hybrid Networks}

 \begin{figure}[h] 
 \centering
	\includegraphics[width=.8\textwidth]{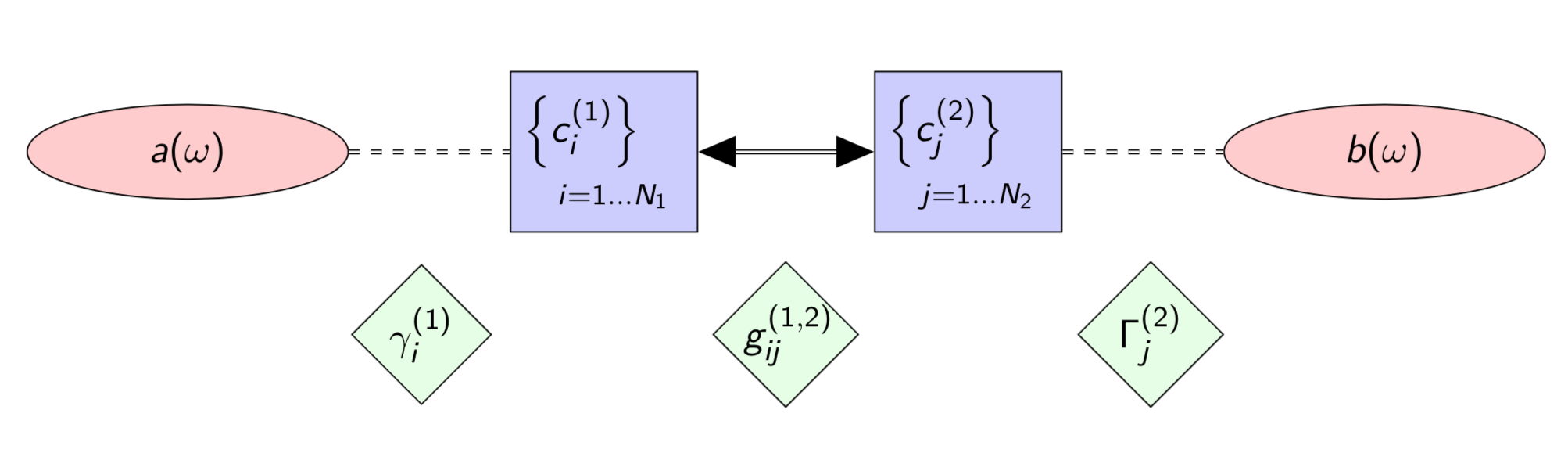} 
	\caption{A hybrid network of two manifolds, each with $N_k$ discrete states described by operators $c^{(k)}_i (\omega)$ where the index $k$ here has only two values (${k=1,\,2}$). The discrete states are coherently coupled at rates $g^{(1,2)}_{ij}$ and incoherently coupled to left (input, external) and right (output, internal) continua $a$ and $b$ with decay rates $\gamma^{(1)}_i$ and $\Gamma^{(2)}_j$, respectively. Each discrete state is part of a manifold: a set of discrete all coupled to the same state and continua (but not necessarily at the same rate).}	\label{hybridschem}
\end{figure}

We now begin to approach the case of a general two-sided quantum network. The fully general problem is intractable analytically, but luckily there are several simplifications we can make that correspond to the network representing realistic photo detecting systems. To illustrate, consider the case of two parallel networks in series: each discrete state connected to each discrete state in the other manifold (but not necessarily at the same rate) with each of the two manifolds of purely virtually coupled discrete states coupled to their own continuum.  One could imagine generating different networks from this one by removing a coupling $g_{ij}$ between discrete states, permuting which discrete states are disconnected, removing an additional coupling, permuting, and so on. However, this is unphysical! Two discrete states in parallel cannot be prevented from coupling to the same discrete state except by selection rules, this would violate unitarity! But the discrete states are also coupled to the same continuum which has the same quantum numbers associated with it (we have removed other degrees of freedom in going to the $1D$ problem), so they must satisfy the same selection rules. The same argument applies in reverse: no discrete state within a manifold can individually stop being coupled to a continuum without the rest of the discrete states doing so as well. This argument also applies to manifolds embedded in a larger network away from any continuum, the requirement that all states satisfy the same selection rules is very strong. As a result, in our analysis we can ignore partially connected networks, they are unphysical! It also gives us a helpful way to organize discrete states, into manifolds of purely virtually coupled discrete states (after diagonalization) that are all coupled to the same set of discrete states and/or continua.

We can now focus on a very large class of quantum networks, hybrid systems consisting of manifolds in series (Fig. \ref{hybridschem}). It will be helpful to denote couplings between discrete states $i$ and $j$ or manifolds $k$ and $\ell$ as $g_{ij}^{(k,\ell)}$, and denote decay rates and discrete states within a manifold with a superscript ($\gamma_i^{(k)}$, $\Gamma_i^{(k)}$, and $c_i^{(k)}(\omega)$, with $\gamma_i^{(k)}=0$ for $k>1$ and $\Gamma_i^{(k)}=0$ for $k<M$, where $M$ is the number of manifolds). This allows us to rewrite the general equation (\ref{quantlangspect}) in a more explicit form

\bea\label{quantlangspectexp}
-i\Delta_i c_i^{(k)} (\omega)= &- \sum\limits_{j=1}^{N_k}\frac{\sqrt{\gamma_i^{(k)}\gamma_j^{(k)}} + \sqrt{\Gamma_i^{(k)}\Gamma_j^{(k)}}}{2} c_j^{(k)}(\omega)  - i \sum\limits_{j=1}^{N_{k-1}} g_{ij}^{(k-1,k)} c_j^{(k-1)}- i \sum\limits_{j=1}^{N_{k+1}}g_{ij}^{(k,k+1)} c_j^{(k+1)}\nonumber \\
&-\sqrt{\gamma_i^{(k)}}a_{\rm in}(\omega)  -\sqrt{\Gamma_i^{(k)}}b_{\rm in}(\omega)
\eea  where the superscripts denote labels of manifolds and we've implicitly defined $g_{ij}^{(0,1)} = g_{ij}^{(N,N+1)} = 0 \,\forall i,j$. We denote the number of discrete states in each manifold $N_k$ such that $\sum\limits_{k=1}^M N_k = N$. In general (\ref{quantlangspectexp}) is still only numerically solvable but we can now note two cases that yield analytic solutions.

\begin{figure}[h]
\centering
            \includegraphics[width=.8\textwidth]{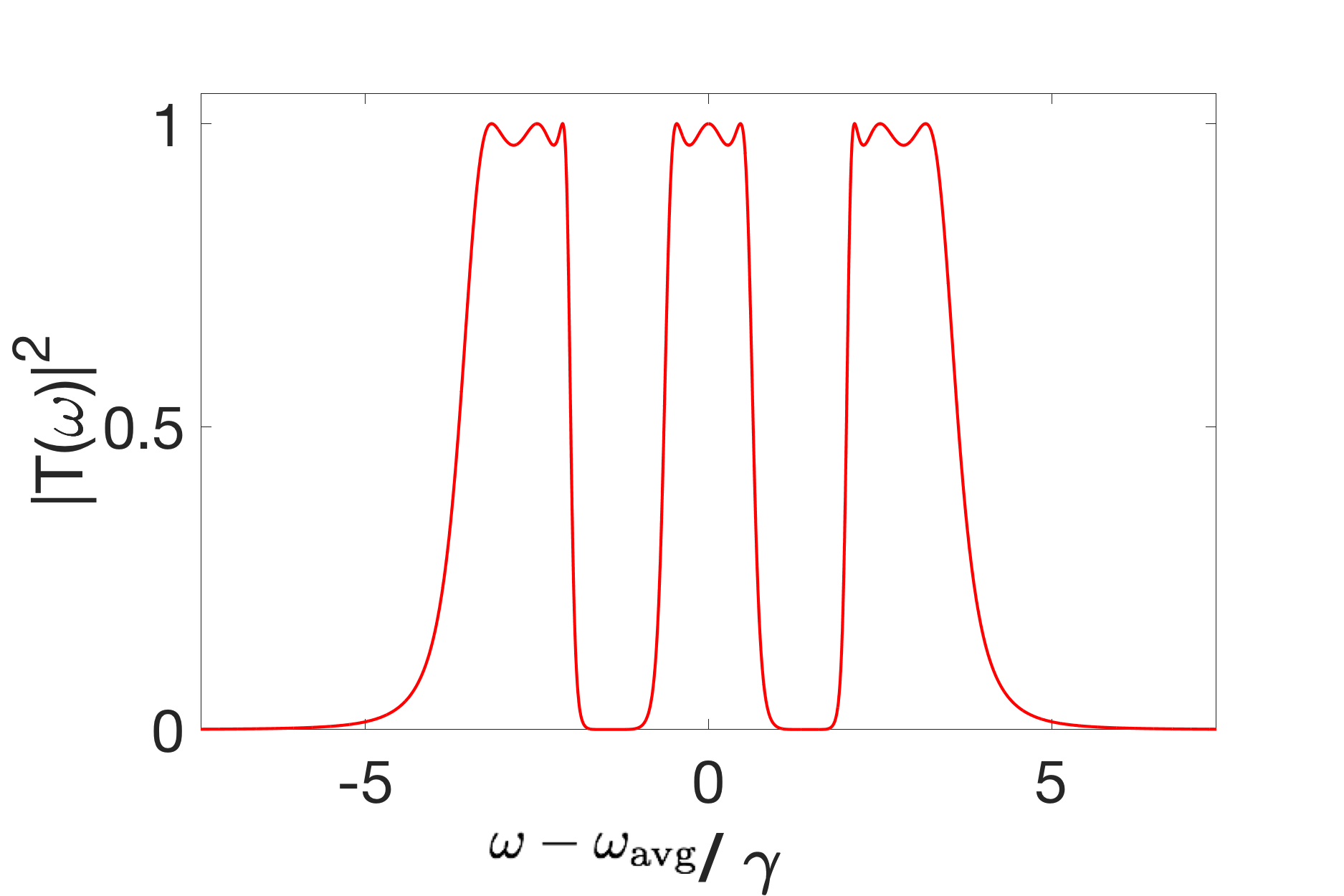}
            \caption{Transmission probability for a hybrid network consisting of $3$ manifolds in series, each with $3$ discrete states with balanced decays and uniform critical coupling. Within each manifold, the discrete states have frequency spacing by $2.5\gamma$. Using the same special conditions as for series network, perfect transmission is achieved for $9$ frequencies, with the multi-layered structure of $|T(\omega)|^2$ encoding the manifold structure.}    
            \label{HybridTrans333}
        \end{figure}
        
\begin{figure}[h]
\centering
            \includegraphics[width=.8\textwidth]{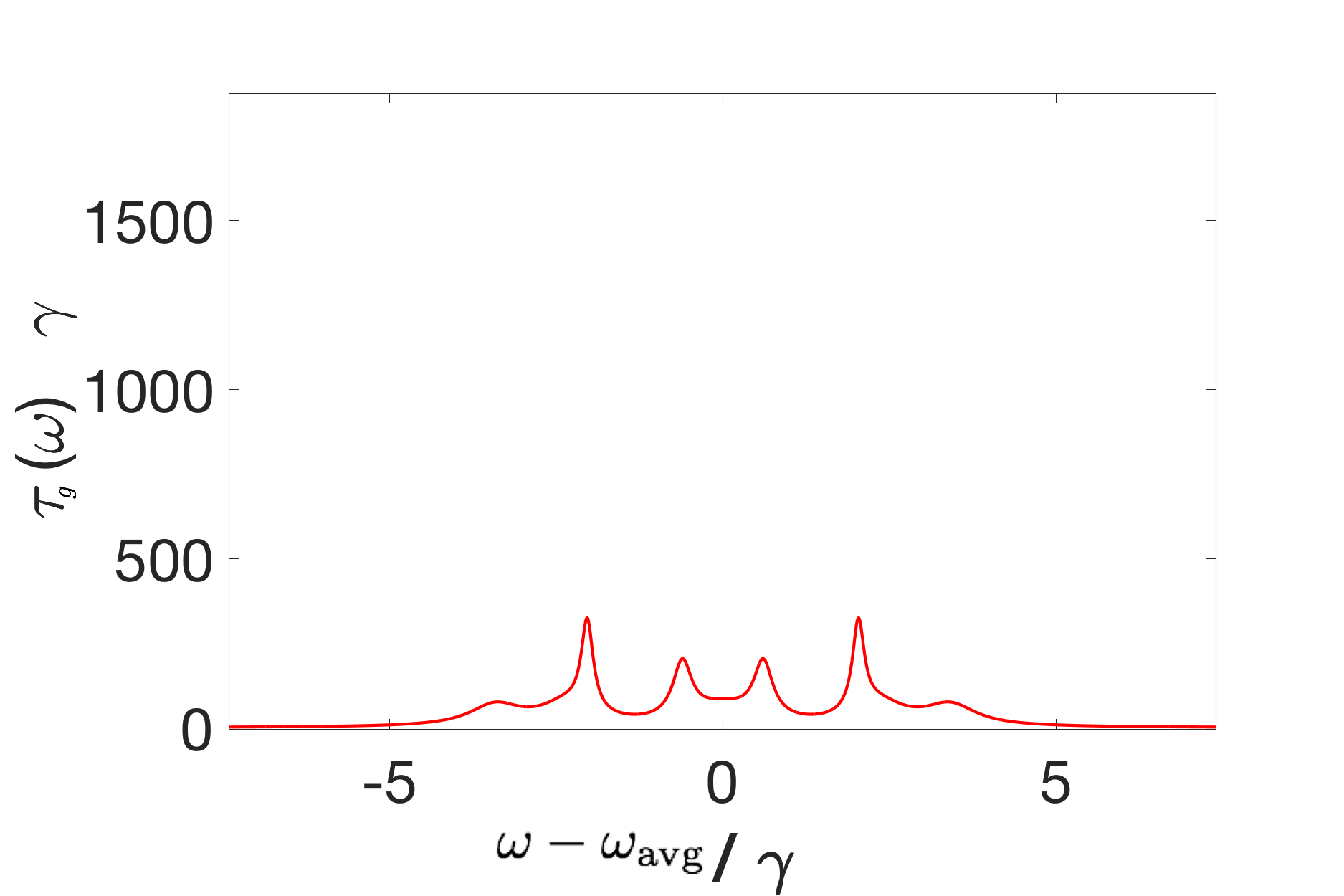}
            \caption{Group delay for a hybrid network consisting of $3$ manifolds in series, each with $3$ discrete states with balanced decays and uniform critical coupling, the transmission function of which is given in Fig. \ref{HybridTrans333}. Within each manifold, the discrete states have frequency spacing by $2.5\gamma$.}    
            \label{HybridTau333}
        \end{figure}

The first is the case of critical coupling between members of each adjacent manifold and, additionally, uniformly unbalanced decays: we first define an effective decay rate for internal couplings within the system so that $g_{ij}^{(k,\ell)}= \frac{\sqrt{\gamma_i^{(k)}\Gamma_j^{(\ell)}}}{2}$ (effectively specializing to the critical coupling case for series networks), and then consider the special case of $\Gamma_i^{(k)}/\gamma_i^{(k)}=k^{(k)}\,\forall i,k$ (inhomogeneous decays that are uniformly unbalanced within each manifold). This leads to a reflection coefficient of the form

\bea\label{refseriesc1}
&\\
R(\omega) &= 1-\cfrac{2 h^{(1)}}{h^{(1)}-i+\cfrac{k^{(1)} h^{(1)}h^{(2)}}{-i+\cfrac{k^{(2)} h^{(2)}h^{(3)}}{\dots+\cfrac{k^{(N-1)} h^{(N-1)}h^{(N)}}{\sqrt{k^{(N)}}h^{(N)}-i}}}}\nonumber
\eea where we have defined a new function $h^{(k)} = \sum\limits_{i=1}^{N_k} \frac{\gamma_i^{(k)}}{2\Delta_i^{(k)}}$. (The appearance of a lone $\sqrt{k^{(N)}}$ at the end of (\ref{refseriesc1}) is due to the final purely virtual coupling to the output continuum, since we've absorbed the rest of the decay rate into the function $h^{(k)}$.) This function encodes the zeroes and singularities we found for parallel networks, which previously gave rise to frequencies of constructive interference and completely destructive interference.

While (\ref{refseriesc1}) is tractable, it is of limited applicability to real systems. More relevant is the second solvable case of homogenous coupling and decays within manifolds: $g_{ij}^{(k,\ell)}= g^{(k,\ell)}$, $\gamma_i^{(k)}=\gamma^{(k)}$, and $\Gamma_i^{(k)}=\Gamma^{(k)}$. This leads to a reflection coefficient of the form

\bea\label{refseriesc2}
& \\
R(\omega) &= 1-\cfrac{\gamma f^{(1)}}{\frac{\gamma}{2}f^{(1)}-i+\cfrac{(g^{(1,2)})^2 f^{(1)}f^{(2)}}{-i+\cfrac{(g^{(2,3)})^2 f^{(2)} f^{(3)}}{\dots+\cfrac{(g^{(N-1, N)})^2 f^{(N-1)} f^{(N)}}{\frac{\Gamma}{2} f^{(N)}-i}}}}\nonumber
\eea where again we have defined a new function $f^{(k)} = \sum\limits_{i=1}^{N_k}  \frac{1}{\Delta_i^{(k)}}$. This provides a nice model for multi-mode systems in series (for instance, a linear network of identical multimode optical cavities; in this case, the critical coupling strength $g=\frac{\sqrt{\gamma\Gamma}}{2}$ corresponds to optical cavities perfectly coupled through their decays).

        We see in both (\ref{refseriesc1}) and (\ref{refseriesc2}) a combination of the structures we observed in  (\ref{reflectk}) and (\ref{reflecthomo}) for parallel networks and (\ref{refseries}) for series networks: correlations in amplitudes between discrete states within a given manifold manifest via a function $h^{(k)}$ or $f^{(k)}$ with $N_k$ poles and $N_k-1$ zeroes (potentially perfectly transmitted and reflected frequencies, depending on the hybrid network's resonance structure and couplings), and causal ordering of the manifolds manifests in a continued fraction structure. This latter property makes them easily analyzable using the Wallis-Euler recurrence relations, allowing us to find $R(\omega)$ and from there $T(\omega)$ and the other quantities of interest.

\begin{figure}[h]
\centering
            \includegraphics[width=.8\textwidth]{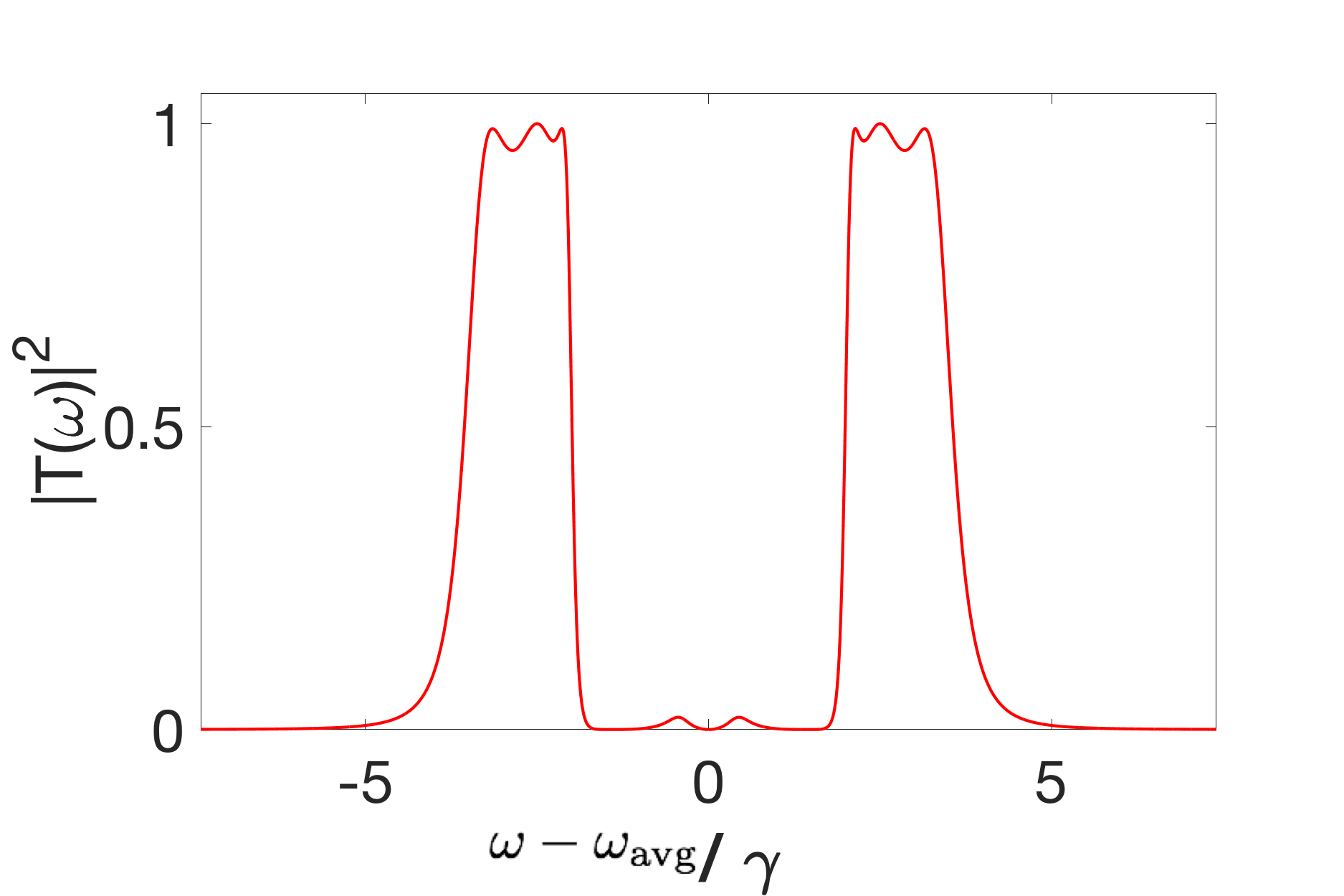}
         \caption{\small Transmission probability for a hybrid network consisting of $3$ manifolds in series, the first with $2$ discrete states and the second and third with $3$ discrete states, with balanced decays and uniform coupling ${g=\frac{\sqrt{\gamma\Gamma}}{2}}$ (the critical value for networks without detuning). Within the first manifold, the discrete states have frequency spacing $5\gamma$. Within the second and third manifolds, the discrete states are detuned by $\frac{5}{2}\gamma$, resulting in relative detuning between the manifolds. Since the couplings are no longer tuned to the critical parameters for a detuned system, perfect transmission is only achieved at $6$ frequencies.} 
                     \label{HybridTrans233p}
\end{figure}

\begin{figure}[h]
\centering
            \includegraphics[width=.8\textwidth]{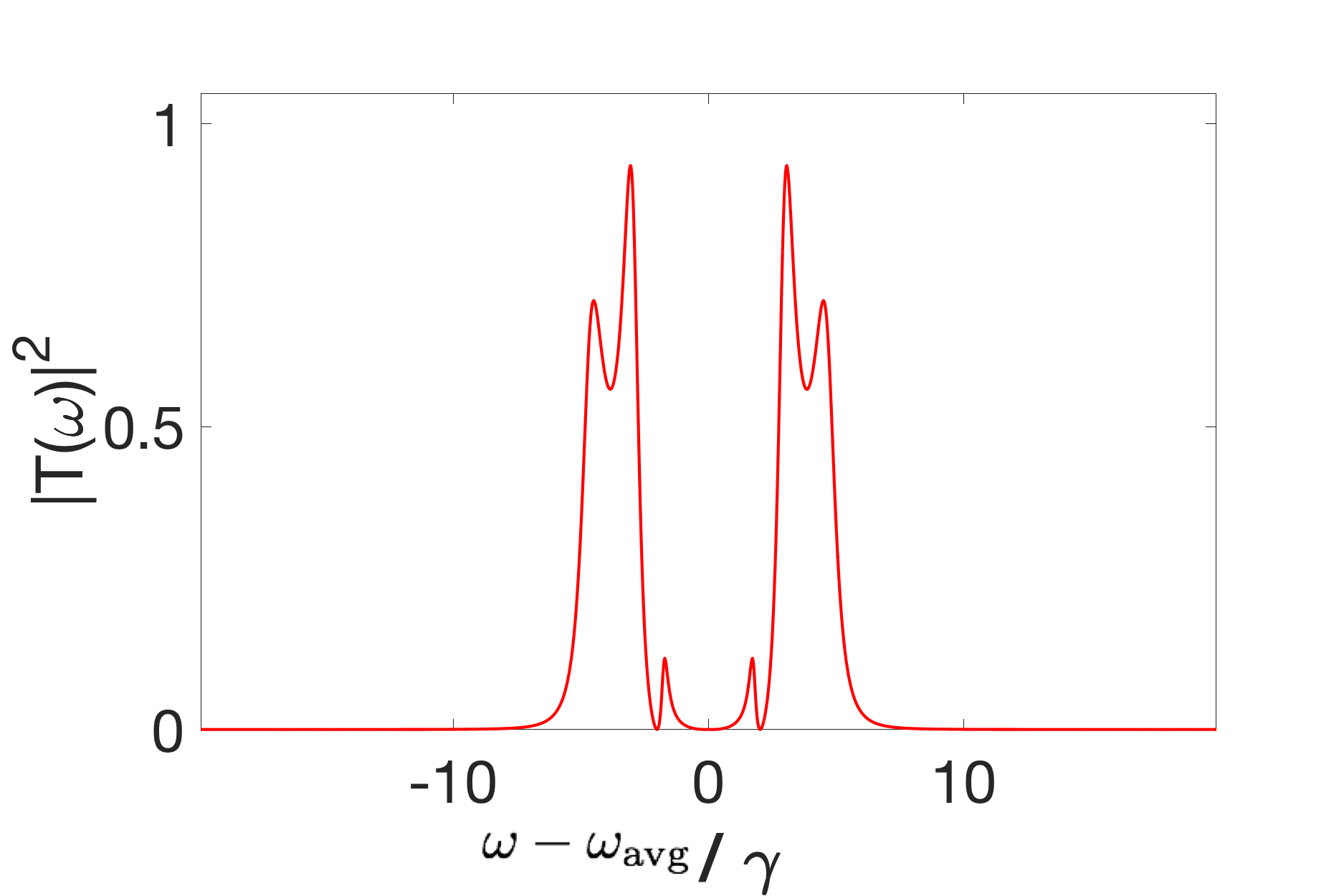}
            \caption{Transmission probability for a hybrid network consisting of $3$ manifolds in series, the first two with $2$ discrete states and the third with $3$ discrete states, with balanced decays ${\gamma=\Gamma}$ and uniform coupling ${g=\sqrt{\gamma\Gamma}}$. Within the first and second manifolds, the discrete states are spaced by ${\frac{5}{2}\gamma}$. Within the third manifold, the discrete states are spaced by $7\gamma$. Now no frequencies are perfectly transmitted, and finding couplings such that perfect transmission is achieved becomes less trivial.}    
            \label{HybridTrans223}
        \end{figure}

\lettersection{Perfect transmission} Focusing on the case of homogenous coupling, we can use the same trick of examining the convergence of an infinite series of identical manifolds around one of resonant frequencies $\omega_i$ to find the two critical conditions as we did for series networks, which again are $\gamma=\Gamma$ and $g=\frac{\sqrt{\gamma\Gamma}}{2}$. We also observe that the form of $T(\omega)$ for hybrid systems exhibits a combination of features of series and parallel networks  (Figs. \ref{HybridTrans333}, \ref{HybridTrans233p}, and \ref{HybridTrans223}). This results in layers of structure, with the small dips and peaks corresponding to intra-manifold structure layered on top of the larger dips and peaks of the inter-manifold structure. 

When at least one discrete state from each of the manifolds have the same resonance frequency, letting $\gamma=\Gamma$ and $g=\frac{\sqrt{\gamma\Gamma}}{2}$ ensures perfect transmission at $M$ frequencies. For a general hybrid network, the number of peaks of unity (perfect transmission) of $T(\omega)$ are bounded above by $M \times {\rm Min}\{N_k\}\leq N$, with $M$ the number of manifolds and $N_k$ the number of discrete states in the $k$th manifold. Here the latter equality is reached for networks that are completely in parallel or completely in series, with critical parameters in either case\footnote{The bound is somewhat stronger for critically coupled networks, with the number of perfectly transmitted frequencies bounded by ${(M-\frac{1}{2}(1+(-1)^M))\times{\rm Min}\{N_k\}}$ instead. This is due to the on-resonance broadening that occurs for critically coupled systems, which prevents the splitting of one peak to two when the number of manifolds $M$ is even.}.

\lettersection{Spectral Bandwidth} For hybrid networks where the first and last manifold have the same number of discrete states $N_1=N_M$, the spectral bandwidth is bounded above by the parallel network bandwidth for that number of discrete states $\tilde{\Gamma}\leq \frac{N_1 2\gamma\Gamma}{\gamma+\Gamma}$ with equality reached in the strong coupling limit. (When the first and last manifold have different number of discrete states $N_1\neq N_M$, the bandwidth is bounded above by $\frac{2\gamma\Gamma}{\gamma+\Gamma}({\rm Min}\{N_1,N_M\}+X)$ for a homogeneously decaying network with $X$ a network-dependent number that is always less than $1/2$.) 

\begin{figure}[h]
\centering
            \includegraphics[width=.8\textwidth]{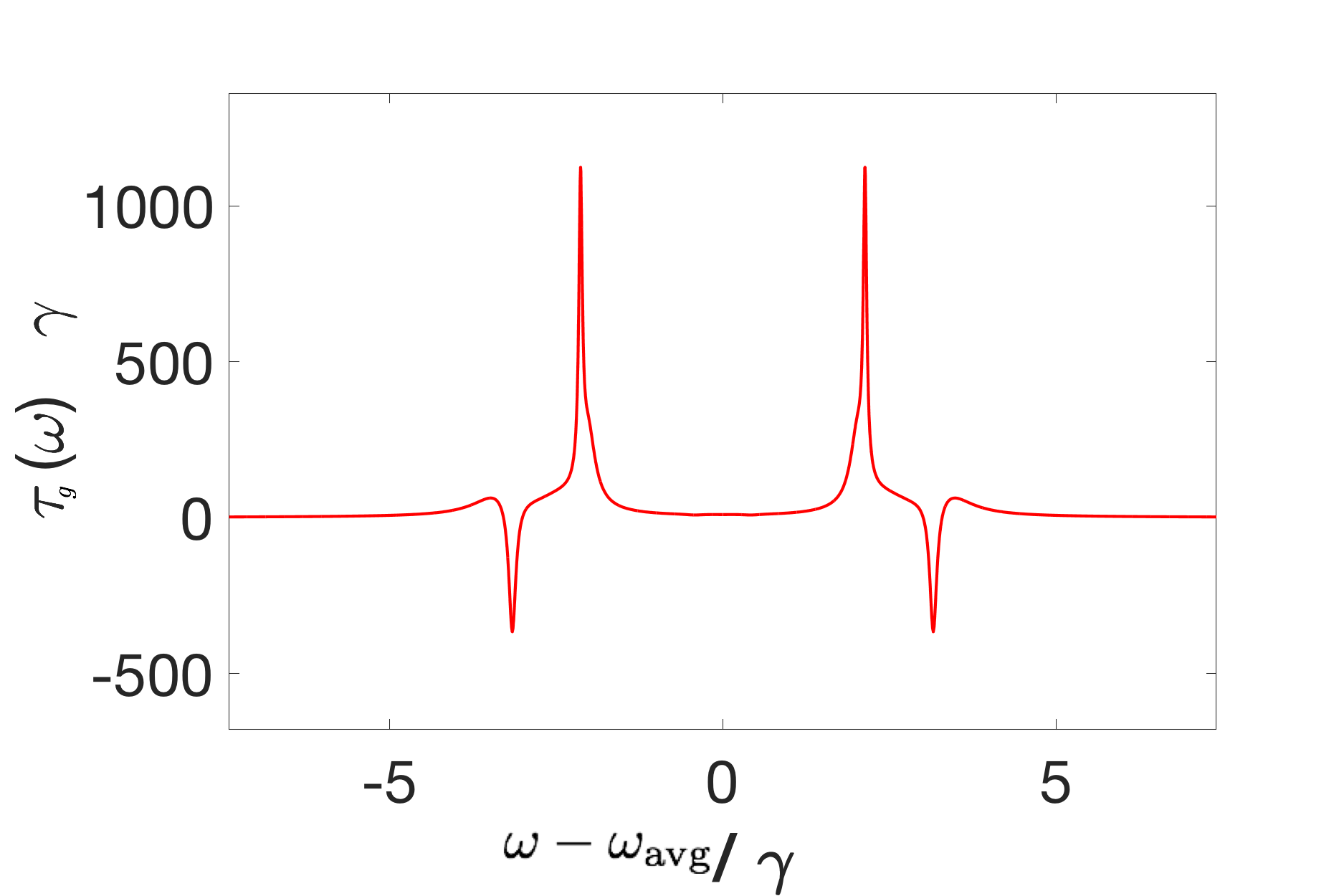}
         \caption{\small Group delay for a hybrid network consisting of $3$ manifolds in series, the first with $2$ discrete states and the second and third with $3$ discrete states, with balanced decays and uniform critical coupling.} 
            \label{HybridTau233}
\end{figure}

\begin{figure}[h]
\centering
            \includegraphics[width=.8\textwidth]{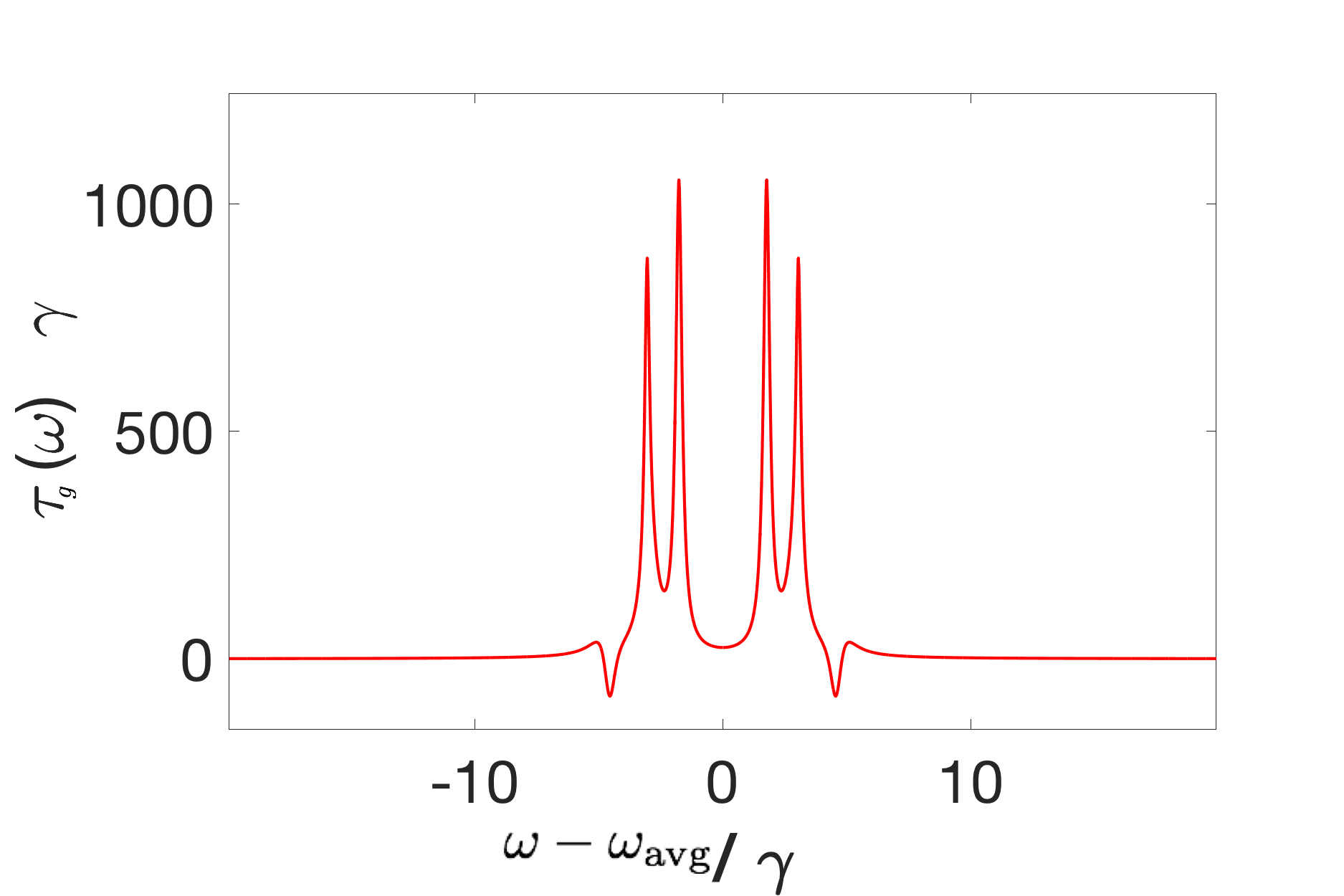}
            \caption{Group delay for a hybrid network consisting of $3$ manifolds in series, the first two with $2$ discrete states and the third with $3$ discrete states, with balanced decays ${\gamma=\Gamma}$ and uniform coupling ${g=\sqrt{\gamma\Gamma}}$, the transmission probability for which is shown in Fig. \ref{HybridTrans223}.}   
            \label{HybridTau223}
        \end{figure}

\lettersection{Group Delay} We observe the same structural properties of the group delay (Fig. \ref{HybridTau333}) as we did for other networks; the frequencies with the largest delays are those where oscillations in the transmission function are most dense. We find that networks with non-identical manifolds can give rise to group delays that are not strictly positive (Figs. \ref{HybridTau233} and \ref{HybridTau223}). 


\section{Other Networks}

\subsection{General Two-Sided Networks}

We now begin to extrapolate from the above analyses to a larger class of two-sided networks. Both the series networks and hybrid networks we've discussed have the key property of asymptotic irrelevance of the causal ordering of discrete states in the strong coupling limit. This means that the asymptotically strong-coupling behavior of these networks is entirely described by fully parallel networks as in (\ref{RGen}), and we can make several statements that will apply to any network that has the same behavior (that is, a network that resembles (\ref{RGen}) in the strong coupling limit). From the structure of (\ref{RGen}) alone, we can bound the total number of peaks and troughs of $|T(\omega)|^2$; we find there are at most $2N-1$ peaks and $2N-2$ troughs in the strong coupling limit and since the number of maxima never decreases with increasing $g$, we can extrapolate these bounds to weakly coupled systems as well. In transitioning from $2N-1$ peaks or constructive interference to $N$ peaks of perfect transmission, we observe a pair-wise merging of peaks (with one extra peak leftover when $N$ is odd). We also find that the number of dips of zero transmission (perfect reflection) is at most $\sum\limits_{k=1}^{M} (N_k - 1) \leq N-1$ where the latter equality is only reached for a completely parallel network or in the strong coupling limit.

We suspect that the above argument for upper bounds applies more generally than just to the specific networks consisting of manifolds in series; that is, there is a large class of networks that reduce to a parallel description in the strong coupling limit\footnote{It's worth reiterating here that \emph{all} networks reduce to (\ref{RGen}) in the very-weak coupling limit ${2g_{ij}\ll\sqrt{\gamma_i\gamma_j}+\sqrt{\Gamma_i\Gamma_j} \,\forall i,j}$}. However, it is unproven! We cannot generalize from the above analysis to a fully arbitrary two-sided network as there also are networks with different topologies; for instance, we can consider networks with loops of discrete states that exist outside the main chain of manifolds that connect the two continua. Since it is not necessary for a photon to pass through the loop of discrete states to make it through the network, these networks behave differently in the strong coupling limit. We can make a further distinction between disconnected loops (dead-ends where photons have to back track) and connected loops (chains of manifolds that provide an alternate route to the output continua). It is possible that connected loop networks behave more like the loop-free networks discussed above in the strong-coupling limit, but with their specific loop-structure encoded in $T(\omega)$ in unexpected ways.

\subsection{Additional Continua}

  \begin{figure}[h] 
 \centering
	\includegraphics[width=.6\textwidth]{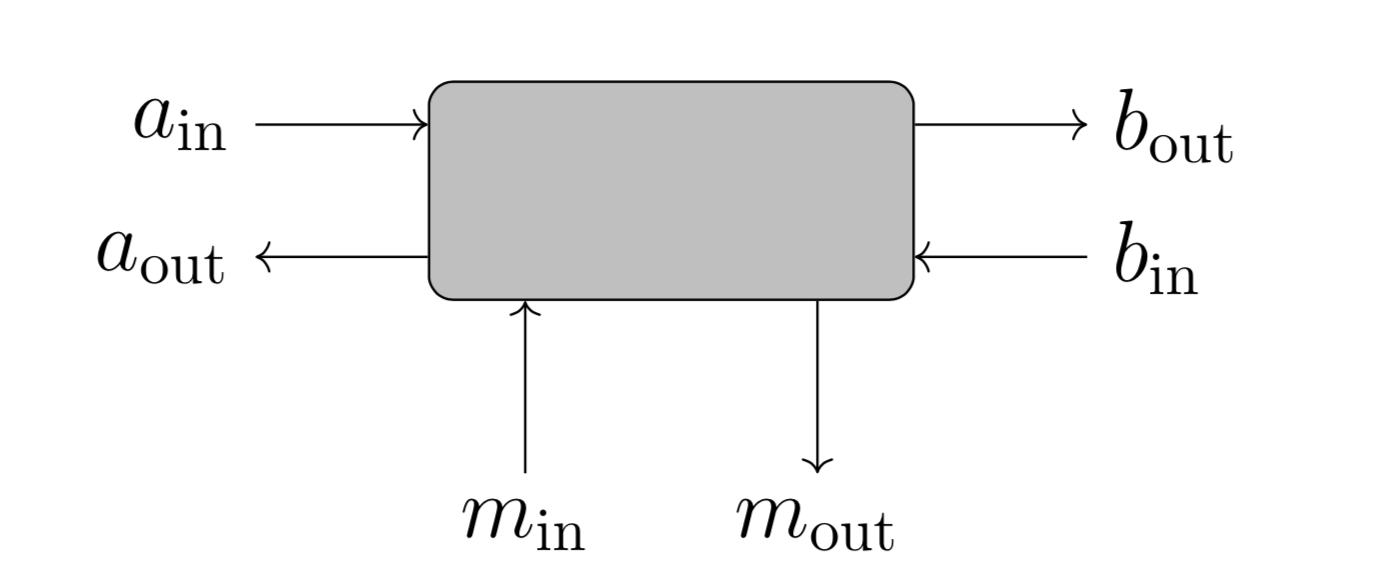} 
	\caption{Input-output field operators for a black box network with one additional side channel. This will give rise to losses that make perfect transmission impossible, as well as dark counts if the side channel contains excitations (i.e. thermal).}
	\label{3ports}
\end{figure}

We have thus far only considered quantum networks with two continnua, and have shown that perfect transmission through a general network structure is possible. We will briefly analyze more general multi-port quantum networks (Fig. \ref{3ports}) to illustrate how they lead to inefficiencies and dark counts. 

Introducing a third continuum coupled to our network of discrete states at rates $\mu_i$, we rewrite (\ref{quantlangspect}) in the form 

\bea\label{quantlangspectside}
-i\Delta_i c_i (\omega)= - \sum\limits_j\,\left(\frac{\sqrt{\gamma_i\,\gamma_j} + \sqrt{\Gamma_i\,\Gamma_j}+\sqrt{\mu_i\,\mu_j} }{2} + i g_{i\,j}\right) c_j(\omega) \nonumber \\
-\sqrt{\gamma_i}\,b_{in}(\omega)  -\sqrt{\Gamma_i}\,a_{in}(\omega)-\sqrt{\mu_i}\,m_{in}(\omega),
\eea where we have introduced a new input field annihilation operator $m_{in}(\omega)$ for the additional continuum (satisfying the canonical commutation relations), satisfying the same form of input-output relations as in (\ref{Nstatebound}) 

\bea
m_{out}(\omega) &= m_{in} (\omega)+ \sum_i\,\sqrt{\mu_i}\,c_i (\omega)\label{out fields:3}.
\eea

When there are side channels and all decay rates in the system are homogenous ($\mu_i=\mu$, $\gamma_i=\gamma$, and $\Gamma_i=\Gamma$), it is impossible to achieve perfect transmission at any input frequency without adding additional excitations (active filtering). To see this, consider taking expectation values and imposing $m_{in}(\omega)\rightarrow0$ over all frequencies. Flux conservation requires that, for a perfectly transmitted frequency $\omega'$, we also have $m_{out}(\omega')\rightarrow0$. From (\ref{out fields:3}), we see the only way to achieve this is for $\sum_i c_i(\omega') \rightarrow 0$ but from (\ref{Nstatebound}), we see that this results in $a_{out}(\omega') = a_{in}(\omega')$ and $b_{out}(\omega') = b_{in}(\omega')$---the frequency $\omega'$ is instead perfectly reflected by the system.

It is still possible that side channels with inhomogeneous coupling $\mu_i$ could yield a system that perfectly transmit some light at a frequency $\omega'$ satisfying both conditions $a_{out}(\omega') = a_{in}(\omega')-\sum_i\,\sqrt{\gamma_i}c_i(\omega') = 0$ and $\sum_i\,\sqrt{\mu_i} c_i(\omega')=0$. After all, the $c_i(\omega')$ with resonant frequencies above and below $\omega'$ generically differ in phase by $\pi$ and could in principle cancel out. However, this would require \emph{incredible} fine-tuning of the system. For a uniformly inhomogeneous parallel network such as (\ref{reflectk}) with a side channel, we can easily see that this approach fails even in the low-loss limit near resonance (where it would be most likely to succeed): we find $m_{out}(\omega_j) \approx \frac{2\mu_j}{\gamma_j+\Gamma_j} a_{in}(\omega_j)$. In general, output channels outside your control lead to loss.

Similarly, we can determine from (\ref{quantlangspectside}) and (\ref{out fields:3}) that, in general, input channels outside your control lead to extra dark counts; photons that are in initially populated side channel can end up in the monitored continuum. If the side channel is at a finite temperature, there will be thermal photons that could populate the system. The probability of this vanishes for $k_b T\ll \hbar \omega_i \,\forall i$, but will generally be non-negligible for high-temperature systems. The specific contribution depends on the specific form of the full transfer matrix, which now will include a function $D_m(\omega)$ that governs the probability of thermal excitations in the side channel to end up in the monitored continuum. 

Of course, the internal mode $\bin$ may be occupied by thermal excitations as well, but the contribution to dark counts by these will depend strongly on the amplification mechanism. Furthermore, if only frequencies such that $|T(\omega_i)|=1$ are amplified, we find that thermal excitations $\bin$ do not contribute at all to dark counts as they always leak out of the system, ending up in the continuum mode $\aout$ populated by reflected input photons.

\section{Results}

We have studied the behavior of quantum networks and have found that they provide a diverse structure of transmission functions for modeling the first stage of single-photon detectors: transmission of a single excitation from the input continuum, through the system, to a monitored output continuum. Inefficiencies and dark counts can be modeled through the incorporation of additional continua (side channels). While we do not find fundamental limits to transmission efficiency, spectral bandwidth, or frequency-dependent group delay across all the studied networks (series, parallel, and hybrid), we do observe that some networks are better suited to certain applications than others, as we will now discuss.

 \lettersection{Perfect Transmission} Often the most important metric for a photodetector is photo detection efficiency, which for a quantum network is tantamount to ensuring $|T(\omega_i)|=1$ is achieved for some frequency or frequencies $\omega_i$. The main conclusions of this chapter are as follows:
 \begin{itemize}
 \item[(i)] Ensuring the decay rates to the input and output continua are balanced ($\gamma=\Gamma$) guarantees perfect transmission for at least one frequency in almost all quantum networks without loops, side channels, and detunings  [including all not covered by (ii)]. 
  \item[(ii)] Ensuring the couplings between manifolds are uniform and critical ($g=\frac{\sqrt{\gamma\Gamma}}{2}$) guarantees perfect transmission for at least one frequency in almost all quantum networks without loops, side channels, and detunings [including all not covered by (i)].
  \item[(iii)] For an arbitrarily detuned quantum network without loops and side channels, we are \emph{always} able to find conditions for the couplings and decay rates such that perfect transmission occurs for at least one frequency.
 \end{itemize}

That finding conditions such that perfect transmission $|T(\omega_i)|=1$ is always possible indicates that perfect photo detection is possible in a wide variety of physical systems. 

 
 
 
If one additionally wants a broadened transmission spectrum at a \emph{particular} perfectly transmitted frequency---so that a broad range of frequencies is detected almost perfectly---we similarly find a variety of ways to accomplish it. One way is to use a parallel network distributed over a small range of states. However, this will result in a number of dark states which will not be detected at all. Instead, it is better to use a series network that meets both the critical coupling and balanced decay conditions, resulting in a maximally broadened on resonance transmission function as seen in Fig. \ref{CircleAndSquare}.

\lettersection{Spectral Bandwidth} We do not find fundamental limits to the minimum or maximum bandwidth of a network $\tilde{\Gamma}$ (or, conversely, to the interaction time between a network and incident light $\tilde{\Gamma}^{-1}$), but that it is generally proportional to the decay rates to both the input and monitored continua ($\tilde{\Gamma}\propto \frac{\gamma\Gamma}{\gamma+\Gamma}$, here assumed to be homogenous across states). We also find that some network structures are more suited to high-bandwidth applications than others; for networks with equivalent decay rates, we generally find that $\tilde{\Gamma}_{\rm series} \leq \tilde{\Gamma}_{\rm simple}\leq \tilde{\Gamma}_{\rm parallel}$. For a series network, equality with the upper limit  of $\tilde{\Gamma}_{\rm simple}=\frac{2\gamma\Gamma}{\gamma+ \Gamma}$ is reached only in the strong-coupling limit, and the lower limit ($\tilde{\Gamma}=0$) corresponds to a completely de-coupled or infinitely-detuned network (so that a photon can never pass through). For a parallel network, the bandwidth is always given $\tilde{\Gamma}_{\rm parallel} = \sum_i = \frac{2\gamma_i\Gamma_i}{\gamma_i+\Gamma_i}$ regardless of detuning, so the lower limit simply corresponds to a single discrete state (reproducing the simple model). Unlike both series and hybrid networks where the spectral bandwidth decreases with detuning, the spectral bandwidth of a parallel network is independent of detuning. This makes parallel networks the better candidate for implementation of broadband single-photon detection where the frequencies that need to be detected are distinct (so that spectral hole-burning is a non-issue). 

Considering hybrid networks, we find that their bandwidths are bounded above and below by parallel and series networks, respectively; given a series network with the same coupling strengths and manifold-number, and a parralel network with the same manifold resonance structure and decay rates, the bandwidth of a Hybrid network is bounded $\tilde{\Gamma}_{\rm series} \leq \tilde{\Gamma}_{\rm hybrid}\leq \tilde{\Gamma}_{\rm parallel}$. The upper limit for a hybrid network is approached only in the limit of strong coupling, and the lower limit is approached when there is only a single discrete state in each manifold. 

 \lettersection{Group Delay} The maximum magnitude of frequency-dependent group delay $\tau_g(\omega)$ increases with both the couplings between discrete states and the density of oscillations in the transmission function $T(\omega)$. In particular, the bounded-box structure of a uniformly coupled series networks yields large delays near $\omega=\pm 2g$. We observe that a negative group delay may occur for series networks with relative detuning between the discrete states. (For more on negative group delay, see Ref.~\cite{solli2002}.) For general networks, the group delay can vary immensely over the range of frequencies where $|T(\omega)|^2$ is non-negligible, with the effect being strongest for hybrid networks (where the resonance structures can be most dense). This means we can expect dispersion effects to be substantial in many photo detection platforms. For applications such as frequency discriminating delay-lines, we may expect hybrid networks to be the best performing candidate as they allow for the finest control of group delay structure.

 \lettersection{Tradeoffs} For an arbitrary series or hybrid network with arbitrary decay rates, perfect transmission requires a particular choice of the couplings. This critical value is generically of order $\sqrt{\frac{\gamma\Gamma}{2}}$. However, the spectral bandwidth is maximized when $g\gg\sqrt{\frac{\gamma\Gamma}{2}}$. Furthermore, the magnitude and location of the maximum group delays depends strongly on the coupling $g$ (and especially so in the high-$N$ limit). So for series and hybrid networks, there is a clear tradeoff between efficient transmission and the spectral bandwidth (which saturates in the high-$g$ limit)\footnote{Of course, this assumes a large spectral bandwidth is a desired. There is always an obvious tradeoff between the spectral bandwidth and the amount one learns about the incoming light; the more frequencies that can be detected efficiently, the less a single click tells you about the incoming light. Whether a high bandwidth or a low bandwidth is preferred will ultimately depend on the application of a SPD.}, with the group delay changing as well. We contrast this with the case of parallel networks where the three quantities are completely independent: perfect detection requires balanced decays ($\gamma=\Gamma$), the bandwidth can be directly scaled by scaling both decay rates together, and the frequency-dependent group delay $\tau_g(\omega)$ alone depends on the detuning between discrete states, which is an independent quantity. So for photodetectors where all three quantities must be determinable independently, parallel networks are the best choice. The exception to this are situations where a negative group delay is required, which parallel networks never exhibit. Then the use of a hybrid or series network is unavoidable, as are their accompanying tradeoffs. 

\lettersection{Final Remarks} We have studied a variety of quantum networks to uncover tradeoffs and limits fundamental to single-photon detection that may arise in the first step of photo detection (transmission). The spectral bandwidth is the only main quantity of interest where there appear to be fundamental limits for particular classes of networks (parallel, series, and hybrid), but even these scale with the relevant decay rates for the networks. Given the freedom to adjust coupling strengths and decay rates, arbitrary group delay, transmission efficiency, and spectral bandwidth are attainable for any network so long as they are attained individually (together there may be tradeoffs for series and hybrid networks).

We do not find fundamental limitations to dark counts and losses in this analysis of transmission: in general uncontrolled input channels lead to dark counts and uncontrolled output channels lead to loss. These can always be mitigated by cooling the system ($k_b T\ll\hbar\omega_i\,\forall i$) and reducing coupling to side channels ($\mu_i,\,\nu_i\ll\gamma_i+\Gamma_i \,\forall i$). These side channels can be as simple as material absorption, where a photon cab be absorbed by the material of the photodetector and contributes to a heating process and where the material of the photodetector acts as a thermal bath, generating potential dark counts.

Other sources of noise, such as from signal amplification (for more on this specifically, see Ref.~\cite{proppamp}), classical parameter fluctuations, as well as the noise inherent in a non-ideal quantum measurement (as described by an arbitrary photo detection POVM) can also be included in our model to give a fully quantum description of the entire single-photon detection process, as will be done in the coming sections. 

\section{An Additional Toy Model POVM}

In this chapter, we have not yet discussed in detail what happens to the excitation after it ends up in the output mode $\bout$, except that it can give rise to a macroscopic photodetector click. Following Ref.~\cite{spectralPOVM}, we can construct an extremely simplified POVM from $T(\omega)$ by assuming \emph{any} excitation in the output continuum will lead to a click (that is, that the continuum is monitored). This single-step POVM will not be realistic\footnote{In particular, the inclusion of a monitored continuum of states leads to a divergent dark count rate.}, but it has pedagogical value in highlighting the role of $T(\omega)$ in our work. 

Assume that we finalize the photo detection process by ascertaining at some time $t$ whether an excitation is indeed in the continuum $\bout$ and also that the amplification is ideal and lossless so that, if a photon makes it to the monitored continuum, it will be detected. We can then define normalized filtered photon states (following Ref.~\cite{spectralPOVM})

\bea\label{state}
\ket{T\,\phi_t}=\frac{1}{\sqrt{\pi\,\tilde{\Gamma}}} \int_{0}^\infty\,d\omega \,T^*(\omega)\,e^{i\,\omega\,t}\,\hat{a}^{\dagger}(\omega)\,\ket{\textnormal{vac}}. 
\eea

From the quantum jump method \cite{molmer1996,pseudomodes1}, we know that a quantum jump from the final manifold of discrete states to the monitored output continuum will occur in an infinitesimal time $dt$ with conditional probability $\frac{\tilde{\Gamma}\,dt}{2}$. We then infer the POVM element for detecting a photon at a particular time $t$ after the photodetector has been on for a time $dt$

\bea\label{povm}
\hat{ \Pi}_t=\frac{\tilde{\Gamma}\,dt}{2}\,\ket{T\,\phi_t}\bra{T\,\phi_t}.
\eea

(To reiterate, here $t$ refers to a possible time of detection analogous to the time $T$ from the previous chapter, and does not refer to the time evolution of the input state.) The probability of getting a click for a normalized input photon $\hat{\rho}$ is $\textnormal{Tr}\left(\hat{\Pi}_t\hat{\rho}\right)\leq1$. So the input state that will be detected with maximum probability is $\hat{\rho}=\ket{T\,\phi_t}\bra{T\,\phi_t}$, yielding an infinitesimal probability of detection of $\frac{\tilde{\Gamma}\,dt}{2}$. This is because this assumes that the detection events at times separated by a time $dt$ correspond to different measurement outcomes which is highly idealized; any realistic photo detection outcome corresponds to detection within an integrated time-window. Nonetheless, (\ref{povm}) provides a POVM description of a continuous measurement process, and can be used to construct a partition of unity.

To take a finite time-window into account, we consider a time-integrated POVM element $\hat{ \Pi}_\tau$, where a click corresponds to an excitation entering the final output continuum sometime between $t=T_0$ when we first turn on the detector and $t=T_0+\tau$. Then the time-integrated POVM element is
\bea\label{povm2}
\hat{ \Pi}_\tau=\int_{T_0}^{T_0+\tau} \,dt\,\frac{\tilde{\Gamma}}{2}\,\ket{T\,\phi_t}\bra{T\,\phi_t}\nonumber\\
 \approx \int_{-\infty}^{\infty}\,d\omega\,|T(\omega)|^2 \ket{\omega}\bra{\omega} \,\,\,\,\,\,(\tau\gg\tilde{\Gamma}^{-1})
\eea where, for $\tau\rightarrow\infty$, the projectors $\ket{\omega}\bra{\omega}$ are truly monochromatic because no timing information is obtained. With the time-integrated form in (\ref{povm2}), we can see that the conditions for perfect transmission discussed above correspond to perfect detection in the limit of $\tau\gg\tilde{\Gamma}^{-1}$ as we used previously in (\ref{povmlongtime}). Physically, a non-infinitesimal integration time $\tau$ means a photon incident on the photodetector at a time $t=-\tau$ has had sufficient time to propagate through the network before we end the integration at time $t=0$ and check whether there has been a click. This is in agreement with our observation that, for the simple model in (\ref{1state}), $\tau_g(\omega_0)=\tilde{\Gamma}^{-1} $: simply, we must at least allow enough time for an on-resonance photon to travel through the network (interact with the device) to achieve perfect detection of a monochromatic on-resonance photon. (\ref{povm2}) is also illustrative of what the POVM element represents: the information a detector click reveals about what led up to it.

We can also use (\ref{povm2}) to construct the POVM element for not getting a click in the finite time $\tau$, which is simply $\hat{ \Pi}_{0}=\hat{ 1}-\hat{ \Pi}_\tau$, such that the full POVM $\{\hat{ \Pi}_{0},\,\hat{ \Pi}_{\tau}\}$ forms a partition of unity for the relevant Hilbert space; that is, the Hilbert space spanned by single-photon states and the vacuum. (See Ref.~\cite{vanenk2017} for inclusion of general photon Fock states in the photo detection POVM.) This POVM corresponds to a photodetector that is reset after each integration time $\tau$. Similarly, one can divide up the the full detection window into $N$ time intervals $\tau_i$ without resetting the device in between so that the full POVM is $\{\hat{ \Pi}_{0},\,\hat{ \Pi}_{\tau_i}\}_{i=1...N}$, reproducing a continuous measurement POVM in the limit $\tau_i\rightarrow 0$. In this case, the measurement outcomes are not orthogonal \cite{spectralPOVM}; even if we know a photon is incident at a definite time $t'$ we cannot predict with certainty when it will be detected.

\chapter{Amplification}

In this chapter, we show that detection of single photons is not subject to the fundamental limitations that accompany quantum linear amplification of bosonic mode amplitudes, even though a photodetector does amplify a few-photon input signal to a macroscopic output signal. Alternative limits are derived for \emph{nonlinear} photon-number amplification schemes with optimistic implications for single-photon detection. Four commutator-preserving transformations are presented: one idealized (which is optimal) and three more realistic (less than optimal). Our description makes clear that nonlinear amplification takes place, in general, at a different frequency $\omega'$ than the frequency $\omega$ of the input photons. This can be exploited to suppress thermal noise and dark counts past what is possible with linear amplification up to a fundamental limit imposed by nonlinear amplification into a single bosonic mode. 

\section{Quantum amplification and noise}The fundamental relations between quantum noise and quantum amplification are most straightforwardly derived in the Heisenberg picture.
Thus, the standard way \cite{caves1982} to describe linear phase-preserving quantum amplification of a bosonic mode amplitude $a$ is through Caves' 
relation for the annihilation operator $\hat{a}$,
\bea\label{caves}
\ha_{{\rm out}}=\sqrt{G}\ha_{{\rm in}}+\sqrt{G-1}\hbd_{{\rm in}},
\eea
where $\hbd$ is the creation operator corresponding to an independent auxiliary bosonic mode $b$.
Here the input field amplitude of mode $a$ is amplified by a factor of $\sqrt{G}$, but there is a cost: extra noise arising from the additional mode $b$\footnote{This could be a mode internal to the detector.}.
If this mode contains (thermal) excitations, mode $a$ after amplification will contain excitations, too, and their number is multiplied by $G-1$. Even if mode $b$ is in the vacuum state, it still adds noise \cite{caves1982}. It is clear that this extra noise is due to the additional creation operator term proportional to $\sqrt{G-1}$ in Eq.~(\ref{caves}), but since that term is necessary so as to preserve the standard bosonic commutation relation $[\ha_{{\rm out}},\had_{{\rm out}}]=\mathds{1}$ this tradeoff between linear amplification and added noise is fundamental. Indeed, phase-preserving linear amplification in proposed number resolving platforms using superconducting qubits have noise bounded by the Caves limit \cite{metelmann2014,clerk2010} and similar bounds are found for linear amplification used in quantum circuits \cite{yurke2004}.

Recently, there has been some effort to describe  all parts of the photodetection process, including amplification \cite{yang2019},  
fully quantum mechanically \cite{young2018b,young2018,vanenk2017,dowling2018}. 
One conclusion that may be drawn from that research is that there is no severe amplification-driven tradeoff between efficiency and (thermally induced) dark counts. In particular, even though a few-photon signal must be amplified to a macroscopic level [forcing us to consider $G\gg 1$], thermal fluctuations in internal detector modes apparently do not get amplified by the same factor of $G$. Experiments \cite{marsili2013,wollman2017uv}  on superconducting nanowires demonstrate that over a wide range of detected wavelengths dark count rates can indeed be extremely low (on the order of one dark count per day). How can we reconcile these results with that of the previous paragraph? 

The answer, as we will show, is that amplification  is not necessarily linear. That is, in the Heisenberg picture, the transformation of the bosonic annihilation operator can be nonlinear while still preserving the bosonic commutation relation.
And, perhaps surprisingly, that way of amplifying can decrease the amount of noise added.
Previous models of nonlinear amplification were developed in a different context, that of amplifying superpositions of different number states \cite{yuen1986,ho1994,yuen1996,dariano1992,dariano1996,bjork1998}. The relevant figure-of-merit there was the signal to noise ratio of the output relative to that of the input, and noise from the auxiliary modes was ignored. Here, in contrast, we are interested exclusively in the noise in the output and especially in the contributions arising from thermal fluctuations in the auxiliary modes, as functions of the gain factor $G$.
In order to explore this issue, we construct physical transformations that implement nonlinear amplification satisfying the correct mathematical requirements which, to the authors' knowledge, have not appeared in the literature. 

\section{Nonlinear amplification}The idea is that for detecting single photons it is sufficient  to have an output field whose total  number of excitations is given by $N_{\rm out}=N_{\rm in}+G n_a$ with $n_a$ the number of input photons we would like to detect, and $N_{\rm in}$ the (fluctuating) number of excitations present in the output mode prior to amplification, which is itself {\em not} amplified.
A physically allowed but highly idealized unitary transformation that accomplishes this is easiest 
written down in the Schr\"odinger picture as \bea\label{simple}
\ket{n}_a \ket{M}_1\ket{N}_2 \longmapsto
\ket{n}_a\ket{M-Gn}_1\ket{N+Gn}_2.
\eea This is valid for any number $n$ of input photons\footnote{The transformation in Eq.~(\ref{simple}) can be realized only when ${M\geq Gn}$. 
There is always such a restriction on amplification relations; the energy transferred to reservoir 2 must come from somewhere.}, even though in practice we will be interested mainly in small values of $n$, say, $n=0,1,2$).

All states here are number (Fock) states of bosonic modes and as written this transformation is superposition-preserving.
The transformation involves two energy reservoirs: energy is transferred from the first  reservoir to the second with the amount of energy transferred determined by the number $n$ of input photons in mode $a$ (with nothing happening at all when $n=0$).
  The assumption is that excitations of the two reservoirs have identical energies, $\hbar\omega'$, such that energy is conserved. The input mode can have any frequency $\omega$.
The second reservoir ideally starts out with $N=0$ excitations---corresponding to the zero temperature limit---such that in the end it would contain exactly $Gn$ excitations if the input field contained $n$ photons. Clearly, this ideal transformation would represent perfect (noiseless) amplification of a photon number state (and $G$ will have to be an integer for this to work).

  \begin{figure}[h] 
  \centering
	\includegraphics[width=.8\linewidth]{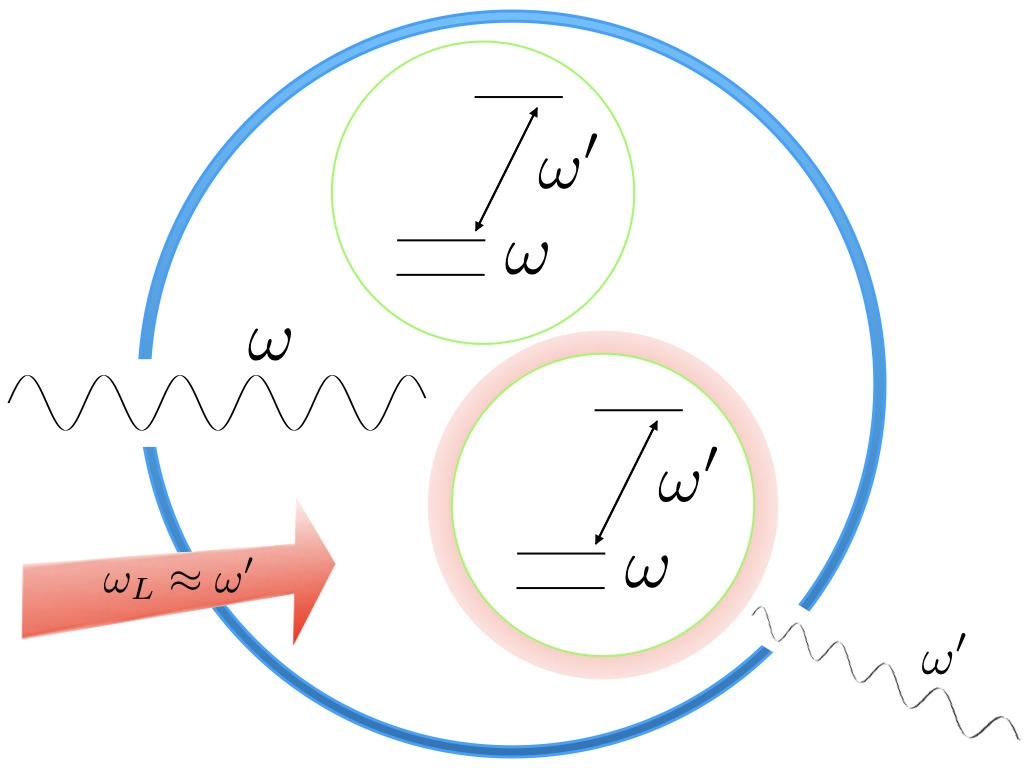} 
	\caption[]{An input photon with frequency $\omega$ undergoes amplification into a macroscopic signal via electron-shelving \cite{dehmelt1975,wineland1980,bergquist1986}: when an on-resonance photon is absorbed, an atom (modeled here as a three-level system) enters the first excited state. This is followed by a laser beam tuned to the second transition frequency ${\omega_L\approx \omega}'$, which induces fluorescence. If there are multiple input photons, they are absorbed by multiple atoms and the fluorescence signal is increased proportionally. The number of fluorescence modes may be reduced by placing a high-Q cavity around the atom so that amplification moves towards the ideal transformation given in Eq.~\ref{simple}.}
	\label{idealamp}
\end{figure}

Now we wish to describe this ideal process in the Heisenberg picture so as to make a direct comparison with Eq.~(\ref{caves}).
In that picture, the ideal transformation that is linear in the number operator for the excitations in the second reservoir should be\footnote{\label{construction}The construction of physically implementable transformations such as Eq.~(\ref{simpleb}) (and transformations including higher powers of the input photon
number operator) are highly constrained by two conditions: that the spectrum of the operator-representation
be  $\mathbb{N}^0$ (the natural numbers including zero) and that the
commutator be preserved \cite{ho1994}.  From these conditions, in Eq.~(\ref{simpleb}) we need \emph{at least} one term with a number operator on the right with a prefactor of one. Additional reservoirs with arbitrary prefactors are allowed but they will carry additional noise and decrease the SNR.}
\be\label{simpleb}
\hbd_{\rm out}\hb_{\rm out}=\hbd_{\rm in}\hb_{\rm in}+G\had_{\rm in}\ha_{\rm in}.
\ee
We are now going to do three things: (A) we will construct expressions for $\hbd_{\rm out}$ and $\hb_{\rm out}$ 
such that Eq.~(\ref{simpleb}) is reproduced and such that their commutator 
$[\hbd_{\rm out},\hb_{\rm out}]=\mathds{1}$;  (B)
we will add non-ideal features that make the model more realistic, and (C) we will include fluctuations in the initial number of excitations in the $b$ mode and calculate the signal-to-noise ratio (SNR) for the final number of excitations in the $b$ mode, both for the ideal limit and the more realistic models. A comparison with linear amplification will then show how nonlinear amplification improves upon the former.

To start with part of task (B), we adjust the idealized Schr\"{o}dinger picture to get rid of two features that make the process in Eq.~(\ref{simple}) obviously inapplicable to real detectors, but such that the Heisenberg picture Eq.~(\ref{simpleb}) is still valid. 
First note that the $n$ photons in the process in Eq.~(\ref{simple}) are not destroyed, whereas in a standard detector they are. We fix that by introducing another quantum system $S$ with a continuum of energies $E$ that can absorb the energy $n\hbar\omega$ of the $n$ photons. This modifies Eq.~(\ref{simple}) by adding a step (potentially at a later time)
\be\label{dump}
\ket{n}_a\ket{E}_S
\longmapsto\ket{0}_a\ket{E+n\hbar\omega}_S.
\ee
Since this extra step does not affect the state of reservoir 2, the crucial expression Eq.~(\ref{simpleb}) is unchanged when the continuum of energies absorbing the energy is traced over. The second change concerns phase: in the Schr\"{o}dinger picture we can insert random phase factors $\exp(i\phiclub)$ on the right-hand side of Eq.~(\ref{simple}). Critically, this is not a global phase but a local phase factor, that will be different even for the same Fock state. Thus, the amplification process destroys superpositions of different number states 
(e.g., coherent states will not be coherently amplified) and the process is now also irreversible (as any amplification process in a real detector is). It destroys any entanglement between the different modes as well\footnote{Phase randomization is necessary for optimal amplification and measurement of photon number due to number-phase uncertainty. Indeed, amplification of photon number deamplifies phase and vice versa, see \cite{dariano1996}.}.

For task (A) we would like to use the polar decompositions of the creation and annihilation operators. That is, in analogy to the polar decomposition of a complex number, $z=\exp(i\phi)\sqrt{|z|^2}$, we would like to write
\bea\label{bb}
\hat{b}_{{\rm out}}&= \hat{S}\,\sqrt{(\hbd\hb)_{{\rm in}}+G(\had\ha)_{{\rm in}}},
\eea
where $\hat{S}$ is a unitary operator which could be written in the suggestive form $\exp(i\hat{\phi})$ for some hermitian operator $\hat{\phi}$.
In a finite-dimensional Hilbert space of dimension $s+1$ there is no problem defining $\hat{S}$: it is a  shift operator that acts on number states $\ket{N}$ of the bosonic mode as
\bea\label{S}
\hat{S}\ket{N}=e^{i\phiclub}\ket{N-1}\,\,\,{\rm for}\,\,s \geq N>0,
\eea
with $\hat{S}\ket{0}=\ket{s}$ and $\phiclub$ the random phase we introduced earlier.  Since Fock space is  infinite-dimensional, we 
use the Pegg-Barnett trick \cite{pegg1989} 
of truncating the Hilbert space at a high excitation number $s$ and only in the end (when calculating physical quantities) taking the limit $s\rightarrow\infty$. 
It is easy to verify that the relation Eq.~(\ref{bb})
yields the commutator $[\hb_{{\rm out}},\hbd_{{\rm out}}]=\mathds{1}_{{\rm in}} - (s+1) \ket{s}\bra{s}$, in which the extra Pegg-Barnett term won't contribute to any physical quantity, while ensuring a traceless commutator, necessary in finite dimensions\footnote{Note the dimension of both input and output mode Hilbert spaces is ${s+1}$; they necessarily match in the Heisenberg picture.}. 

Construction of Hamiltonians that implement nonlinear photon number amplification was done elsewhere (see \cite{dariano1992,bjork1998}), but they are far more complicated than the Hamiltonians implementing linear amplification; in the interaction picture phase-preserving linear amplification is implemented with the simple interaction Hamiltonian
\bea\label{linHam}
\hat{H} = i\hbar \kappa (\ha \hb - \had \hbd)
\eea 
which amplifies via two-mode quadrature squeezing (as introduced in Ref.~\cite{caves1985}) with a (non-integer) time-dependent gain factor $G=\cosh(\kappa t)$ with $\kappa$ the coupling strength and $t$ the interaction time \cite{caves2012}. In contrast, an interaction Hamiltonian implementing nonlinear amplification even for a modest gain of $G=5$ has $40$ terms 
(fewer terms for smaller $G$, more terms for larger $G$) \cite{bjork1998}. Since we are only interested in the output signal and not the details of the evolution of other operators, we will not construct a system Hamiltonian; indeed, the same transformation on the output signal could be given by different Hamiltonians corresponding to different physical implementations of photon number amplification that give the same overall gain factor.

The nonlinear expression Eq.~(\ref{bb}) does not seem to have appeared in the large literature on bosonic amplification (for a review, see, e.g., \cite{clerk2010}). 
\cite{yuen1986,ho1994,yuen1996} did discuss photon-number amplifiers (especially in the high-photon number limit) decades ago, but no attempt was made there to find commutator-preserving operator equations.

\section{More realistic models for amplification}\label{Models} Continuing with task (B), 
in a more realistic description the reservoirs consist of many modes.
So, instead of having just one bosonic output mode $b$ we really should describe many output reservoir modes.
For example, we may have $G$ modes $b_k$ [recall that for non-linear amplification $G$ must be an integer] each one of which satisfies
\bea\label{bk}
\hat{b}_{k\,{\rm out}}&=\hat{S}_k\,\sqrt{(\hbd\hb)_{k\,{\rm in}}+(\had\ha)_{{\rm in}}},\,\,\,\,\,k=1\ldots G.
\eea
Here the macroscopic signal monitored and analyzed consists of the {\em sum} of all detected excitations (since each mode by itself contains just a microscopic number of excitations we cannot simply assume to be able to count those individual numbers: then we would not need amplification at all!). That is, we consider as our macroscopic output signal
\bea\label{Iout}
\hat{I}_{{\rm out}}=
\sum_{k=1}^G (\hbd\hb)_{k\,{\rm out}}
=\sum_{k=1}^G (\hbd\hb)_{k\,{\rm in}} + G (\had\ha)_{{\rm in}}.
\eea
Another extension is to ``avalanche" photodetection where one small-scale amplification event triggers the next and the process repeats, giving rise to a macroscopic signal. Iterating the transformation Eq.~(\ref{simpleb}) of single mode amplification $N$ times with a gain factor $g$ in each step gives a total gain factor $G=g^N$ and an input-output relation 
\be\label{simplebn}
(\hbd\hb)_{\rm N \,out}=\sum_{k=1}^{N}g^{N-k} (\hbd\hb)_{k\,{\rm in}} + G (\had\ha)_{{\rm in}}
\ee 
where mode $\hat{b}_k$ here contains the output of the $k$th amplification step, and the last mode $b_N$ contains the signal. 

Another extension, relevant for $n>1$, describes multiplexing: the idea is that $n$ photons are most conveniently detected by $n$ detectors that each detect one (and only one) photon, along the lines of \cite{nehra2017,yu2018}. We will not describe this model in any detail, except to state that
amplification would in that case be described by
$Gn$ modes, each containing exactly one extra excitation. 

Lastly, we combine both multi-mode and multi-step extensions above by repeating the process in Eqs.~(\ref{bk}) and (\ref{Iout}) of amplification into several ($g$) modes $N$ times, again with a total gain factor defined $G=g^n$ and an input-output relation for the macroscopic signal
\bea\label{Ioutn}
\\
\hat{I}_{{\rm out}} = \sum_{k_N=1}^G (\hbd\hb)_{k_N\,{\rm out}}
= \sum\limits_{n=1}^{N}\sum\limits_{k_n=1}^{g^n} (\hbd\hb)_{k_n\,{\rm in}} + G(\had\ha)_{{\rm in}}\nonumber
\eea
where the mode $\hat{b}_{k_n}$ is the $k_n$th mode in the $n$th step.

Note that in our nonlinear amplification models the amplified signal ends up in a different bosonic mode or modes: indeed, a photodetector typically converts the input signal (light) to an output signal of a physically different type, e.g., electron-hole pairs (which may sometimes be approximated as composite bosons; see also \cite{keldysh1968,devoret2000,laikhtman2007,combescot2007}).

\section{Number fluctuations}We turn to task (C) and calculate the noise in photon number introduced by the amplification process and by the coupling to reservoirs. 
For the reservoir we monitor, we write 
\bea\label{bvar}
\expect{(\hbd \hb)_{\rm{in}}}=\overline{n}_b;\,\expect{(\hbd \hb)^2_{\rm{in}}}=\overline{n}^2_b + \Delta n_b^2
\eea 
and make no further assumptions about its initial state.

We assume that there is some (unknown) number of photons in the input mode $a$ that we want to measure. We thus consider input states that are diagonal in the photon number basis, with some nonzero photon number fluctuations $\Delta n_a$. 
(Thanks to the randomized phase assumption we can use this assumption without loss of generality for our nonlinear models.)
So, we write
\bea\label{avar}
\expect{(\had\ha)_{\rm{in}}}=\overline{n}_a;\,\expect{(\had\ha)^2_{\rm{in}}}=\overline{n}^2_a +\Delta n_a^2.
\eea 
In the following we always assume the initial states of modes $a$ and $b$ to be independent, such that
\be
\expect{f(\ha,\had)g(\hb,\hbd)}=\expect{f(\ha,\had)}
\expect{g(\hb,\hbd)}
\ee
for any functions $f$ and $g$.

 For linear phase-insensitive amplification Eq.~(\ref{caves}), we find the following variance in the number of excitations in the amplified signal:
\bea
\label{cavesvar}
\sigma_{(\had\ha)_{\rm out}}^2 &=&G^2\Delta n_a^2 + (G-1)^2 \Delta n_b^2 +\nonumber\\
&& G(G-1) (2 \overline{n}_a\,\overline{n}_b + \overline{n}_a+\overline{n}_b + 1).
\eea
We observe that in addition to amplification of the thermal fluctuations in the auxiliary mode $b$ [the second term in Eq.~(\ref{cavesvar})], even at zero temperature where $\Delta n_b^2=0$ there is inherent noise from the amplification process itself [the second line is strictly positive for $G>1$]. 

We should also consider linear phase-sensitive amplification \cite{caves1982}, described by 
\bea\label{phase}
\ha_{{\rm out}}=\sqrt{G}\ha_{{\rm in}}+\sqrt{G-1}\had_{{\rm in}}.
\eea
Here, compared to Eq.~(\ref{caves}) the $\hbd$ term is replaced by the $\had$ term, such that the commutator $[\ha_{{\rm out}},\had_{{\rm out}}]$ is still preserved.
This gives  a variance
\bea
\label{phasevar}
\!\!\sigma_{(\had\ha)_{\rm out}}^2 &=& (6G(G-1) + 1) \Delta n_a^2 +\nonumber\\
&& 2G(G-1) (\overline{n}^2_a+\overline{n}_a+1).
\eea 
There is  again extra amplification noise  for $G>1$ [the second line], much like what we found for phase-insensitive amplification.

We compare these two results for linear amplification to the result for the nonlinear amplification process described by Eq.~(\ref{bb}). The variance in excitation number is
\bea\label{bbvar}
\sigma_{(\hbd\hb)_{\rm out}}^2 &=\Delta n_b^2 + G^2 \Delta n^2_a.
\eea 
Here the number fluctuations in the auxiliary mode are {\em not} amplified and there is no additional amplification noise either; if the input field is in a number state such that $\Delta n_a=0$, the gain factor $G$ will not effect the noise at all. This is a result of the reservoir and auxiliary mode being coupled conditionally on the presence of the input field; the transformation Eq.~(\ref{simple}) \emph{always} transfers exactly $Gn$ excitations to the auxiliary mode, leaving the noise in the auxiliary mode unchanged when $\Delta n_a=0$. In contrast, linear amplification \emph{never} adds a definite number of excitations to the auxiliary mode because the input mode and the auxiliary mode are directly coupled; even at zero temperature there is inherent noise from the amplification of the auxiliary mode amplitude by a factor $\sqrt{G-1}$ [see Eq. (\ref{caves})]. For linear amplification this is necessary so as to preserve the spectrum of the number operator [which non-linear amplification through Eq.~(\ref{simpleb}) does by construction].

 Already we are able to see that the scheme of amplification into a single mode is optimal; any transformation that would reduce the prefactor of $\Delta n_b^2$ in Eq.~(\ref{bbvar}) below unity would fail to realize a well-behaved annihilation operator (for details, see again\footnotemark[3])! 

For nonlinear amplification into many modes described by the more realistic model in Eqs.~(\ref{bk}) 
and (\ref{Iout}), we find 
\bea\label{MNvar}
\sigma_{\hat{I}_{\rm out}}^2 &=G\Delta n_b^2 + G^2 \Delta n^2_a,
\eea 
where for simplicity we assumed all reservoir modes to be independent with the same number fluctuations and we have defined $\hat{I}_{\rm out}$ as the integrated signal in Eq. (\ref{Ioutn}). This shows amplifying according to Eq.
(\ref{bk}) is suboptimal; even though it still beats both linear amplification limits Eqs.~(\ref{cavesvar})
and  (\ref{phasevar}) the noise in the reservoir modes is still amplified. Similarly, amplification using multiple fermionic degrees of freedom will be sub-optimal; a similar multi-mode description will be necessary\footnote{One way around this limitation is for the incident photons to only interact with a single symmetrized collective degree of freedom of many fermions, on which a measurement is then made. In this idealized case, this collective degree of freedom plays the role of a single bosonic mode and amplification could still be described by Eq.~(\ref{bb}) and photon number amplification is improved past the limit for linear fermionic amplification \cite{yurke2004}.}.

Defining the total gain $g^N=G$ with $N$ the number of steps, we find for our multi-step models that 
\bea\label{iter}
\sigma_{(\hbd\hb)_{\rm out}}^2=\frac{G^2-1}{g^2-1}\Delta n^2_b+G^2\Delta n^2_a
\eea
for amplification of $g$ excitations into a single mode and 
\bea\label{itermult}
\sigma_{\hat{I}_{\rm out}}^2=G\frac{G-1}{g-1}\Delta n^2_b+G^2\Delta n^2_a
\eea
for amplification of a single excitation into $g$ modes. (Note Eqs.~(\ref{iter}) and (\ref{itermult}) reduce to Eqs.~(\ref{bbvar}) and (\ref{MNvar}) for $g=G$.)

\section{Signal-to-noise ratios}We can now write down explicit tradeoff relations between amplification and number fluctuations in terms of signal-to-noise ratios for all types of amplification discussed here, for the case where the number of input photons is {\em fixed} to be $n_a$  (and so $\Delta n_a=0$).
Using the standard signal-to-noise ratio as the number of excitations in the amplified mode minus the background, divided by the standard deviation in the number of excitations, we get 
\bea\label{all}
\textnormal{SNR}_{{\rm PhaseInsensitive}} &\leq&\frac{G}{G-1}\frac{n_a}{\Delta n_b}\label{PhaseInsensitive}\\
\textnormal{SNR}_{{\rm PhaseSensitive}}  &\leq&\frac{2G - 1}{\sqrt{2G(G-1)}} n_a\label{PhaseSensitive}\\
\textnormal{SNR}_{{\rm SingleMode}}&=& \frac{Gn_a}{\Delta n_b}\label{SM}\\
\textnormal{SNR}_{{\rm GModes}}&=&\frac{Gn_a}{\sqrt{G}\Delta n_b}=\frac{\sqrt{G}n_a}{\Delta n_b}.\label{GM}
\eea
The linear amplification mechanisms have increasingly worse signal-to-noise ratios as $G$ increases\footnote{The signal-to-noise ratios Eqs.~(\ref{PhaseInsensitive}) and (\ref{PhaseSensitive}) for linear amplification become infinite at ${G=1}$ simply because there is no noise when both ${G=1}$ and ${\Delta n_a = 0}$.}, albeit saturating in the limit $G\rightarrow\infty$.
In contrast, the signal-to-noise ratios for the nonlinear amplification mechanisms improve with increasing $G$, with amplification into a single-mode performing best, with an improvement by a factor of $G$ over an unamplified signal\footnote{We find the linear dependence on $G$ resulting from single-shot single-mode amplification holds for transformations describing higher-order amplification of photon number operator, again subject to the constraints of [23].}. Amplification into $G$ modes performs worse than single mode amplification, showing an improvement over an unamplified signal by a factor of $\sqrt{G}$. In the large $G$ limit, this can be considered the well-known $1/\sqrt{G}$ reduction in the standard error; while there are $G$ noisy modes, they do each carry the signal we wish to measure so the total noise is reduced by a factor of $\sqrt{G}$.


  \begin{figure}[h] 
  \centering
	\includegraphics[width=.8\linewidth]{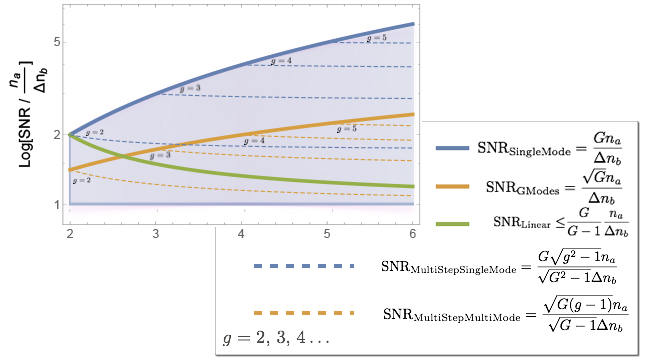} 
	\caption[]{The signal-to-noise ratio results we have derived are plotted logarithmically. Here, the linear amplification scheme plotted is the phase insensitive one (Eq. \ref{PhaseInsensitive}). The dashed lines correspond to the multi-step amplification schemes in Eq. \ref{ISM} and Eq. \ref{IMM} and perform worse than their single-step counterparts. The fundamental limit is reached only by the single mode amplification scheme (Eq. \ref{SM}) implementing ideal amplification. The trade-space is further filled in by considering additional noise modes that are not occupied by excitations deterministically.}
	\label{ampsum}
\end{figure}

Similarly, we consider multi-step amplification models
\bea
\textnormal{SNR}_{{\rm MultiStepSingleMode}}&=&\frac{G\sqrt{g^2-1}n_a}{\sqrt{G^2-1}\Delta n_b}\label{ISM}\\
\textnormal{SNR}_{{\rm MultiStepMultiMode}}&=& \frac{\sqrt{G(g-1)}n_a}{\sqrt{G-1}\Delta n_b}.\label{IMM} 
\eea 
These intermediate noise limits fill in the space between the optimal SNR in Eq.~(\ref{SM}) and linear amplification as shown in Fig. \ref{ampsum}\footnote{The space is further filled in by considering nonlinear amplification where $G$ excitations are distributed into ${G'>G}$ modes so that ancillary modes contribute only to the noise and not to the signal. In this case, we find that the SNR goes to $0$ as ${G'\rightarrow \infty}$; the effect of additional noise modes is always to reduce the SNR and move us away from the optimal SNR in Eq.~(\ref{SM}).}. (Indeed, multi-step multi-mode amplification performs slightly \emph{worse} than both linear mechanisms for $g=2$!) 

In general, nonlinear amplification enables SNRs that, as functions of the total gain factor $G$, do not saturate with the gain $G$ and outperform what is possible with linear amplification. This is most apparent in the low temperature limit where the lack of zero-temperature noise for non-linear amplification (due to amplification taking place at a different frequency) is most prominent.


  


\section{Single-photon pre-amplification}While this chapter focuses on the amplification part of the photodetection process,  we now very briefly consider the pre-amplification process. We certainly cannot treat that part in full generality here and we adopt several simplifications in order to arrive at an important result concerning the suppression of thermal noise. First, we assume that we can decouple the amplification stage from the pre-amplification filtering  [by having an irreversible step in between the two] such that filtering does not interfere negatively with the absorption/transduction part \cite{young2018}. We then focus on just the time/frequency degree of one incoming photon.  A single absorber with some resonance frequency $\omega_0$ able to absorb that single photon will act as a frequency filter. If the pre-amplification filtering is passive (easy to implement,  but we certainly can go beyond this\footnote{See, for example, \cite{raymer2010}. The result is that, instead
of certain frequencies, it is certain spectral ``Schmidt modes'' that are detected perfectly.}) and unitary (i.e., lossless: we consider this because we are interested in the fundamental limits of photodetection. Internal losses only degrade performance.), then frequency filtering is described by the linear transformation 
\be\label{transfer}
\ha_{{\rm out}}(\omega)=T(\omega) \ha_{{\rm in}}(\omega)+R(\omega)\hat{c}_{{\rm in}}(\omega)
\ee 
where $c_{{\rm in}}(\omega)$ is yet another internal bosonic detector mode at the same frequency as the input mode\footnote{
Photon-number resolved photodetection can be achieved
by multiplexing an $n$-photon signal to many (${N\gg n}$) single-photon detectors \cite{nehra2017}, each satisfying Eq.~(\ref{transfer}) independently. However, this means an additional noise mode will be added with each splitting of the signal, decreasing the integrated signal-to-noise ratio.  To avoid added noise a nonlinear multi-photon filtering process could be used, but for this a full S-matrix treatment must be used, see \cite{fan2010,caneva2015,xu2015}.}. Here $T(\omega)$ and $R(\omega)$ are ``transmission'' and ``reflection'' coefficients which
satisfy 
$|T(\omega)|^2+|R(\omega)|^2=1$ and which are determined by the resonance structures internal to the photodetector. 
The amplification process that follows the initial absorption of the photon energy is applied to
the operator $\hat{a}_{\rm out}(\omega)$ of Eq.~(\ref{transfer}), so that (explicitly displaying the different frequencies of the modes now) ideal amplification (single-mode and single-shot) is described by
\be\label{simplec}
\!\hbd_{\rm out}(\omega')\hb_{\rm out}(\omega')=\hbd_{\rm in}(\omega')\hb_{\rm in}(\omega')+G\had_{\rm out}(\omega)\ha_{\rm out}(\omega).
\ee
This makes rigorous the idea that one {\em can} amplify at any frequency, enabling the mantra that one {\em should} amplify at high (optical) frequencies \cite{Dowling}. Namely, thermal fluctuations at a frequency $\omega'$ may be suppressed by choosing the reservoir mode frequency $\omega'$ such that
$\hbar\omega'\gg kT$. This suppression is exponential: $\Delta n_b^2\propto\bar{n}_b\propto \exp(-\hbar\omega'/kT)$. 
Note that number fluctuations in the internal mode $c_{{\rm in}}(\omega)$ at the input frequency will be amplified by the subsequent amplification process.
However, one can in principle construct ideal detectors for light with a particular frequency $\omega_0$ \cite{young2018}, such that $|T(\omega_0)|=1$ and hence $R(\omega_0)=0$, avoiding internally generated dark counts at that particular frequency. 
If instead of a single monochromatic mode a small range of frequencies is amplified with differing probabilities (that is, a particular temporal mode), matching the amplification spectrum to the filtering spectrum is sufficient for reducing internally generated dark counts, as we will discuss in more detail in the next chapter. 

\section{Connections}The models for amplification considered here apply to other types of quantum measurement as well.
For example, electron-shelving \cite{dehmelt1975,wineland1980,bergquist1986} is a well-known method to perform atomic state measurements. Here one particular atomic  state (e.g., one of the hyperfine ground states of an ion) is coupled resonantly to a higher-lying excited state which can then decay back by fluorescence only to that same ground state.
A laser tuned to that transition can then induce the atom to emit a macroscopic amount (visible by eye) of fluorescent light.
In the language accompanying Eq.~(\ref{simple}), the laser beam forms the first reservoir, while the second reservoir consists of vacuum modes that are filled with fluorescent light as described by Eq.~(\ref{bk}). The gain factor $G$ (the number of fluorescence photons) is determined by the ratio of Einstein's coefficients for spontaneous and stimulated emission and the total integration time.  
By placing the atom/ion inside a high-Q optical resonator (with resonant frequency $\omega'$) we would reduce the number of output modes and thereby get closer to the optimum. 
The idea of placing a detector inside a resonant cavity is, of course, not new \cite{unlu1995}, but that idea is usually associated with increasing the coupling to light. Although we do have that effect as well, the main purpose here is to reduce the number of output modes, and thereby increase the SNR (Fig.~\ref{idealamp}). 

A method similar to electron shelving (but using a collective metastable state of an atomic gas) was proposed in \cite{imamoglu2002} for implementing high-efficiency photon counting with an, in principle,  quantum nondemolition measurement. Our model describes that detector too and clarifies why such a measurement is possible; nonlinear amplification enables low-noise photon counting and does not destroy the input photons without an additional step as in Eq. (\ref{dump}).

In \cite{yuen1986}, a transformation similar to Eq.~(\ref{simple}) is given: $n$ photons are directly converted to $Gn$ photons in a single mode. Though this transformation is unphysical (there is no way to preserve the commutator), a SNR is calculated that increases linearly with $G$ like our Eq.~(\ref{SM}). However, the SNR found in \cite{yuen1986} diverges for a photodetector with unit efficiency, which is not the case once fluctuations in the reservoir mode are properly taken into account as our results clarify.

In \cite{yang2019}, Yang and Jacob propose an interesting model for amplification that makes use of a first-order phase transition for a collection of $N$ interacting spin-1/2 particles. These spins are coupled both to an input photon and to an output bosonic mode. The SNR (as we define it here) for that model
scales as $\sqrt{N}$ while the gain $G$ of that model is linear in $N$. Thus, the SNR scales with $\sqrt{G}$ just as our Eq.~(\ref{GM}): the number of spins in \cite{yang2019}'s model plays a similar role as our number of amplification modes.


\section{Remarks}We discussed various linear and nonlinear  amplification schemes for bosonic modes. For detecting few photons, we found that the latter add considerably less noise, leading to better signal-to-noise ratios, as exemplified in Eqs.~(\ref{all})--(\ref{IMM}). Unlike for linear amplification, number fluctuations in internal detector modes are not amplified, while the number of photons that we want to detect {\em is} amplified. All amplification schemes explicitly preserve the bosonic commutation relations.

While amplification into a single-mode may not be feasible in practice, it provides the fundamental lower limit to noise in photon-number measurements across amplification mechanisms. In practice,
one may have many output modes and thus may find a SNR  closer to Eq.~(\ref{GM}), which is worse by a factor of $\sqrt{G}$ than the fundamental limit (but still better by a factor of $\sqrt{G}$ than linear amplification), or one may have multiple amplification steps as in Eq.~(\ref{ISM}), or both as in Eq.~(\ref{IMM}). To test this,
we suggest that measurement of the gain dependence of the SNR for a given photodetector should provide a rough but useful indication of the underlying amplification mechanism.

\chapter{Measurement and Applications}

The model of a SPD as an isolated two-level system is highly idealized. In a more realistic system, photodetection is an extended process wherein a photon is transmitted into the detector, interacting with the system and triggering a macroscopic change of the photodetector state (amplification) which can then be measured classically. Many theories of single-photon detection have been developed over the past century, \cite{glauber1963,kelley1964,scully1969,yurke1984,MandelWolf95,ueda1999,schuster2005,helmer2009,clerk2010,young2018,dowling2018,leonard2019} and indeed there are numerous implementations of SPD technology \cite{Allen39,mcintyer81,haroche2013,marsili2013,frog,wollman2017uv}. Across all systems, we have identified the three stages of transmission, amplification, and measurement as universal.  In the previous two chapters, we have studied amplification and transmission in isolation, discussing the tradeoffs and limits that emerge for each stage individually. In this chapter, we will combine our results with a final quantum measurement and derive a POVM that incorporates all three stages quantum mechanically, extending our model to include fluctuations of system parameters. Then, in the remainder of this chapter we will discuss applications enabled by such a photo detecting system.

\section{Realistic SPD POVMs}

The time-dependent two-level system from the previous section enabling arbitrary wavepacket projection is incorporated into the three-stage model as the trigger for the amplification mechanism. This will enable us to still project onto arbitrary single-photon wavepackets, a result we will prove at the end of this section. We will assume in this analysis that the system is left on for a sufficient time such that the subnormalization of $\Psi(t)$ is minimal and ${\cal W}\approx 1$ (\ref{calW}).

  \begin{figure}[h!] 
  \centering
	\includegraphics[width=\linewidth]{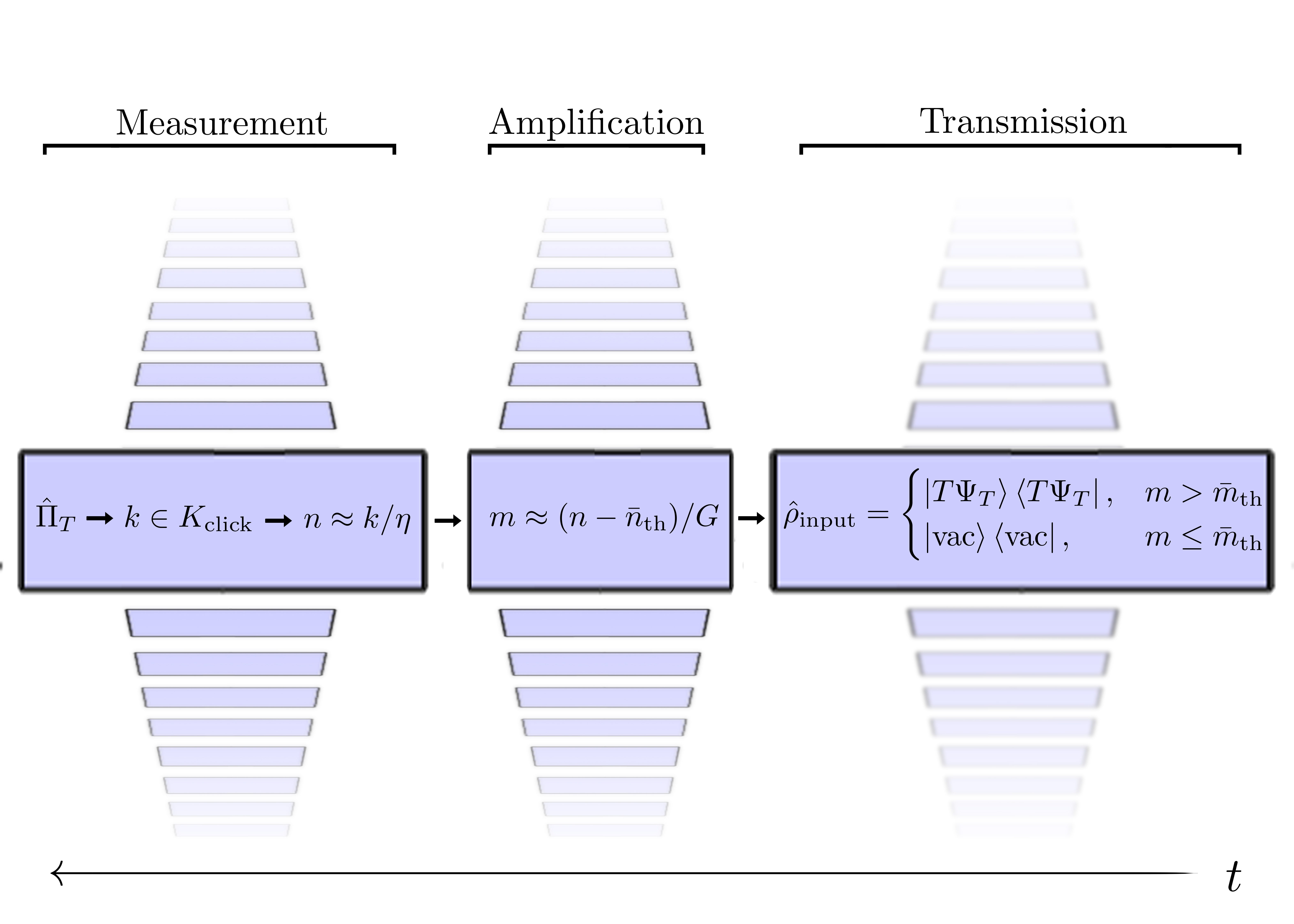} 
	\caption[]{A POVM description of the three-stage model of photodetection, where the chain of inference (left to right) moves opposite the arrow of time, connecting a macroscopic ``click'' outcome to the state of the input field. A ``click'' outcome represented by the POVM element $\hat{\Pi}_T$ indicates ${k\in K_{\rm click}}$ excitations were measured in the final measurement with detection efficiency $\eta$. This retrodicts that ${n\approx k/\eta}$ (and, strictly, ${n\geq k}$) excitations were present post-amplification, either from thermal fluctuations or an amplification process. In turn, this indicates that ${m\approx (n-\bar{n}_{\rm th})/G\leq n}$ excitations were likely incident to the amplification process trigger, with $\bar{n}_{\rm th}$ the expected number of thermal excitations already in the amplification target mode. If $m$ is larger than the expected number of thermal excitations in the amplification trigger mode $\bar{m}_{\rm th}$, we conclude that one input photon of the form $\ket{T\Psi_T}\bra{T\Psi_T}$ was likely present. In addition to the states written explicitly, there are other possible states where $k$, $n$, and $m$ deviate from their most likely values. These states (denoted by parallelograms) contribute to the POVM element, as the internal state of the photodetector is in general highly mixed. Nonetheless, in the end a SPD POVM element $\hat{\Pi}_T$ only projects onto the two input states $\ket{T\Psi_T}\bra{T\Psi_T}$ and $\ket{\rm vac}\bra{\rm vac}$, remaining relatively pure.}
	\label{timeFlip}
\end{figure}

In the spirit of what a POVM does (connect present outcomes to probabilistic statements about quantum states in the past), we will begin our endeavor at the very end of the photodetection process following Fig. \ref{timeFlip}. Consider a macroscopic measurement performed at time $T$ with a binary response triggered by $k$ excitations measured in the amplified signal\footnote{There is of course latency in any detector, but this is not interesting: one merely shifts ${T\rightarrow T+\delta T}$ in each step so that the final POVM matches the timing of a detection event to the quantum state projected onto.}. Such a POVM can be written as a projector onto Fock states in the Hilbert space internal to the system

\bea\label{clickexample}
\hat{\Pi}_{T}=\sum\limits_{k\in K_{\rm click}} \ket{k,\,T}\bra{k,\,T}. 
\eea Here, we have defined a set $K_{\rm click}$ that sets the threshold for how many amplification excitations must be measured to trigger a macroscopic detection event. At this stage, we can already see that the internal state the POVM projects onto is highly mixed, but this will not directly translate to an impure measurement on the Hilbert space of input photons. Indeed, this is what we would expect; we do not need to know \emph{precisely} the internal state of the photodetector in order to use it to efficiently detect the  presence of a single photon.

The macroscopic measurement performed on the amplified signal will, in general, be inefficient. We model this in a standard way \cite{Burnett1998}, using a beamsplitter with frequency independent transmission amplitude $\sqrt{\eta}$. We can then rewrite the POVM element (\ref{clickexample}) so that it projects onto Fock states in the amplification target mode prior to the measurement

\bea\label{kmeasurement}
\hat{\Pi}_{T}=\sum\limits_{k\in K_{\rm click}} \sum\limits_{n=k}^\infty  \Pr(n|k) \ket{n,\,T} \bra{n\,T}
\eea where we have defined 

\bea\label{kmeasureprob}
\Pr(n|k)= {n \choose k}  \eta^k (1-\eta)^{n-k}
\eea the probability to detect $k$ excitations given that there were $n$ excitations in the output mode of the amplification process. An inefficiency $1-\eta$ affects photodetection by changing which post-amplification Fock states are projected onto: the larger $1-\eta$ is, the more post-amplification states $\ket{n,\,T} \bra{n,\,T}$ will be projected onto intermediately by the same POVM element $\hat{\Pi}_T$, making it harder to distinguish between signal and noise. 

We now move one step further back in the chain of inference (Fig. \ref{timeFlip}) so the POVM element $\hat{\Pi}_T$ projects onto the number of excitations $m$ input to the amplification trigger. Amplification is a generic feature of photodetection; without a macroscopic change in the internal state of a photodetector, there is no way to correlate detector outcomes with the presence of a single photon \cite{yang2018,yang2019,zubin2020} (that is, without invoking additional single-excitation detectors in an argument \emph{circulus in probando}). There are many interesting methods for implementing amplification \cite{caves1982,imamoglu2002,yurke2004,clerk2010,metelmann2014,yang2018,yang2019,zubin2020}, but the fundamental quantum limit to amplification of any bosonic Fock state is achieved by a \sch picture transformation \cite{proppamp}
\bea\label{simpleAmp}
\ket{m}_{\rm trig} \ket{M}_{\rm res}\ket{N}_{\rm targ} \longmapsto
\ket{m}_{\rm trig} \ket{M-Gm}_{\rm res}\ket{N+Gm}_{\rm targ}\nonumber \\
\eea such that exactly $G$ excitations are transferred from the reservoir mode to the target mode for each excitation in the trigger mode. In using this expression we do impose a restriction that there must be $M>Gn$ excitations in the reservoir mode, but restrictions of this type are to be expected (the energy for amplification must come from somewhere) and we will be most interested in few photons ($n=0,1,2$) in this analysis. In most physical platforms $G$ will fluctuate\footnote{For instance, electron shelving \cite{dehmelt1975} is exactly described by (\ref{simpleAmp}) in the high-Q cavity limit when at most a single excitation is present with the laser acting as reservoir. Here the fluctuations in $G$ will be sub-Poissonian due to photon bunching.}, as will other (classical) system parameters which we will return to at the end of this section. (Exceptions do exist; for Hamiltonians that implement deterministic amplification schemes [with small integer values for $G$] see Ref.~\cite{bjork1998}.) However, even with a definite gain factor $G$ and number of input excitations $m$, we will still not end up with exactly $n=N+Gm$ excitations if the target mode is initially in a thermal state with mean occupation number $\bar{N}$ (as opposed to a Fock state with exactly $N$ excitations). We now assume this, writing the state of the target mode in the Fock basis 
\bea\label{thermalstate}
\hat{\rho}_{\rm targ}^{(th)} = \sum\limits_{N=0}^\infty  P^{\rm th}_{N,\,T}\Proj{N,\,T},
\eea with the probability for $N$ thermal excitations given by
 \bea\label{thermal}
 P^{\rm th}_{N} = \frac{1}{1+\bar{N}}\left(\frac{\bar{N}}{1+\bar{N}}\right)^N;\,\bar{N}=\frac{1}{e^{\frac{\hbar\omega'}{k_b T}}-1},
 \eea  where $\omega'$ and $k_B T$ are the frequency and the thermal energy of the target mode. Assuming the ideal amplification scheme in (\ref{simpleAmp}), we now write the POVM element $\hat{\Pi}_k$ in terms of the number $m$ excitations that trigger the application mechanism
\bea\label{PhiPOVMtraced:1}
&\!\hat{\Pi}_{T}= \sum\limits_{k\in K_{\rm click}} \sum\limits_{n=k}^\infty  \Pr(n|k)\sum\limits_{m=0}^{{ \rm Int}_- [ \frac{n}{G}]}  P^{\rm th}_{n-Gm} \ket{\Psi_T}\bra{\Psi_T}^{\otimes m} \nonumber\\
\eea where we define $\ket{\Psi_T}\bra{\Psi_{T}}^{\otimes 0} = \vacl$, $P^{\rm th}_{N}=0$ for $N<0$, and the function ${\rm Int}_-$ rounding down to the nearest integer. We can now see the benefit of having a large gain factor G; it shifts the probability distribution over $n$ that corresponds to non-zero excitations in the trigger mode, minimizing its overlap with the probability distribution for zero excitations. In this way, one can dramatically reduce the background noise (dark counts) without decreasing signal by changing the minimum value of the set $ K_{\rm click}$\footnote{The set $ K_{\rm click}$ need not have a maximum.}. In (\ref{PhiPOVMtraced:1}) we have reintroduced the state $\ket{\Psi_T}$ defined in (\ref{simpleState}) as the state projected onto by the trigger mechanism. As in the previous section, we will assume a time-dependent resonance frequency $\Delta(t)$ and decay rate $\kappa(t)$ so that arbitrary pulse-shaping is possible. 

The POVM element in (\ref{PhiPOVMtraced:1}) now projects onto quantum states \emph{internal} to the photodetector. We need to connect the internal continuum of states coupled to the amplification trigger to the external continuum containing the photons we wish to detect (the transmission stage in Fig. \ref{timeFlip}). This is accomplished by introducing a two-sided quantum network to serve as the transmission stage \cite{proppnet}. Any such network is completely described by a single complex frequency-dependent transmission coefficient $T(\omega)$ (related to a reflection coefficient at each frequency via $|T(\omega)|^2 + |R(\omega)|^2 = 1$ and $R(\omega)T^*(\omega) + R^*\omega)T(\omega) = 0$). We now invoke the single-photon assumption so that there is at most a single excitation input to the quantum network. Any other excitations present in the internal continuum will be from internally-generated thermal fluctuations reflected by the quantum network back to the trigger mechanism. In this way, we can construct a POVM element that projects onto product-states $\ket{\psi_{\rm ex}}\bra{\psi_{\rm ex}}\otimes\ket{\psi_{\rm in}}\bra{\psi_{\rm in}}$ of the external and internal continua


\bea\label{PhiPOVMprod}
\!\!\!\!\!\!\!\!\!\!\!\!\!\!\!\!\!\!\!\!\!\!\!\!\hat{\Pi}_{T}&=& \sum\limits_{k\in K_{\rm click}} \sum\limits_{n=k}^\infty  \Pr(n|k) \left( P^{\rm th}_{n} \vacl\otimes \vacl \phantom{\sum\limits_{m=1}^{{ \rm Int}_- [ \frac{n}{G}]} } \right. \nonumber \\
&+& \sum\limits_{m=1}^{{ \rm Int}_- [ \frac{n}{G}]}  P^{\rm th}_{n-Gm} \rho^{2m}\vacl\otimes\ket{R\Psi_T}\bra{R\Psi_T}^{\otimes m}\phantom{\sum\limits_{m=1}^{{ \rm Int}_- [ \frac{n}{G}]} } \nonumber \\ 
&+&\left.\sum\limits_{m=1}^{{ \rm Int}_- [ \frac{n}{G}]}  m P^{\rm th}_{n-Gm} \tau^2 \rho^{2(m-1)}\ket{T\Psi_T}\bra{T\Psi_T} \otimes\ket{R\Psi_T}\bra{R\Psi_T}^{\otimes m-1} \right). 
\eea The first line corresponds to dark counts generated from thermal excitations post-amplification and the second line corresponds to dark counts generated by thermal excitations that then trigger the amplification mechanism. Only the third line contains a projection onto a photon to be detected. (The multiplicative factor $m$ in the third line is combinatorial in origin: $m$ total excitations in the trigger mode with $m-1$ generated from thermal fluctuations.) In writing (\ref{PhiPOVMprod}) we have defined transmitted and reflected normalized single-photon states and coefficients 

\bea\label{singlephotonstate}
\ket{T\Psi_T}&=&\frac{1}{\tau} \int_{-\infty}^\infty\,d\omega \tilde{\Psi}(\omega)T^*(\omega)e^{i\,\omega\,T}\hat{a}^{\dagger}(\omega)\ket{\textnormal{vac}} \nonumber \\
\ket{R\Psi_T}&=&\frac{1}{\rho} \int_{-\infty}^\infty\,d\omega \tilde{\Psi}(\omega)R^*(\omega)e^{i\,\omega\,T}\hat{b}^{\dagger}(\omega)\ket{\textnormal{vac}}\nonumber\\
\tau&=&\sqrt{\int d\omega |\tilde{\Psi}(\omega)|^2  |T(\omega)|^2}\nonumber\\
\rho&=&\sqrt{\int d\omega |\tilde{\Psi}(\omega)|^2  |R(\omega)|^2} 
\eea where $\hat{a}^{\dagger}$ and  $\hat{b}^{\dagger}$ are the creation operators for the external and internal continua and we have defined a Fourier-transformed wavepacket for the amplification trigger mode $\tilde{\Psi}(\omega) = {\rm FT}[\sqrt{\kappa(t)}\Psi(t)]$. We can now see how pre-amplification dark counts (the second line of (\ref{PhiPOVMprod})) can be suppressed: by reducing the overlap of $|\tilde{\Psi}(\omega)|^2$ and $|R(\omega)|^2$, that is, by only amplifying the frequencies we wish to detect so that $\rho^2\ll 1$. In this case, the POVM element (\ref{PhiPOVMprod}) will be dominated by the $m=1$ term of the third line (the signal to be detected with no thermal excitations), as well as potentially the first line. (To reiterate, these are dark counts post-amplification, but these can be reduced by amplifying at a high frequency such that $\hbar\omega'\gg k_B T$, where $\omega'$ and $k_B T$ are the frequency and thermal energy of the target mode.)

 
 Finally, we trace over the internal continuum, which we assume is in a thermal state with thermal energy $k_B T'$\footnote{At my advisor's request, I would like to apologize for the overuse of the letter ``T'' in this dissertation.} so the POVM projects onto the external continua only
 
\bea\label{POVMExt}
\hat{\Pi}_{T}&=& \sum\limits_{k\in K_{\rm click}} \sum\limits_{n=k}^\infty  \Pr(n|k) \left(\sum\limits_{m=0}^{{ \rm Int}_- [ \frac{n}{G}]}  P^{\rm th}_{n-Gm} P^{'\rm th}_{m} \rho^{2m} \right.\vacl \nonumber \\
&+& \left. \sum\limits_{m=1}^{{ \rm Int}_- [ \frac{n}{G}]}  m P^{\rm th}_{n-Gm} P^{'\rm th}_{m-1} \tau^2 \rho^{2(m-1)}\ket{T\Psi_T}\bra{T\Psi_T} \right) \nonumber \\ 
&\equiv& w_0 \vacl + w_T \ket{T\Psi_T}\bra{T\Psi_T}
\eea where in the last line we have absorbed the sums in front of the two projectors into weights so that the POVM element has the form of (\ref{POVM1}) and with $P^{'\rm th}_{j} $ the probability to have $j$ excitations (now in the non-monochromatic reflected mode defined in (\ref{singlephotonstate})). For a finite detector on-time $T_0> -\infty$, the weights $w_0$ and $w_T$ will be slightly less than in (\ref{POVMExt}) due to wavepacket sub-normalization (\ref{calW}). However, this deviation is negligible provided the detector is left on for a time comparable to the temporal mode's width.

We now reconsider the question of projecting onto an arbitrary wavepacket, including the full quantum description. We find that this is possible to do in principle, \emph{provisio} $T(\omega)$ is nowhere zero (except at infinity), a result we will now prove. That is, we can ensure that the single-photon wavepacket $\ket{T\Psi_T}$ has any desired (smooth) shape and will be projected onto with a high-efficiency and high-purity measurement.

\lettersection{Proof} Consider a photon with complex normalized spectral wavepacket $\tilde{f}(\omega)$. If detection is achieved with a time-dependent two-level system preceded by a quantum network with filtering transmission function $T(\omega)$, the system will project onto a state $\ket{T\Psi_T}$ as defined in (\ref{singlephotonstate}). In the low-noise limit this will be the only state projected onto by the (pure) POVM element. From the Born rule, the probability of detection will be 
\bea\label{ProbDetectProofT}
P_T = w_T \frac{1}{\tau^2} \left | \int_{-\infty}^{\infty} d\omega \tilde{f}(\omega) \tilde{\Psi}(\omega) T^*(\omega) \right |^2
\eea with $w_T$ the overall weight given by (\ref{POVMExt}) and maximum possible detection efficiency, which can be arbitrarily close to unity. It is possible to achieve $P_T = w_T$ (mode-matched detection) in (\ref{ProbDetectProofT}) if and only if
\bea\label{ProbDetectProofSol}
\tilde{\Psi}(\omega) =\frac{\tilde{f}^*(\omega)}{T(\omega)} e^{i\omega T}. 
\eea From (\ref{timeCSolPack}), we know that it is possible to generate an arbitrary temporal wavepacket  ${\rm FT}^{-1}[\tilde{\Psi}(\omega)]=\sqrt{\kappa(t)}\Psi(t)$ from a time-dependent two-level system.  The Fourier transform of a continuous smooth function is itself smooth and continuous. Thus, if the right hand side of (\ref{ProbDetectProofSol}) is a well-defined spectral wavepacket (smooth and continuous), one can find functions $\kappa(t)$ and $\Delta(t)$ such that $\tilde{\Psi}(\omega)$ has the form of  (\ref{ProbDetectProofSol}). \qed

\lettersection{Remark} Arbitrary wavepacket detection (and thus Fourier-limited simultaneous measurements of time and frequency) is in principle possible only when there are no photonic band gaps induced by the filter; if $T(\omega')=0$ for some frequency $\omega'$, there is simply no way to compensate for the lost information about $\omega'$. Photonic band gaps are a generic feature of parallel (and hybrid) quantum networks \cite{proppnet} as well as certain non-Markovian systems \cite{Garraway2006}. Network/reservoir engineering must be employed to ensure any $\omega'$ where $T(\omega')=0$ is not a frequency of interest. 

The POVM $\{\hat{\Pi}_{T},\,\hat{\Pi}_0\}$ with $\hat{\Pi}_{T}$ defined in (\ref{POVMExt}) and $\hat{\Pi}_{0}=\idh-\hat{\Pi}_{T}$ provides a complete description of the single-photon detection process that is fully quantum from beginning to end (Fig. \ref{timeFlip}). However, there is a final element that must be considered to make the description applicable to labratory systems: classical parameter fluctuations. For continuous parameter fluctuations over any system parameter or set of system parameters $X$, these are naturally incorporated 

\begin{table}
\centering
\begin{tabular}{| c | c | c|}
\hline
Symbol& System Parameter & Fluctuations \\
\hline\hline
$K_{\rm click}$& Macroscopic Detection Threshold & $w_0,\,w_T$\\
\hline
$\eta $& Macroscopic Detection Efficiency & $w_0,\,w_T$\\
\hline
$k_B T$& Target Mode Thermal Energy & $w_0,\,w_T$\\
\hline
$\omega' $& Target Mode Frequency & $w_0,\,w_T$\\
\hline
$G$& Amplification Gain & $w_0,\,w_T$\\
\hline
$\tilde{\Psi}(\omega) $& Amplification Trigger Mode & $w_0,\,w_T,\, \ket{T\Psi_T}$\\
\hline
$k_B T' $& Internal Continuum Thermal Energy & $w_0,\,w_T$\\
\hline
$T(\omega)$& Transmission Coefficient & $w_0,\,w_T\, \ket{T\Psi_T}$\\
\hline
\end{tabular}
\caption{The effects of classical fluctuations in system parameters on the final POVM: either the weights and states are changed, or only the weights are changed. The fluctuations over the functions $\tilde{\Psi}(\omega) $ and $T(\omega)$ (and thus $R(\omega)$ by unitarity) could be caused by fluctuations in other system parameters (decay rates, resonances) internal to those functions. A subset of fluctuations in $\tilde{\Psi}(\omega) $ are fluctuations in the time of detection $T$. Importantly, these shift the wavepacket $\Psi(t)$ projected onto, resulting in a mixed measurement with larger temporal uncertainty (jitter) that depends on the ratio of the fluctuations in $T$ to the width of the temporal wavepacket.}\label{table2}
\end{table}

\bea\label{POVMFluct}
\\
\!\hat{\Pi}_{T} = \int dX {\rm Pr}(X) \left(w_0 \vacl + w_T \ket{T\Psi_T}\bra{T\Psi_T}\right) \nonumber
\eea where we have assumed a (known) probability distribution $ {\rm Pr}(X)$. In (\ref{POVMFluct}), the system parameter(s) $X$ could be such that only the weights $w_0$ and $w_T$ depend on $X$, or $X$ could be such that the state $ \ket{T\Psi_T}$ depends on $X$ as well (for a summary, see Tbl. \ref{table2}). In the case of the latter, the POVM will become less pure and will need rediagonalization to determine which states are projected onto\footnote{By varying certain key parameters, it is possible to induce an exceptional-point structure in (\ref{POVMFluct}), for instance, by introducing a discrete probability distribution over resonance frequencies corresponding to classical ignorance about a discrete set of detector settings. Here the exceptional point occurs when the frequencies are made degenerate, which (since the discrete states have identical quantum numbers) is forbidden by unitarity. Here the range over the resonances are distributed is the exceptional-point parameter \cite{Heiss2012}.}. This final POVM not only includes ignorance about the internal state of the photodetector as was depicted in Fig. \ref{timeFlip}, but also classical ignorance about the state of the photodetector due to system-lab interactions.

\section{Applications}

\begin{figure}[h!] 
  \centering
	\includegraphics[width=.8\linewidth]{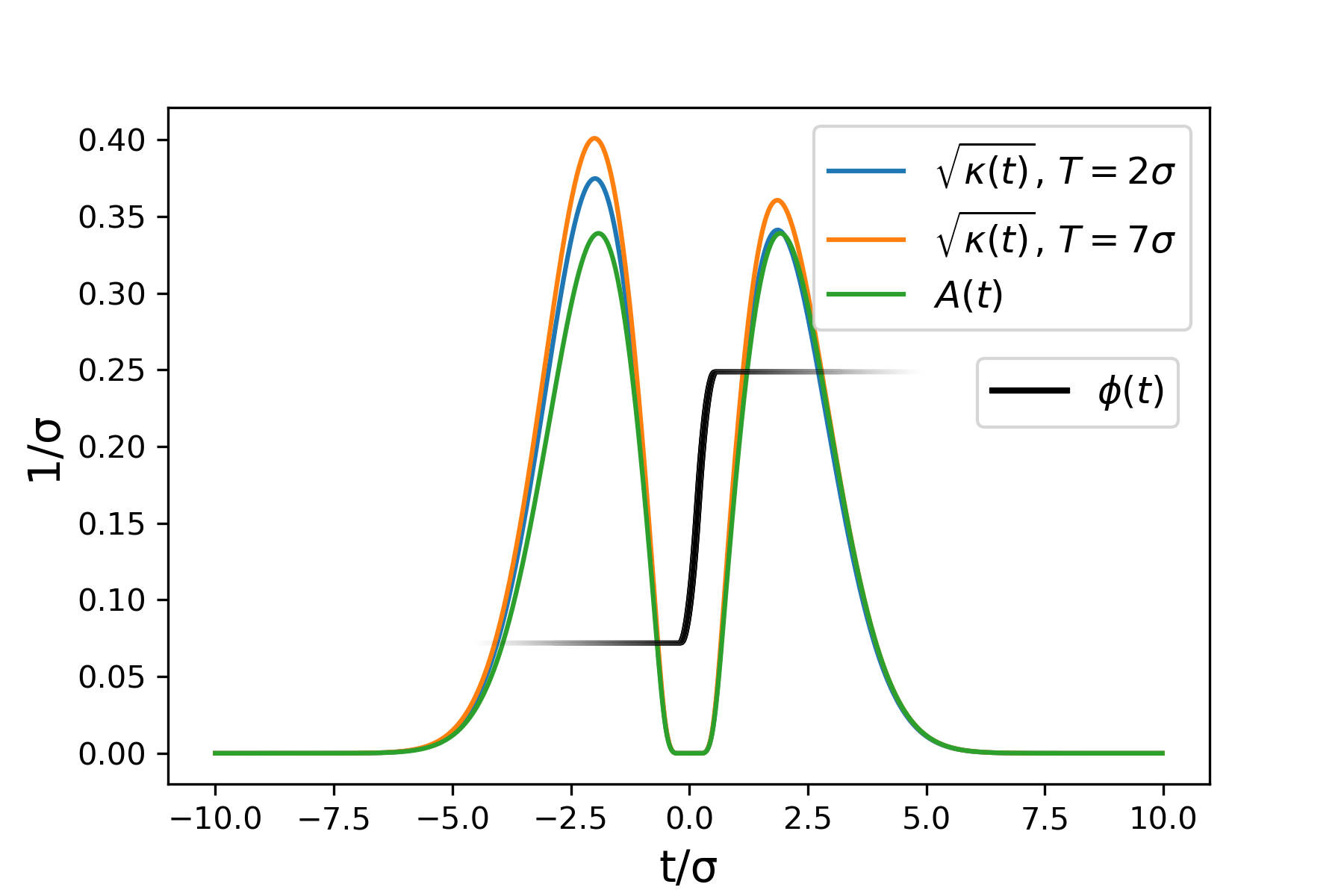} 
	\caption[]{The time-dependent coupling $\kappa(t)$ that generates a wavepacket exactly orthogonal to the minimum uncertainty Gaussian wavepacket in Fig. \ref{dataCouplingSimp} are plotted for times of detection ${T=2\sigma}$ (${{\cal W} = 0.72}$) and ${T=7\sigma}$ (${{\cal W} = 1-10^{-5}}$) and detector on-time ${T_0=-\infty}$. The wavepacket's time-dependent amplitude $A(t)$ (depicted in the rotating frame) is an approximate first-order Hermite-Gaussian pulse, where the singular region of zero as been expanded with half-width of ${z=0.5\sigma}$. In this way, the wavepacket's phase (solid black line, not to scale) can go from $0$ to $\pi$ in a finite time, whereas the Heaviside phase-flip in the exact Hermite-Gaussian pulse requires an unphysical delta-function detuning (second line of (\ref{sepExpress}). Both the phase and amplitude have been convolved with a triangular smoothing function with full-width ${s=0.5\sigma}$, ensuring $A(t)$ is continuously differentiable to first order \cite{Boehme1966,deBoor1972}, which is pre-requisite for solving the Bernoulli equation (\ref{conditionBernouli}). For any finite ${z\geq s>0}$ the phase flip [here implemented with a triangular detuning $\Delta(t)$] occurs while the amplitude is zero. This ensures exact orthogonality of the approximate first-order Hermite-Gaussian to the gaussian pulse. In the limit ${s,z\rightarrow 0}$, an exact first-order Hermite-Gaussian is recovered. This smoothing procedure generalizes to higher order Hermite-Gaussian pulses, forming a mutually unbiased basis for efficient detection of higher dimensional qudits \cite{reddy2015}.}
	\label{OrthogonalState}
\end{figure}

Using the time-dependent two-level system, we are able to project onto orthogonal quantum states (Fig. \ref{OrthogonalState}). This enables efficient detection of photonic qubits, an essential component of any quantum internet \cite{kimble2008,lukens2017}. More generally, temporal modes provide a complete framework for quantum information science \cite{reddy2015}, with efficient detection of orthogonal modes (and their superpositions to create mutually unbiased bases) a key ingredient. Fully manipulable temporal modes also play a key role in error-corrected quantum transduction \cite{vanenk1997}, where a time-reversed temporal mode can restore an unknown superposition in a qubit. Here, efficient detection of arbitrary temporal modes is essential so that quantum jumps out of the dark state are efficiently heralded. 

High-purity measurements that project onto orthogonal single-photon wavepackets also enable super-resolved measurements \cite{bonaszek2017}. Suppose we have two single-photon sources emitting almost identical pure states differing slightly in either emission time or central frequency
\bea
\ket{\tilde{\phi}_1} &=\frac{\ket{\phi_1} + \sqrt{\epsilon} \ket{\phi_2}}{\sqrt{1+\epsilon}}\nonumber\\
\ket{\tilde{\phi}_2} &=\frac{\ket{\phi_1} -\sqrt{\epsilon} \ket{\phi_2}}{\sqrt{1+\epsilon}}
\label{superres}
\eea with $\braket{\tilde{\phi}_1}{\tilde{\phi}_2}$ real, $\epsilon\ll 1$, and $\braket{\phi_1}{\phi_2}=0$.
Alternatively,  we may imagine a single source of light but the light we receive may have either been slightly Doppler-shifted or it may have been slightly delayed.

 Suppose now that we receive one photon that could equally likely be from either source so that our input state is
\bea
\hat{\rho} = \frac{1}{2} \ket{\tilde{\phi}_1}\bra{\tilde{\phi}_1} +  \frac{1}{2}\ket{\tilde{\phi}_2}\bra{\tilde{\phi}_2}. \label{inputstatemixed}
\eea If we can measure both $\hat{\Pi}_1=\eta \ket{\phi_1}\bra{\phi_1}$ and $\hat{\Pi}_2=\eta \ket{\phi_2}\bra{\phi_2}$ (that is, if we have separate photodetectors with these (pure) POVM elements, or a single non-binary-outcome photodetector), then we find the probability of clicks
\bea
P_1 &=& {\rm Tr} \left[ \hat{\Pi}_1 \hat{\rho} \right] = \eta \frac{1}{1+\epsilon} \nonumber\\
P_2 &=& {\rm Tr} \left[ \hat{\Pi}_2 \hat{\rho} \right] = \eta \frac{\epsilon}{1+\epsilon}
\label{probclicks}
\eea so that the ratio of clicks gives a direct estimate of $\epsilon$, even for low efficiency $\eta$. Here all that is needed for time-frequency domain super-resolved measurement of $\epsilon$ are SPDs with time-dependent couplings and resonance frequencies as opposed to nonlinear optics \cite{donohue2019}.

In traditional quantum key distribution (QKD) schemes (that is, \emph{not} measurement device independent (MDI)-QKD), specification of the measurement POVM is essential to robust security proofs \cite{norbert2004,qi2006,qi2009}. Here, we have verified several assumptions about an eves-dropper's capabilities common in security proofs: that high-purity measurements are possible, that high efficiency measurements are possible, and (for continuous-variable (CV)-QKD proofs) minimum time-frequency uncertainty measurements are possible. In particular for CV-QKD, an eavesdropper can perform measurements that project onto variable-width spectral modes, disrupting temporal correlations between Alice and Bob (who are assumed to use fixed time-frequency bins) \cite{bourassa2019}. Here, the capacity to adjust the width of the spectral mode $\tilde{\Psi}(\omega)$ provides Alice and Bob a new strategy to mitigate Eve's attack and extract a secure key. 

More generally, detector tomography \cite{dariano2004,lundeen2009,coldenstrodt2009,undeen2009,coldenstrodt2009,Ma2016} is an important tool across implementations of single-photon and number-resolved photodetection. Real-time tomography could be useful in QKD protocols resistant to ``trojan-horse attacks'' \cite{gisin2006} or any SPD platform subject to time-dependent environmental parameter fluctuations: for instance, atmospheric turbulence in MDI-QKD \cite{Hu2018} or interplanetary medium in deep space classical communications \cite{Banaszek2019}. Recently tomography speed-ups have been achieved using machine learning assisted tomography protocols \cite{abj2019}. The POVMs derived in this thesis provide priors which can further speed up detector tomography \cite{Heinosaari2013}. These include approximate effects of environmental fluctuations as outlined in Fig. \ref{table} and a global optimum POVM for single-photon detection (\ref{POVMExt}) which can be used to incorporate detector calibration and optimization into \emph{in situ} tomographic protocols.

\chapter{Conclusions}

Having constructed a fully quantum model incorporating the three stages of photodetection (transmission, amplification, and measurement), we can now discuss the fundamental limits and tradeoffs inherent to photodetection as a whole.

We have constructed single-photon measurements that are fundamentally limited in two ways simultaneously: the first is that they can project onto Fourier-limited (Gaussian) time-frequency states as illustrated in Fig. \ref{dataCouplingSimp}, and the second is that the amplification scheme reaches a Heisenberg-limited (linear in the gain $G$) signal-to-noise ratio, surpassing the standard quantum limit (a signal-to-noise ratio going like of $\sqrt{G}$). Achieving these simultaneously is possible in principle with no drawback. Indeed, the only stringent tradeoff we encounter in this analysis is between efficiency and photon counting rate, which becomes substantial when an SPD is reset at a faster rate than $\sim 1/\Delta t$. (The photon does not have sufficient time to excite the two-level system with high probability before the system is reset.) For other figures of merit, we find that they are either independent, or deteriorate together\footnote{For instance, inefficiency and dark counts both increase with the coefficient $\rho$ in (\ref{singlephotonstate}) when onwae considers an amplification scheme like electron shelving, where the absorption of one excitation precludes the absorption of a second.}. While it does appear from (\ref{POVMExt}) that improving efficiency also increases dark counts, these are decoupled by ensuring the coefficient $\rho\ll 1$---that is, by making $T(\omega)$ broader than $\tilde{\Psi}(\omega)$. While it is commonsense that one should only amplify the frequencies they wish to detect, our work clarifies how enormously important this is. The dark counts produced in this way are insuperable; they cannot be removed post-amplification without removing the single-photon signal as well.

One optimistic conclusion from our work is that, while there are emergent tradeoffs to consider when constructing SPDs due to their multi-stage nature, these are generally avoidable with foresight and engineering. For instance, using a parallel network allows one to simultaneously optimize transmission efficiency $|T(\omega)|^2$, spectral bandwidth (\ref{bandwidthdef}), and frequency dependent group delay (\ref{groupdelaydef}) despite the individual tradeoffs present for other network types. Using a time-dependent amplification trigger, it is possible to still achieve Fourier-limited simultaneous measurements of time and frequency provided no frequencies of interest have been removed from the signal (that is, destructive interference has not resulted in $T(\omega)=0$) as demonstrated in (\ref{ProbDetectProofSol}). While some noise is unavoidable, minimal post-amplification arbitrarily high signal can be reached by amplifying into fewer modes even at room temperature by amplifying at higher frequencies. Furthermore, detection of a particular arbitrary wavepacket changes pre-amplification dark counts; the functions $T(\omega)$ and $\tilde{\Psi}(\omega)$ both determine the projected wavepacket $\tilde{f}(\omega) = T(\omega) \tilde{\Psi}(\omega) e^{-i\omega T}$ and are interchangeable up to complex conjugation. However, they have different roles in determining dark counts. One can significantly reduce pre-amplification dark counts by ensuring that the transmission stage $T(\omega)$ has a broader spectrum than the trigger mechanism $\tilde{\Psi}(\omega)$; pre-amplification dark counts go to zero as the reflection function $R(\omega)$ and trigger mechanism spectrum $\tilde{\Psi}(\omega)$ become orthogonal.

Another conclusion from this work is also rather optimistic. Here we have given a quantum description of an entire single-photon detection process projecting onto arbitrary single-photon states and the only fundamental limitations encountered are Heisenberg and Fourier limits. Incorporating realistic descriptions of amplification and a final measurement reduce efficiency and increase dark counts, but even so a Heisenberg-limited measurement is still achievable in principle. Similarly, incorporating the filtering of a first irreversible step does not impede implementation of Fourier-limited measurements provided no frequencies are completely blocked from entering the trigger mechanism.  Even considering  parameter fluctuations (\ref{POVMFluct}) in the two internal thermal energies $k_B T$ and $k_B T'$, amplification frequency $\omega'$, and amplification gain factor $G$---which are unavoidable in any realistic system---Fourier limited time-frequency measurements are achieved. To our knowledge, this works contains the first proposed quantum procedure for reaching Fourier-limited time-frequency measurements in a realistic quantum system. This is a fundamental limit to SPD performance enforced by foundational quantum theory, and provides a limit to any precision measurements involving time and frequency. Furthermore, probing Heisenberg limits paves the way for future experimental tests of quantum theory.
\bibliographystyle{unsrtnat}
\bibliography{mybib} 
\end{document}